\documentclass[iop,revtex4]{emulateapj}
\pdfoutput=1

\usepackage{multirow}
\usepackage{color}
\usepackage{lscape}

\newcommand{\ind}[1]{_{\mathrm{#1}}}
\newcommand{\DP}{\Delta\Pi_1}
\newcommand{\Dnu}{\Delta\nu}

\slugcomment{Accepted for publication in The Astrophysical Journal, March 2013}

\shorttitle{Red Giants in Eclipsing Binaries}
\shortauthors{Gaulme et al.}

\begin{document}


\title{Red Giants in Eclipsing Binary and Multiple-Star Systems: \\
  Modeling and Asteroseismic Analysis of 70 Candidates from {\it Kepler} Data }


\author{P. Gaulme}
\author{J. McKeever}
\author{M. L. Rawls}
\author{J. Jackiewicz}
\affil{Department of Astronomy, New Mexico State University, P.O. Box 30001, MSC 4500, Las Cruces, NM 88003-8001, USA \\ \emph{gaulme@nmsu.edu}}

\and

\author{B. Mosser}
\affil{LESIA, CNRS, Universit\'e Pierre et Marie Curie, Universit\'e Denis Diderot, Observatoire de Paris, 92195 Meudon cedex, France}

\and

\author{J. A. Guzik}
\affil{Los Alamos National Laboratory, XTD-2, MS T-086, Los Alamos, NM 87545-2345, USA}




\begin{abstract}

Red-giant stars are proving to be an incredible source of information for testing models of stellar evolution, as asteroseismology has opened up a window into their interiors. Such insights are a direct result of the unprecedented data from space missions CoRoT and \textit{Kepler} as well as recent theoretical advances. Eclipsing binaries are also fundamental astrophysical objects, and when coupled with asteroseismology, binaries provide two independent methods to obtain masses and radii and exciting opportunities to develop highly constrained stellar models.
The possibility of discovering pulsating red giants in eclipsing binary systems is therefore an important goal that could potentially offer  very robust characterization of these systems. Until recently, only one case has been discovered with  \textit{Kepler}.
We cross-correlate the detected red-giant and eclipsing-binary catalogs from \textit{Kepler} data to find possible candidate systems. Light-curve modeling  and mean properties measured from asteroseismology are combined to yield specific measurements of  periods, masses, radii, temperatures, eclipse timing variations, core rotation rates, and red-giant evolutionary state.
After using three different techniques to eliminate false positives, out of the 70 systems common to the red-giant and eclipsing-binary catalogs we find 13 strong candidates (12 previously unknown) to be eclipsing binaries, one to be an non-eclipsing binary with tidally induced oscillations, and 10 more to be hierarchical triple systems, all of which include a pulsating red giant. The systems span a range of orbital eccentricities, periods, and spectral types F, G, K, and M for the companion of the red giant. One case even suggests an eclipsing binary composed of two red-giant stars and another of a red giant with a $\delta$-Scuti star.
The discovery of multiple pulsating red giants in eclipsing binaries provides an exciting test bed for precise astrophysical modeling, and follow-up spectroscopic observations of many of the candidate systems are encouraged.  The resulting highly constrained stellar parameters will allow, for example, the exploration of how binary tidal interactions affect pulsations when compared to the single-star case.

\end{abstract}



\section{Introduction}

Binary stars are prime targets to study stellar evolution: for some spectral types more than 50\% of stars are estimated to belong to binary or multiple-star systems \citep[e.g.,][]{duquennoy1991,lada2006}, and Kepler's third law allows us to retrieve their global physical parameters, which is of crucial importance to constrain stellar evolution models. It is possible to determine the masses of each component $(M_1,~M_2)$ for visual or eclipsing binaries, provided that spectral lines are detected for each star to track the Doppler shifts along their orbits, and the product $(M_1 +M_2) \sin i$ for spectroscopic-only binaries, where $i$ is the orbital plane inclination. Moreover, close-in binary systems host unique and poorly-understood physical processes such as mass exchange between the two stars.

The \textit{Kepler} satellite \citep{borucki2010} detected 2616 eclipsing binaries (hereafter EBs) by the end of January 2013, which represented about 1.4\,\% of the targets. These targets were listed by \citet{prsa2011}, then updated by \citet{slawson2011} and \citet{matijevic2012}. Within this list, binary systems were classified into four types.
(i) Detached (D), where each component's radius is smaller than its Roche lobe so stars are spherical. The stars have no major effect on each other, and essentially evolve separately. Most binaries belong to this class. 
(ii) Semi-detached (SD), where the biggest star fills its Roche lobe leading to mass exchange. The mass transfer dominates the evolution of the system. In many cases, the inflowing gas forms an accretion disc around the accretor. 
(iii) Over-contact (OC), where both stars fill their Roche lobes and are in contact. The uppermost part of the stellar atmospheres forms a common envelope that surrounds both stars. 
(iv) Ellipsoidal variation (ELV), where no eclipse is observed but the system is detected in \textit{Kepler}'s light curves through the ellipsoidal shape of the stars. In addition to the \textit{Kepler} list, \citet{coughlin2011} proposed a list of low-mass ($< 1 ~M_{\odot}$) EBs,  of which many in fact  belong to the lists of \citet{slawson2011} and \citet{matijevic2012}.

Red giants are evolved stars that have depleted the hydrogen in their cores and are no longer able to generate energy from core hydrogen burning. The physical processes taking place in their interiors are fairly poorly understood. However, the study of the global pulsations of red giants (hereafter RGs) with asteroseismology is capable of unambiguously determining bulk properties such as mass, radius, temperature, metallicity, and also the evolutionary state of RGs. Indeed, the new era of  space-based missions such as CoRoT \citep{baglin2009} and \textit{Kepler} has dramatically increased the amount and quality of the available asteroseismic data. In particular, global oscillations of several thousands of solar-like stars and RGs have been detected in \textit{Kepler} data, and  analysis of the oscillation eigenmodes now allows robust seismic inferences to be drawn about their internal structure \citep[e.g.,][]{chaplin2011, hekker2009, bedding2010, mosser2010, huber2010, bedding2011}.

Applying modern asteroseismology to EB systems for which photometric and radial velocity data exist leads to the best possible physical characterizations: this is because masses and radii may be measured in two independent ways. Such systems are cornerstones for testing stellar evolution models. Until recently, only one case of an oscillating RG belonging to an EB system had been reported \citep[KIC 8410637,][]{hekker2010}. As only one eclipse of that system was observed,  no estimate of orbital period or eccentricity could be obtained, but the global oscillations of the RG star were clearly detected. In addition, \citet{derekas2011} report the detection of a triple system containing a RG (\object{HD 181068}), and they explicitly mention that solar-like oscillations are not visible, even though most stars with similar parameters in the \textit{Kepler} database do clearly show such oscillations. In general, the \textit{Kepler} mission has succeeded in making breakthroughs in both the fields of binary stars and asteroseismology. For example, \textit{Kepler} helped reveal the presence of tidally-induced pulsations in the binary system KOI 54, which are the result of resonances between the dynamic tides at periastron and the free oscillation modes of one or both of the stars \citep{Welsh_2011}.

Since most stars are observed at a 29-min cadence with \textit{Kepler}, global modes of main-sequence solar-like stars are not accessible; however, global modes of RG stars larger than $3.5~R_\sun$ are accessible \citep[e.g., Table 3 of][]{mosser2012}. Fortunately, an RG catalog from the \textit{Kepler} Team has been compiled and made public for the scientific community.\footnote{http://archive.stsci.edu/kepler/red\_giant\_release.html} In this paper, we  establish a list of RG candidates that likely belong to EBs or multiple-star systems, which we obtain from the EB and RG public catalogs. We test whether these candidates are part of EB or multiple systems and characterize their main physical properties. Note we do not work with radial velocity measurements, for which data acquisition is in progress. We first present the results of a cross-correlation of the EB and RG catalogs  (Section \ref{sec_2}), and the subsequent detailed analysis of their light curves to determine eclipse and  asteroseismic properties  (Section \ref{sec_3}). We identify 70 RGs possibly belonging to EB systems, of which 47 show clear global oscillation modes. Mean properties of the global modes are used to infer RG masses and radii. We study several ways to determine whether the oscillating RGs actually do belong to the EBs with which they are associated, and then describe  details of several important cases in Section \ref{sec_4}. In Section \ref{sec_5} we conclude by  defining the observations that are needed to fully characterize this set of stars and discuss general implications for oscillating stars in binary systems.

\section{Data} \label{sec_2}
\subsection{\textit{Kepler} data} \label{sec_21}
All data used in this paper are photometric measurements obtained by the NASA satellite \textit{Kepler}, launched in March 2009 to search for exoplanets in the habitable zone \citep{borucki2010}. Since its launch, \textit{Kepler} has been monitoring about $156\,000$ stars in a 105~deg$^2$ field of view in between the Cygnus and Lyra constellations. Data are subdivided in {\it quarters} Q, i.e., three-month runs at the end of which the satellite is rotated by 90~deg to maintain the Sun's position on its solar arrays and to keep the radiator pointed to deep space. The commissioning quarter Q0 and quarter Q1 lasted respectively 10 and  35 days, respectively. Here, we only utilize public data that are available through Q13 as of January 2013.

Light curve data released for the public on the MAST database can be obtained in {\it raw} or {\it corrected} form. Corrected fluxes have been processed by the Presearch Data Conditioning (PDC) pipeline that removes signatures in the light curves that are correlated with systematic error sources from the telescope and spacecraft, such as pointing drift, focus changes, and thermal transients. The PDC attempts to correct for such errors while preserving planet transits and other astrophysically interesting signals. Further details of the pipeline are described by \citet{kinemuchi2012}, \citet{Stumpe_2012}, \citet{Smith_2012}, and the \textit{Kepler} handbook.\footnote{http://archive.stsci.edu/kepler/manuals/archive\_manual.pdf} We use corrected PDC fluxes in this work.

In addition, for each object there are {\it target pixel files}, consisting of the flux for all the pixels contained within a predefined mask which are used to create the data found in the photometric light curve files. Each target pixel file contains these pixels as a time series of images in a binary FITS table. We use target pixel files in Section \ref{sec_35} to determine whether the eclipsing signal we observe is from a nearby contaminating object.

The \textit{Kepler} observations are sampled at either {\it long cadence} (29.4244 minutes) or {\it short cadence} (58.89 seconds, for only 512 targets). Most targets in this study were observed  at long cadence; for the few objects that were also observed at short cadence it was never for more than 30 days. Therefore, we  work primarily with long cadence data,  and only consider the short-cadence data to search for high-frequency modes of the main-sequence or subgiant companion star belonging to the considered EB system. For long-cadence data, the Nyquist cut-off frequency is $\nu\ind{nyq} = 283.2\ \mu$Hz, which limits the possibility of performing asteroseismology on any object whose frequency at maximum amplitude $\nu\ind{max} \geq \nu\ind{nyq}$. By assuming an RG effective temperature of 4800~K and 1 $M_\odot$, asteroseismic scaling laws \citep{kjeldsen1995,huber2011,Belkacem_2011} predict that the lower limiting radius is approximately $3.5~R_\sun$.

\subsection{Red-giant and binary-system catalogs} \label{sec_22}
The {\it Kepler} team released a list of $13\, 698$ RGs on September 27, 2011, selected from the target list using color-magnitude estimates and considering effective temperatures $T\ind{eff} = 4800\pm300$ K and surface gravities $\log g = 2.9\pm0.5$ (cgs). This latter criterion cuts off RGs with predicted oscillation peak frequency larger than about 320~$\mu$Hz, slightly above the long-cadence Nyquist frequency. We note that not all of the RGs were continuously observed, and they are typically the first targets to be dropped from the survey when pixel resources become scarce (see the \textit{Kepler} red giant database). The RG \textit{Kepler} magnitudes range from 7.9 to 14.0 which makes them sufficiently bright for ground-based spectrometry to determine atmospheric parameters (as the SDSS III APOGEE/\textit{Kepler} experiment APOKASC at Apache Point Observatory is currently realizing).

\citet{prsa2011} released a catalog of EBs identified in the {\it Kepler} field from the first data releases (Q0--Q1). The catalog was motivated by the exquisite quality of {\it Kepler} data which has led to the discovery of  hundreds of new systems, revolutionized accuracy in modeling EB systems, and provided an estimate of the frequency of occurrence of EBs. Their method uses the Transit Planet Search (TPS) algorithm, first developed to detect exoplanets in the {\it Kepler} data \citep{jenkins2010a,jenkins2010b}, by adapting it to search for eclipse durations consistent with EBs. The targets that present a positive detection are then down-selected to exclude objects already identified as exoplanets, variable or spotted stars that mimic eclipse shapes, blends from background stars, or pointing jitter artifacts. The eclipses are subsequently modeled with EBAI \citep[Eclipsing Binaries via Artificial Intelligence;][]{prsa2008}, which relies on trained neural networks using synthetic eclipse profiles to extract the physical parameters. For contact systems, orbits are assumed to be circular and the main physical parameters fit are the temperature ratio $T_2/T_1$, the mass ratio $q=M_2/M_1$, the fillout factor $F$ which is a function of the potential gravity, and the inclination $\sin i$. For semi-detached and detached binaries, the mass ratio cannot be estimated, and $F$ is replaced with the sum of fractional radii $\rho_1 + \rho_2 = (R_1 + R_2)/a$, where $a$ is the semi-major axis of the binary orbit. The eccentricity $e$ together with the argument of the periastron $\omega$ are further introduced through $e\cos\omega$ and $e\sin\omega$.

A second catalog from \citet{slawson2011} updates the results by including data from Q2, adding some \textit{Kepler} Objects of Interest (KOIs) flagged as possible exoplanets and later determined to be EBs, adding objects with period longer than 44 days (i.e., Q0$+$Q1), rejecting variables stars initially considered to be EBs, and removing EBs that were initially blends situated at the edge of the photometric aperture later re-observed with a re-centered aperture. This updated catalog presents 2165 EBs, composed of 58\,\% detached (D), 7\,\% semi-detached (SD), 22\,\% overcontact (OC), 6\,\% ellipsoidal (ELV), and 7\,\% undetermined systems. These studies conclude that eclipsing binaries represent 1.4\,\% of the {\it Kepler} target list. A third update has been published that employs an automatic morphological classification scheme \citep{matijevic2012}, and additional EB candidates have been released on the \textit{Kepler} MAST EB database, leading to a total of 2616 EBs.

\subsection{Cross-correlation of both lists} \label{sec_23}

The {\it Kepler} database contains $13\, 698$ stars identified as RGs, while the eclipsing binary catalog contains 2616 targets, as of January 2013. The cross-correlation of both catalogs reveals that 70 stars are flagged as both EB and RG (hereafter RG/EB):  56 systems (66\,\%) are classified as D, seven (10\,\%) as SD, 11 (16\,\%) as OC, and two (4\,\%) as ELV; two (3\,\%) are unclassified. Such proportions are close to those of the whole sample, which is initially surprising since we do not expect RGs to belong to close-in systems, in particular SD and OC, where orbital periods range from 0.2 to 6 days (see Table \ref{table_1}). 
For our purpose of considering stars flagged as EB and RG, it is prudent to keep in mind the recommendations of \citet{prsa2011}:
\begin{enumerate}\itemsep 0cm
\item  ``The high star density leads to a non-negligible likelihood of associating an EB event with the wrong star;''
\item  ``The EB interpretation should be taken with extreme caution for stars with high fraction of flux contamination;''
\item  ``Stars with shallow eclipse events should also be regarded with caution even if the flux contamination is modest.''
\end{enumerate}

\begin{figure}
  \centering
	\epsscale{1.15}
   \plottwo{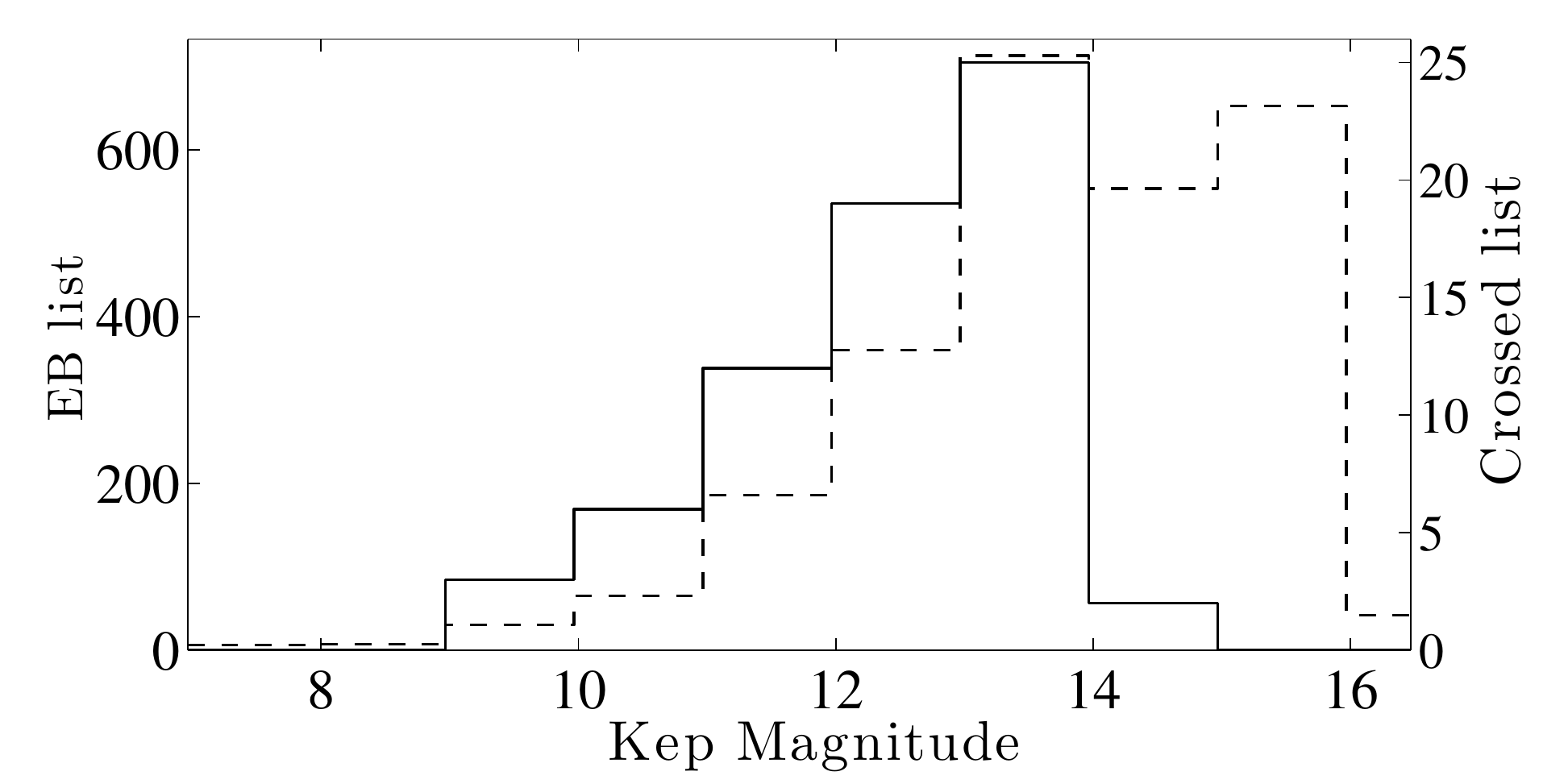}{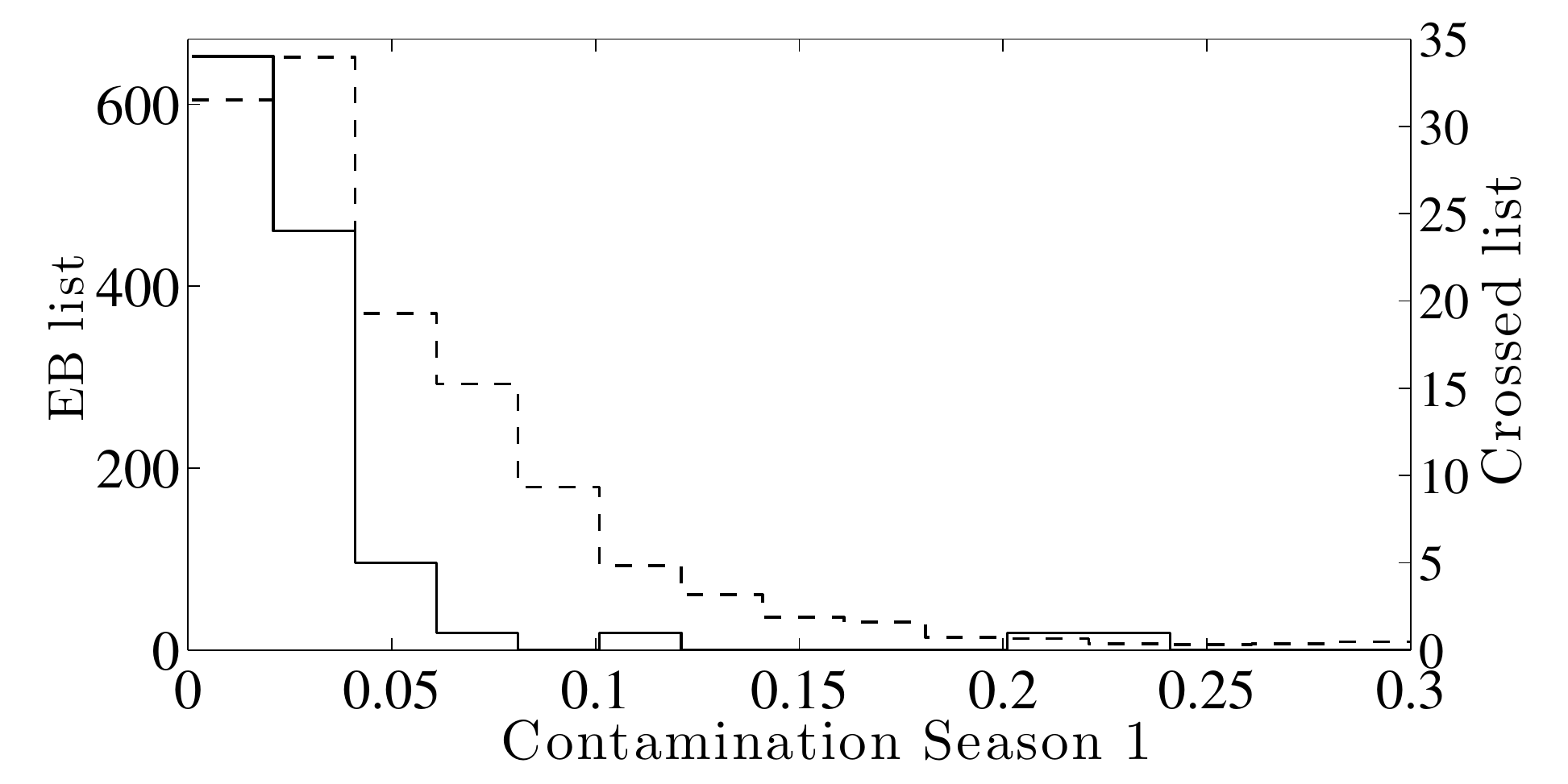}
   \plottwo{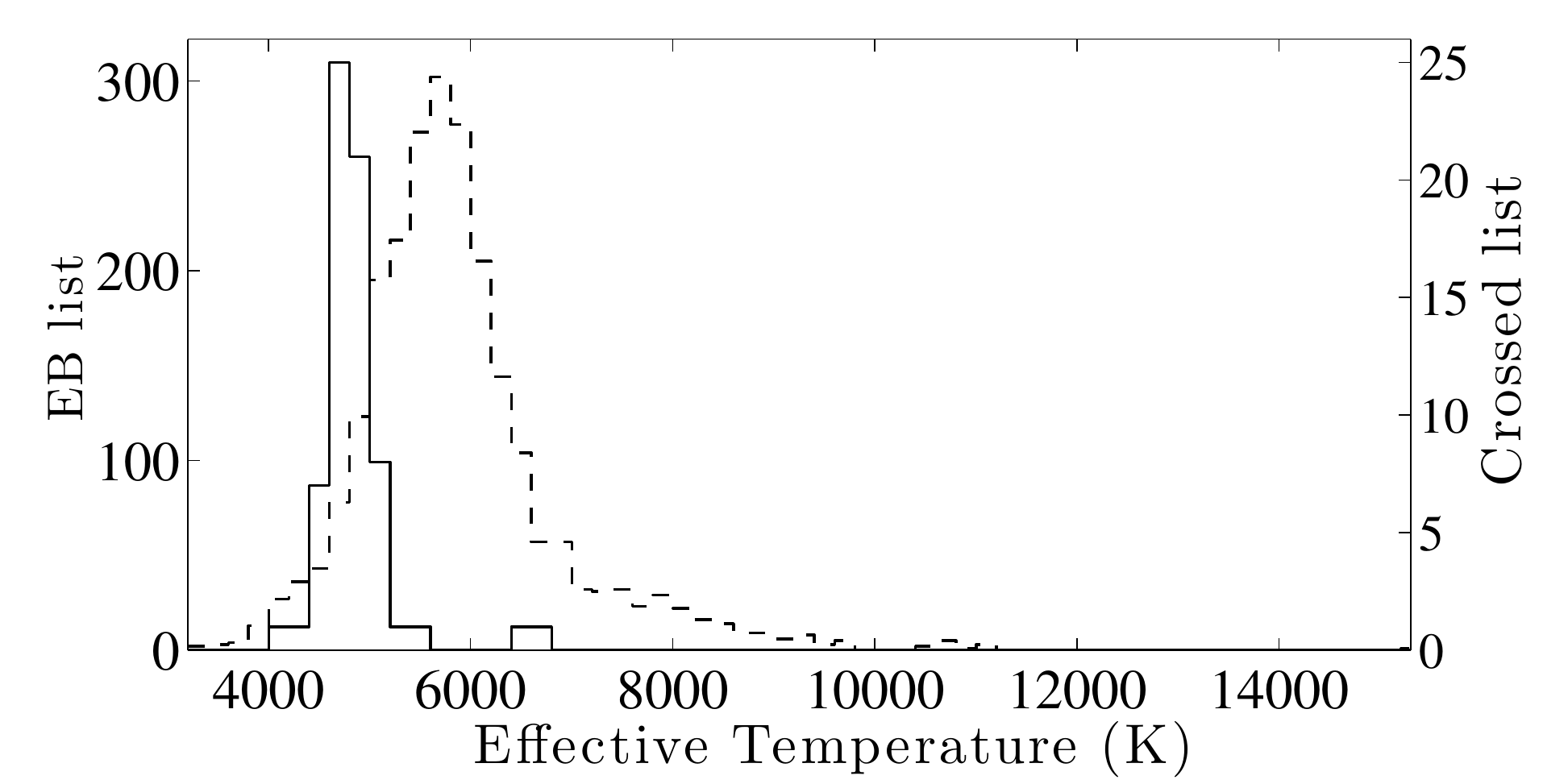}{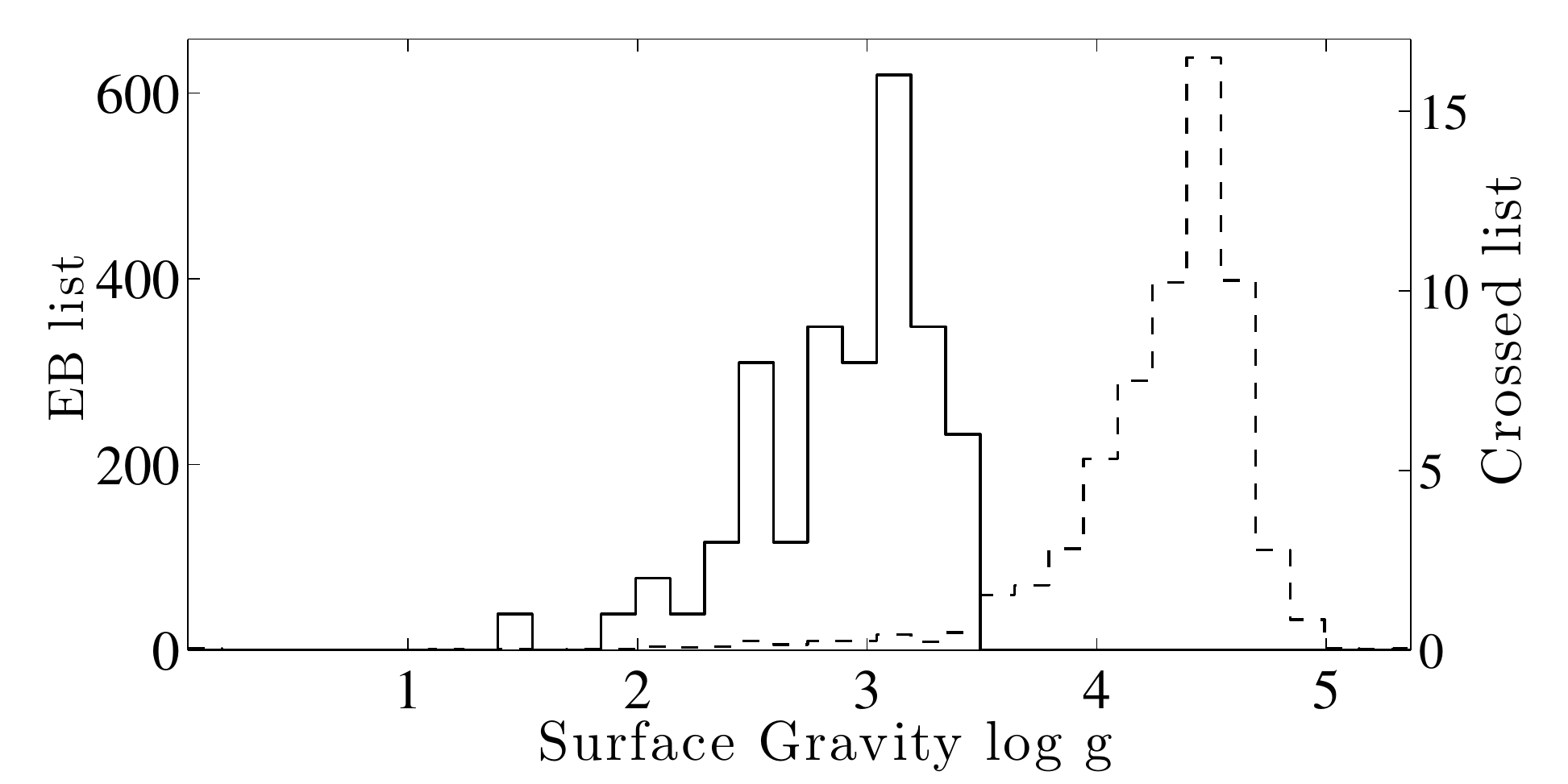}
  \caption{From left to right and  top to bottom, histograms of {\it Kepler} magnitude, contamination factor, effective temperature, and surface gravity for the entire EB catalog (dashed line, left $y$-axis) and for the sample flagged as RG and EB (solid line, right $y$-axis). Data are from the KIC. \label{fig1}}
\end{figure}

\begin{figure}
  \centering
	\epsscale{1.15}
   \plottwo{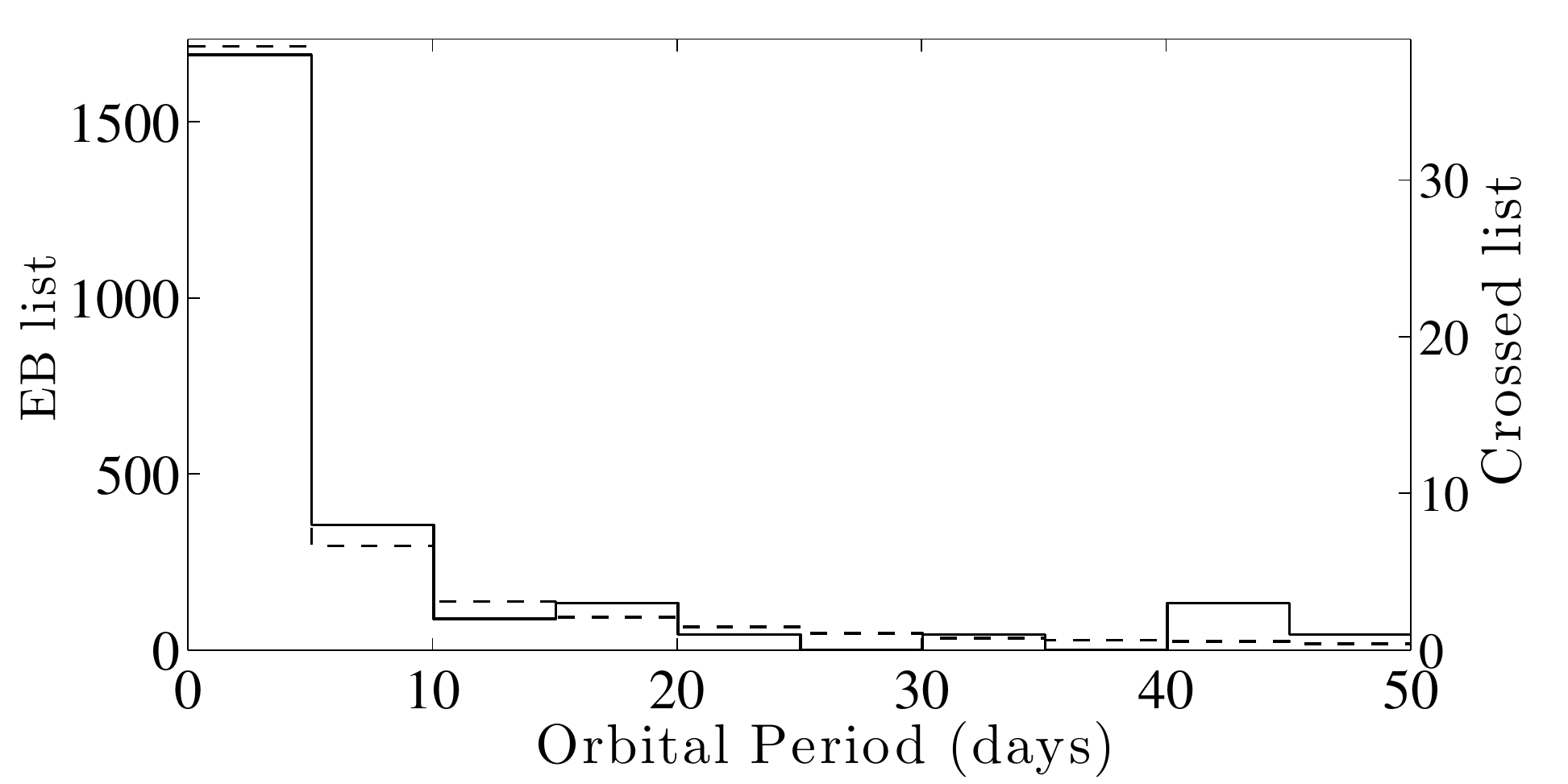}{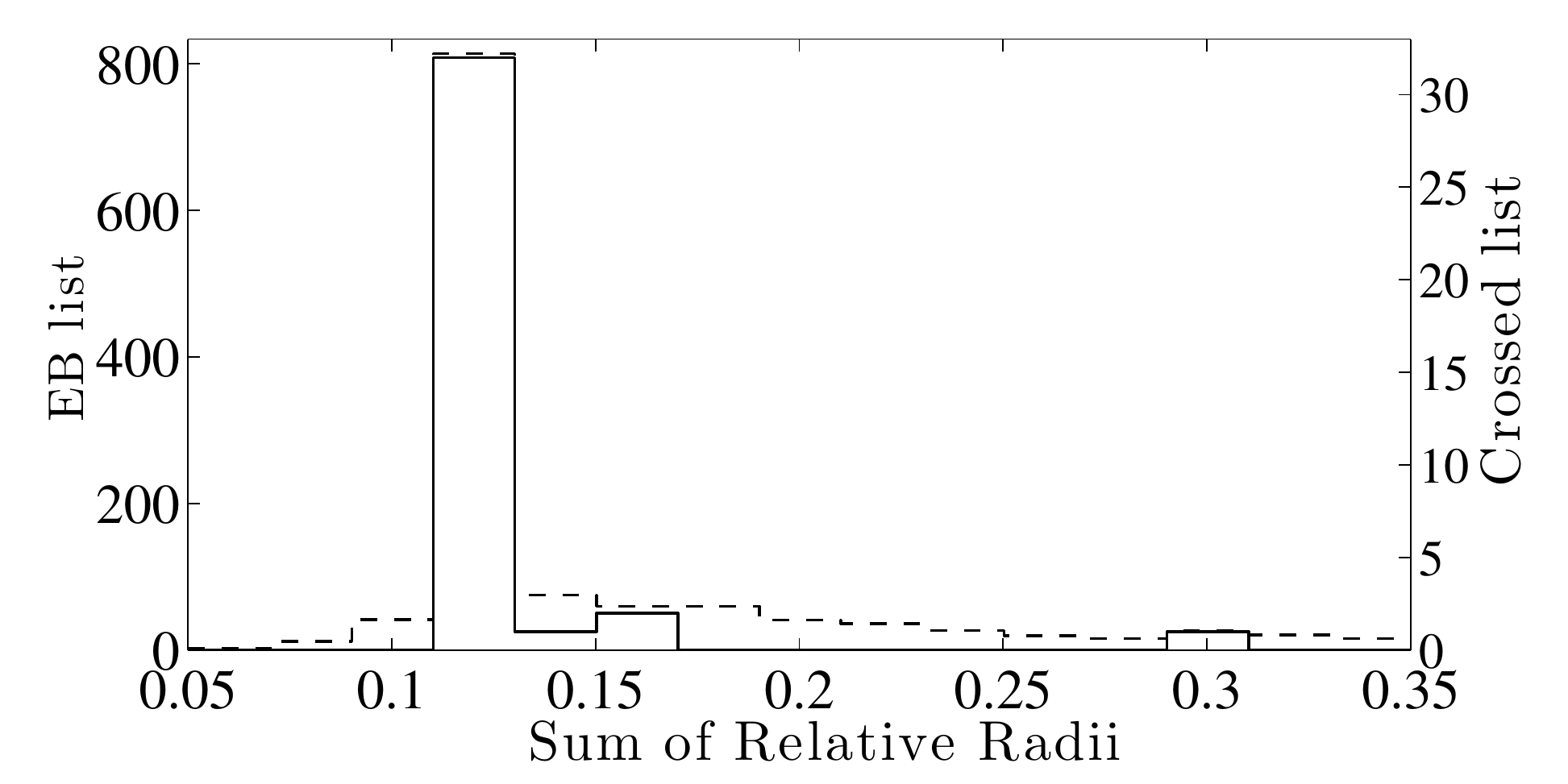}
   \plottwo{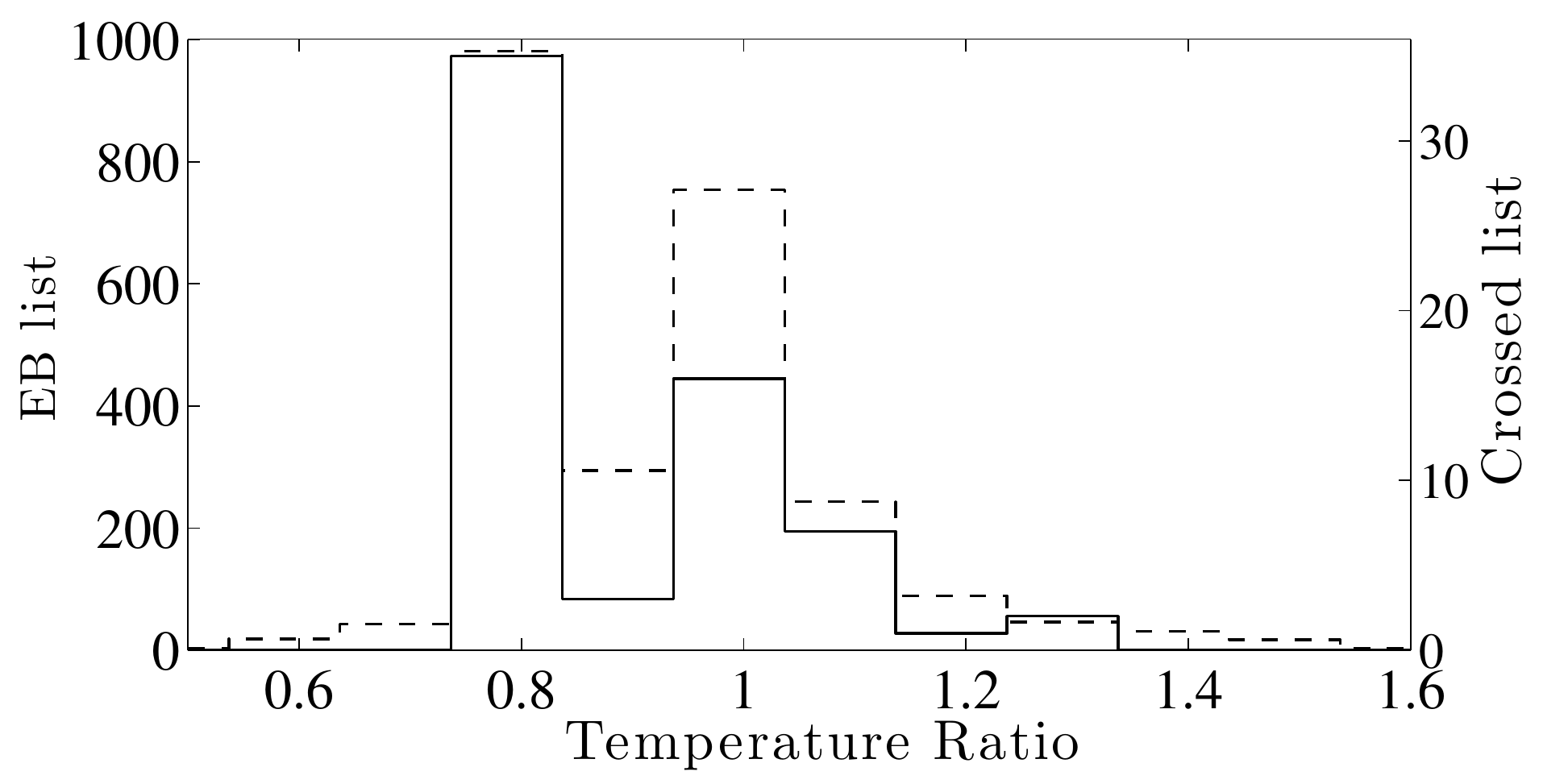}{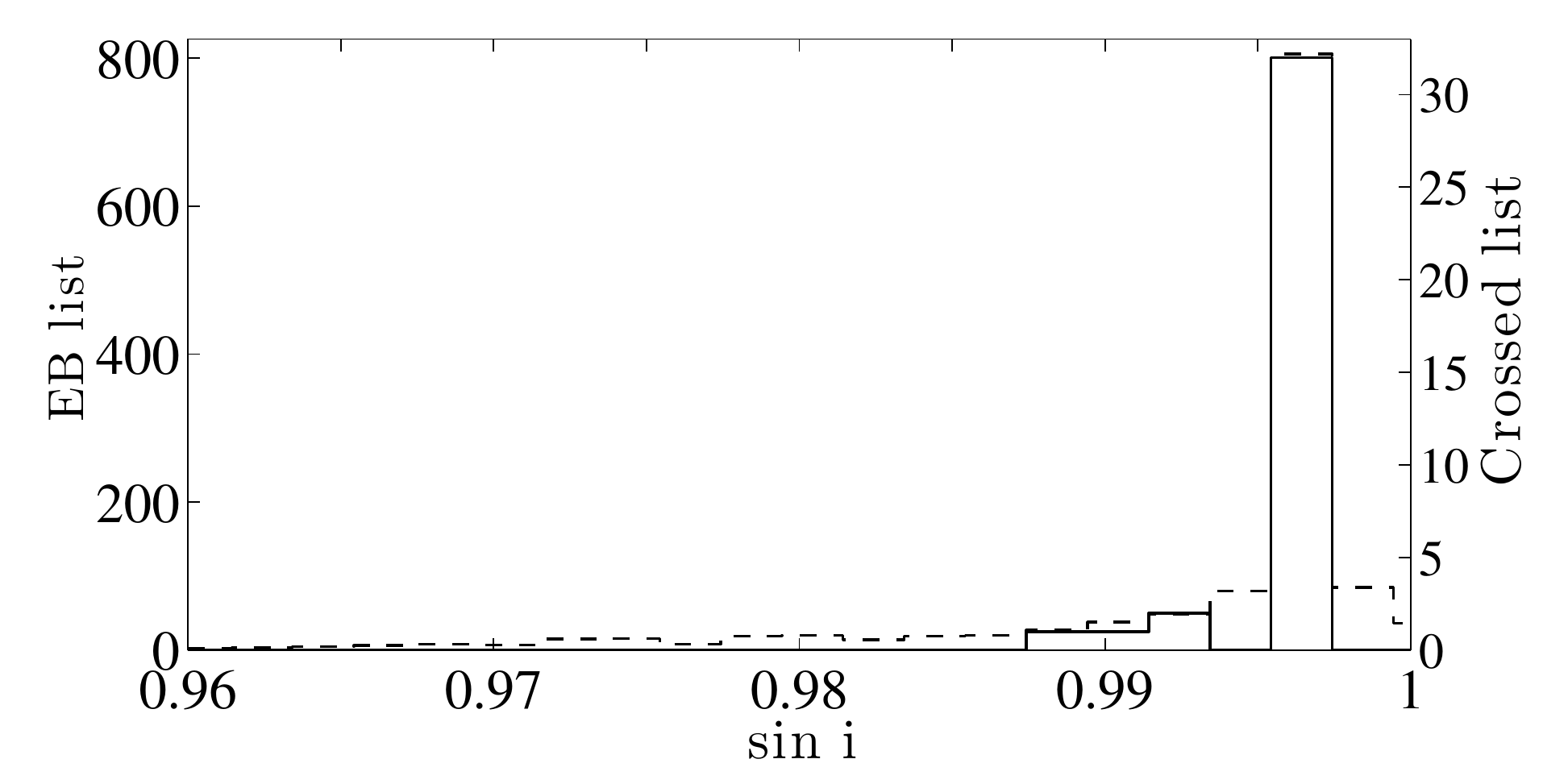}
   \plotone{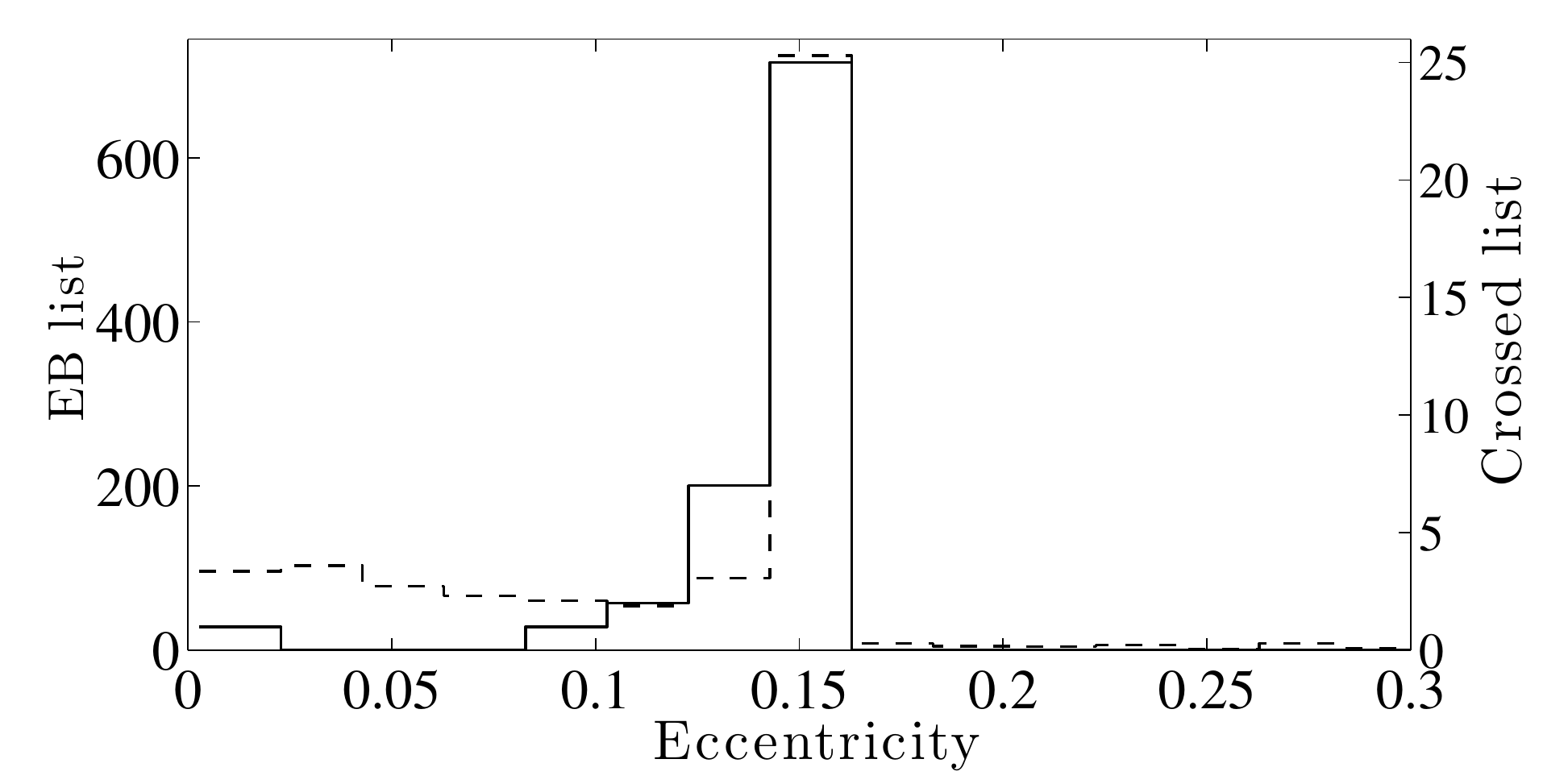}{}
  \caption{From left to right and top to bottom, histograms of orbital period, sum of relative radii, stellar temperature ratio, orbital plane inclination $\sin i$, and eccentricity for the whole EB catalog (dashed line, left $y$-axis) and for the sample flagged as RG and EB (solid line, right $y$-axis). The x-axis is truncated at an eccentricity of 0.6 because so few systems are above this value. Data are from the \citet{slawson2011} EB catalog. \label{fig2}}
\end{figure}

Hereafter, when we mention RG/EBs, we imply \textit{candidate} RG/EB systems. To understand the specificities of RG/EBs with respect to the whole EB catalog and how likely they may correspond with misidentifications, we compare histograms of observing conditions and stellar atmospheric parameters (estimated when systems were supposedly single stars) from the \textit{Kepler} Input Catalog (KIC, \citealt{Brown_2011}), and orbital parameters published by \citet{slawson2011} in Figures \ref{fig1} and \ref{fig2}.

It appears that RG/EBs are slightly brighter on average than the full EB catalog, but this is not significant since RGs were selected with a magnitude limit of 14. For stellar crowding, we only show contamination factors calculated on the apertures corresponding to the second of the four positions of the satellite (i.e., during Q1, Q5, and Q9). We also tested the three other satellite orientations without noting any relevant difference. We see that the contamination factor of RG/EBs is lower than the mean contamination of all EBs, which shows that our candidates are not overly predisposed to the risk of target misidentification or blending. Histograms of RG/EB effective temperature and surface gravity are not representative of the whole EB catalog, but are instead typical of RGs. This means that even if a RG does not belong to an associated EB, there is still one RG per photometric aperture,  whose flux overwhelms that from any secondary star. We note two exceptions at temperatures around 6500 K (see Table~\ref{table_1}) that are classified as OC and are likely misidentified as RGs.

The orbital period histograms of the RG/EBs and the total EB sample in Figure \ref{fig2} are similar, and shows that most RG/EB systems have orbital periods $< 10$ days. The sum of relative radii are also similar in both samples with the exception of the lack of values lower than 0.1 in RG/EBs. The relative deficit of stars with a temperature ratio equal to one among RG/EBs can be related to the minor proportion of contact systems in the RG/EB sample. 
We note that the inclination and orbital eccentricity distributions seem to be locked in the ranges $\sin i = [0.99, 1]$ and $e=[0.1,0.15]$, which suggests that the estimate of these parameters by \citet{slawson2011} could be biased. Measuring orbital eccentricity from photometric data alone, i.e., without radial velocities, is known to be inaccurate.

\begin{figure}
  \centering
	\epsscale{1.25}
   \plotone{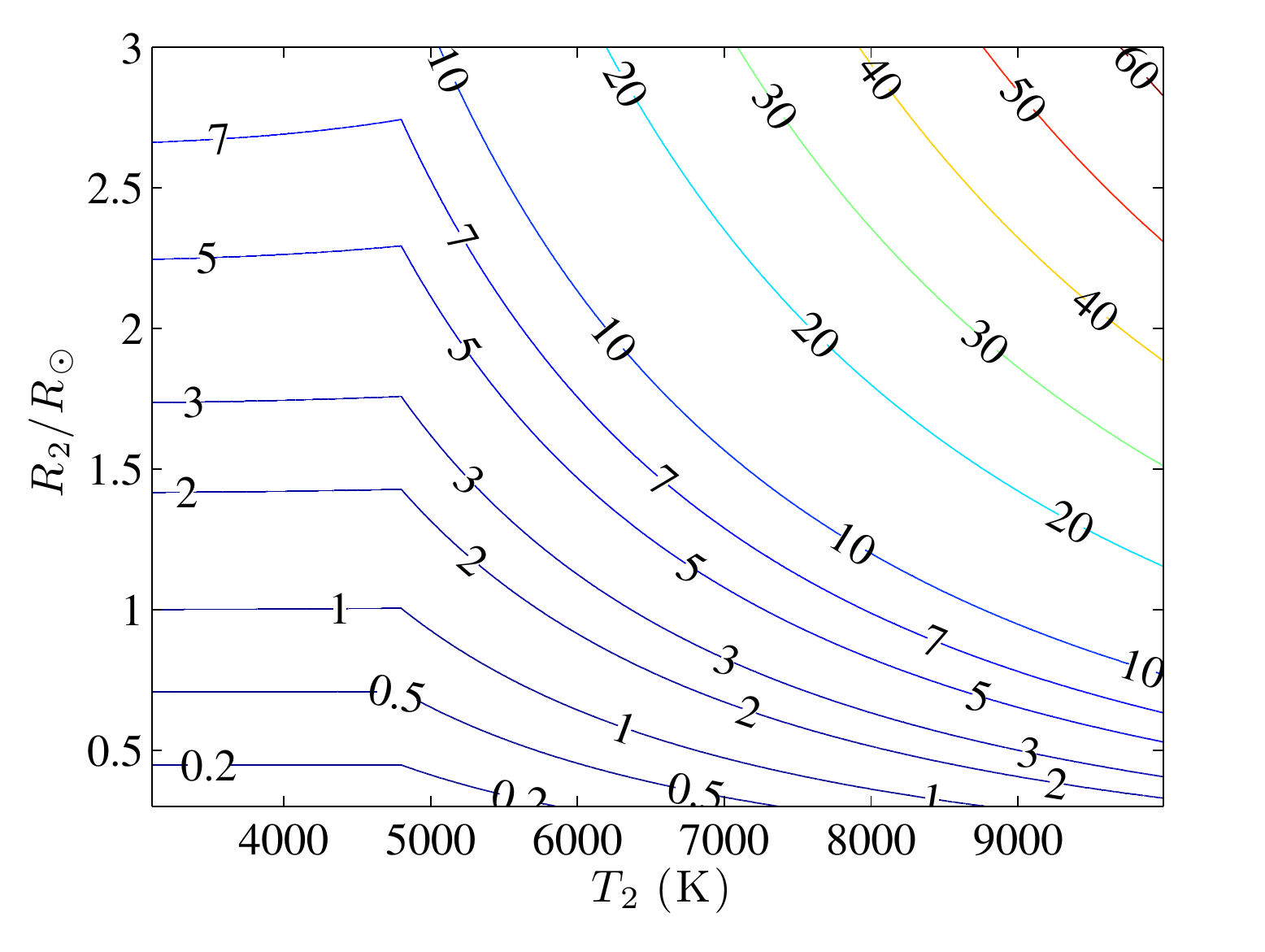}
  \caption{Expected primary eclipse depth (given by the contours, in \%) by assuming a companion star of temperature $T_2$ and radius $R_2$ eclipsing a RG of temperature 4800~K and radius $10\,R_\sun$.\label{fig3}}
\end{figure}

To get an initial rough estimate whether a RG is compatible with an EB, we determine the amplitude of the deepest eclipse measured in the RG/EB light curve. We assume the system is observed from the orbital plane, and that the RG is larger than its companion. We therefore adopt the convention that the deepest eclipse is the secondary eclipse (i.e., companion star going behind the RG) if the companion's temperature is higher than the RG's, and that the deepest eclipse is the primary eclipse otherwise (i.e., when the RG's temperature is higher than that of the companion star). Primary and secondary eclipses will have equal depths when the two stars' effective temperatures are equal. A raw proxy of relative photometric dimming during primary and secondary eclipses can be obtained by neglecting stellar limb darkening and comparing simple luminosities: 
\begin{eqnarray}\nonumber
\left(\frac{\delta I}{I}\right)\ind{primary}  &=& \frac{R_2^2\ T_2^4}{R_1^2\ T_1^4\ +\ R_2^2\ T_2^4} \\
\left(\frac{\delta I}{I}\right)\ind{secondary}  &=& \frac{R_2^2\ T_1^4}{R_1^2\ T_1^4\ +\ R_2^2\ T_2^4}, 
\label{eqn1}
\end{eqnarray}
where the subscripts $1$ and $2$ stand for the RG and the companion star, respectively. Note the numerator of the second equation is $R_2^2 T_1^4$ since since the light dimming is due to the star of radius $R_2$ which hides a surface $\pi R_2^2$ of the star of temperature $T_1$. In Figure \ref{fig3} we present the expected eclipse depth assuming a typical RG with radius $10\ R_\sun$ and effective temperature $4800$~K. We conclude that the range of photometric dimmings $\sim0.2- 20\,\%$, which we measure in all but nine of the RG/EBs, is compatible with a stellar pair that includes a RG and a main-sequence star (see Table \ref{table_1}). We note that primary eclipse depths will be lower than the theoretical proxy in Equation (\ref{eqn1}) when eclipses are grazing due to limb darkening. For the nine systems with very shallow eclipses, their depths range from 0.02\,\% to 0.14\,\%. These cases could either correspond to RGs with a background EB, EB systems with grazing eclipses, or EB systems where the companion star's size is a few percent of the RG's size, as could be possible for brown dwarfs or giant planets.

\begin{figure}
  \centering
	\epsscale{1.25}
  \plotone{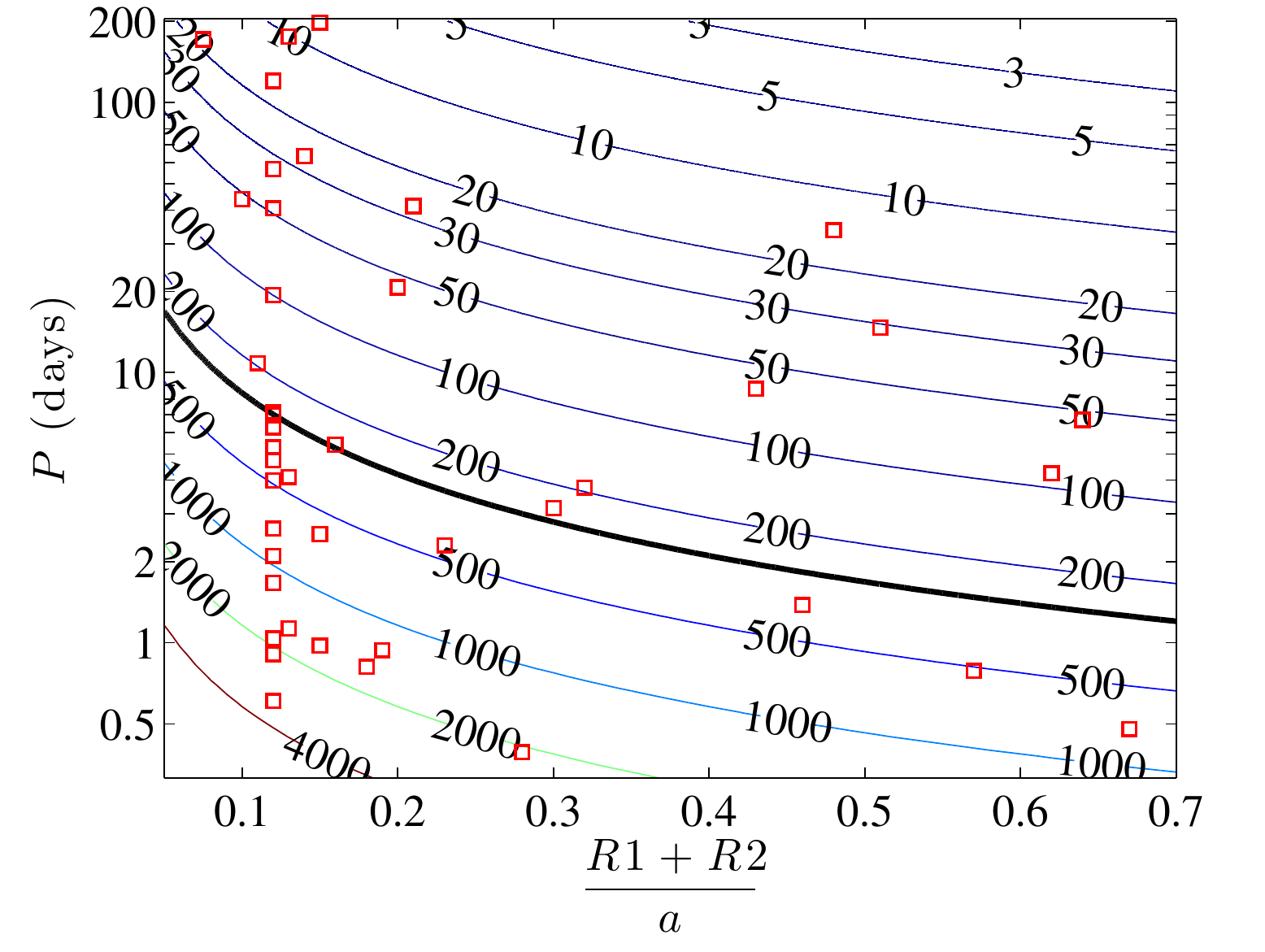}
  \caption{Orbital velocity contours (km s$^{-1}$) of a RG in a binary system as a function of the orbital separation and orbital period. A circular orbit is assumed. We further assume $(R_1,M_1) = (10R_\sun,1.4M_\sun)$ for the RG and $(R_2,M_2) = (1R_\sun,1M_\sun)$ for the companion star. Red crosses represent the values of the RG/EB sample from \citet{slawson2011}. The thick black line indicates the critical velocity for a phase-locked RG. \label{fig4}}
\end{figure}

From the RG and EB catalogs, as well as from the associated KIC parameters $\log g$ and $T\ind{eff}$, it is assumed that each candidate system contains one EB and one RG in the photometric aperture.  However, suspicion of blending arises from quick calculations regarding the orbital period distribution. Let us consider a simple configuration where a RG with radius and mass ($10 R_\sun, 1.4 M_\sun)$ is in a phase-locked binary system with a solar-like companion ($1R_\sun, 1M_\sun$) in a circular orbit. (Such parameters for the RG are close to the median values of RG/EBs obtained with asteroseismology in Section \ref{sec_33}.) In Figure \ref{fig4}, we show the expected orbital velocities that the RG would have in that configuration, by varying both orbital period $P$ and semi-major axis $a$: 
\begin{equation}
v\ind{orb,1} =  \frac{2\pi a}{P} \frac{M_2}{M_1+M_2}.
\label{eqn2}
\end{equation}
By using the values for the sum of fractional radii and orbital periods estimated for 46 out of the 70 systems by \citet{slawson2011}, we see that 14 RGs (mostly of the OC  class)  would have an orbital velocity between 500 and 2600~km\,s$^{-1}$. Even if our assumptions about the companion star's mass and radius are incorrect, this indicates that some systems are not physically possible as they would require an orbital velocity greater than the critical velocity that would begin to tear apart the RG. We consider a velocity to be ``critical'' when the rotation velocity at the equator equals the escape velocity. In the case of a phase-locked system, the rotation period is simply the orbital period. Any additional proper stellar rotation in the direction of the orbital motion (e.g.,~Earth) would lower the critical velocity, while a stellar rotation in the opposite direction of the orbital motion (e.g.,~Venus, which is unlikely) would enhance the critical velocity. Based on this criterion, only 25 out of the 46 systems, for which we have orbital parameters, have a period compatible with the critical velocity threshold. This is the first of several clues that points to the fact that many of the candidates are not true RG/EB systems.


In the following sections, we use several analysis techniques to better characterize the systems and more accurately determine how many of our candidate systems are bona-fide RG/EBs.


\section{Light curve analysis for eclipses and asteroseismology} \label{sec_3}

\subsection{Search for contamination from surrounding stars} \label{sec_35}

The cross correlation between the RG and EB databases resulted in 70 identifications. We consider four possible scenarios:
\begin{enumerate}
\item  The RG is actually one of the EB stars.
\item The RG is aligned within the pixel field of view with the EB and is a part of a multiple system (gravitationally speaking), of which two close-in stars mutually eclipse, with the RG out of the EB's orbital plane. Eclipse Timing Variations (ETVs) may be observed in this case (see Section \ref{sec_41}).
\item The RG is aligned within the pixel field of view with the EB, but is not gravitationally bound. 
\item The RG and the EB fall on different pixels in the aperture and are not gravitationally bound.
\end{enumerate}

The last case can be verified using the target pixel files associated with each star by computing a map of the relative intensity variation ${\rm d}I/I$ to check whether the depth of the eclipse is correlated to the peak intensity source on the detector.

Every quarter when the spacecraft rotates, a star falls onto a different set of pixels, and the pixel mask (aperture) that defines the optimal output light curve often changes.  In the case of a long-period EB, the eclipses only occur in a few of the quarters. We detrended each pixel light curve in a quarter  by a low-order polynomial and then modulated (``folded'') it by the period estimated from the global light curve. The relative intensity drop  ${\rm d}I/I$ was then calculated and an image of this eclipse depth was compared to an average intensity image for the quarter. 

 \begin{figure}
   \epsscale{1.4}
   \centerline{\plotone{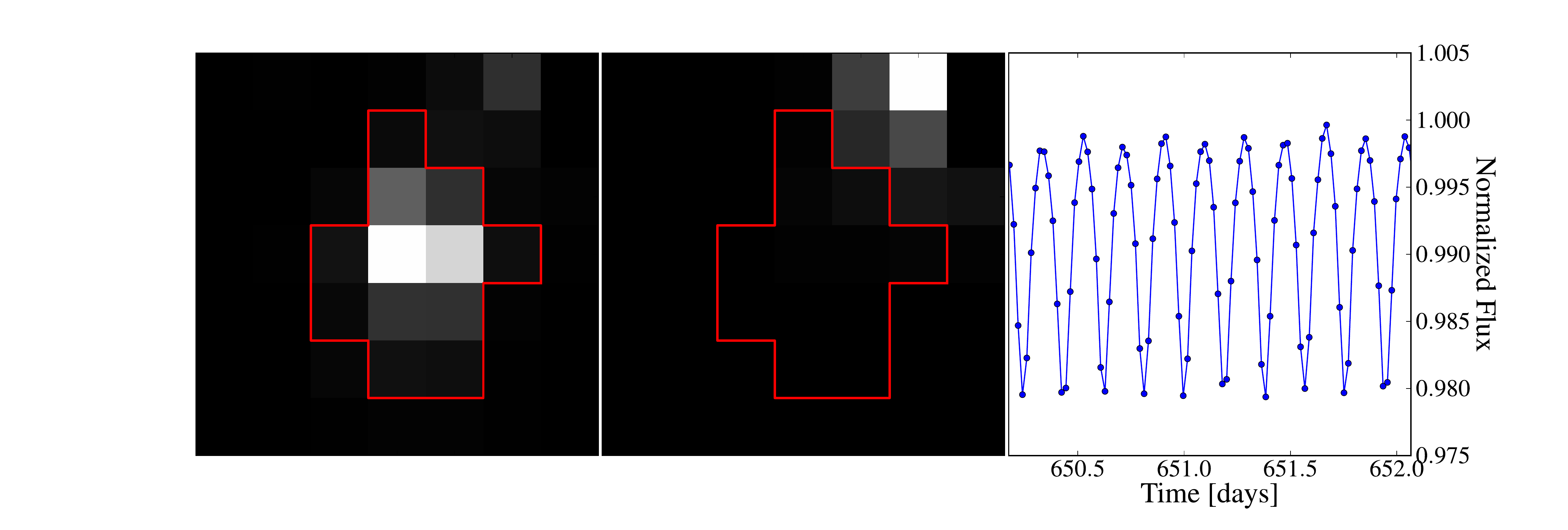}}
   \vspace{-.3\textwidth}
   \centerline{\hspace{.15\textwidth}\large\color{white}{(a)}\hspace{.23\textwidth}\large\color{white}{(b)}\hspace{.22\textwidth}\large\color{black}{(c)}\hfill}
   \vspace{.26\textwidth}
   \epsscale{0.8}
   \centerline{\plotone{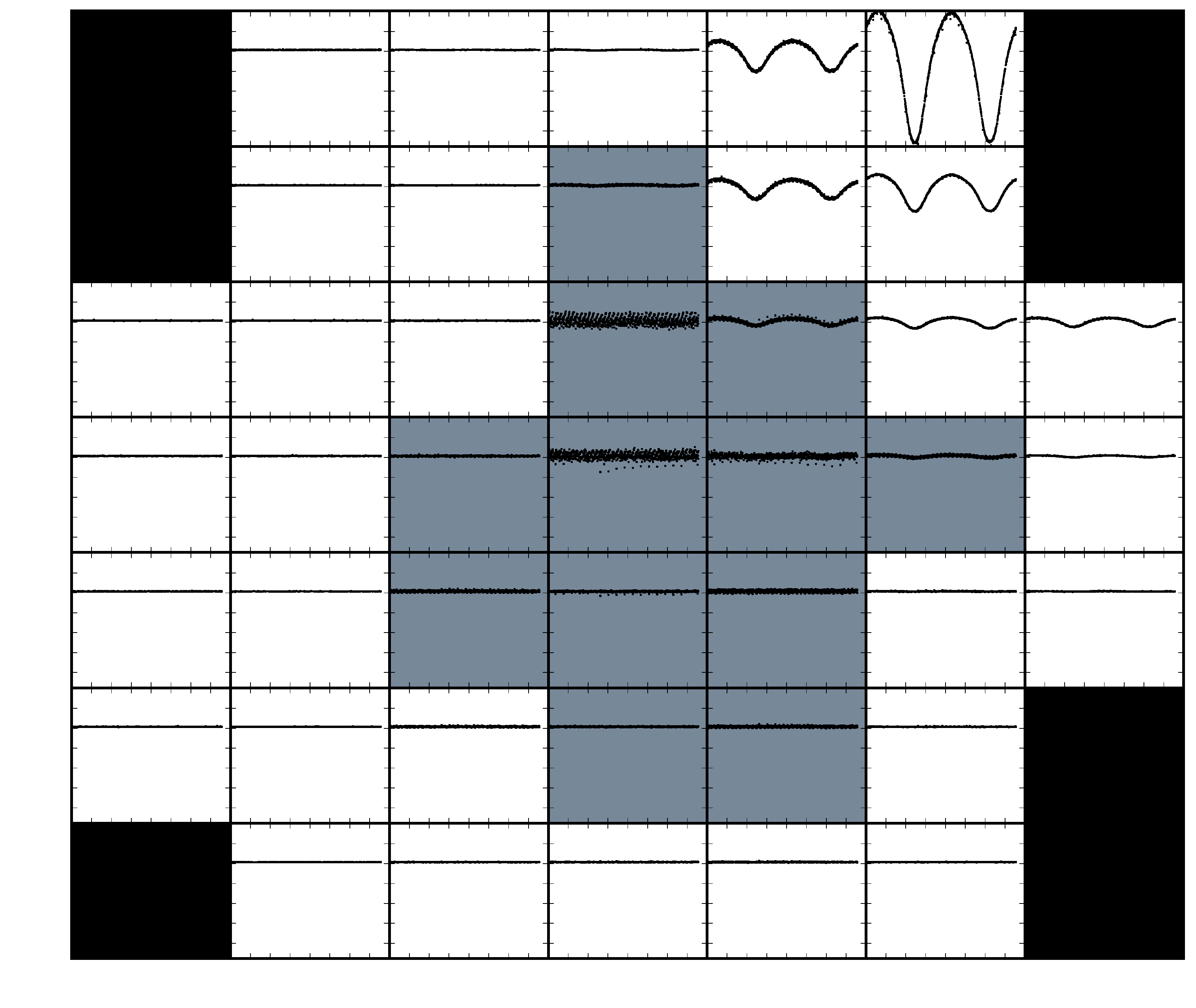}}
   \vspace{-.75\textwidth}
   \centerline{\hspace{.14\textwidth}\large\color{white}{(d)}\hfill}
    \vspace{.73\textwidth}
   \caption{Identification of RG star KIC 4576968 and contaminating EB from the \textit{Kepler} Target Pixel Files. The top panels show (a) the map of the mean intensity, (b) the map of eclipse depth in each pixel, and (c) a segment of the light curve obtained by summing over all pixels (including outside the aperture). White corresponds to highest flux on a linear scale, and the red outlines the aperture used to compute the \textit{Kepler} light curves. The matrix in panel (d) shows the period-folded light curve in each target pixel normalized to the median value. The gray boxes denote the aperture. The strongest eclipse signature is outside of the aperture, while the highest flux pixel is within the aperture. All data are from Q7. Note that the dispersion of the folded light curves is higher in the three pixels where the intensity is maximum because of the presence of RG oscillations.} 
   \label{fig_jean}
\end{figure}

In 40 of the 47 cases, we find that the peak in intensity corresponds to the peak in eclipse depth, indicating scenarios 1 or 3 above cannot be ruled out (and for the nine brightest stars, saturation in the aperture that buries any eclipse signature on individual pixels does not allow this method to be used). We find 7 clear cases where the pixel with the deepest eclipse occurs away from the pixel with the peak mean intensity (and typically outside of the \textit{Kepler} aperture as well). An example is shown in Fig.~\ref{fig_jean} for KIC 4576968, which is classified in the binary catalog as an OC system. The overall peak in intensity of the mean photometry is well inside the defined aperture, yet the maximal eclipse depth is about two pixels outside of this region. The eclipse depth considering only pixels within the aperture is about 0.06\,\%, while outlying pixels reach about 2\,\%. Studying the individual folded light curves for each pixel confirms that the EB and the RG are on different pixels, and likely not part of the same system.

All of these 7 cases occur in short, approximately less than 1 day period EBs, where it is indeed physically impossible for the RG to be one of the eclipsing stars and still remain intact. Five such systems (KIC 2711123, 4576968, 5652071, 7879404, 11968514) show behavior where we are confident that the RG is a background contaminant and not actually a part of the EB, as scenario 4 describes. Two other cases (KIC 7031714, 7955301) show ETVs and are consistent with scenario 2,  but are likely to be false positives. Indeed, if these objects were gravitationally bound and separated by one or more four-arcsecond pixels, the physical separation at the distance of a magnitude-9 giant star would be on the order of several hundred to thousands of AU, which implies an orbital period of several thousands of years. Hence these cases could correspond to a RG with a background triple system.

In addition, for each star, we compute the power spectrum of each pixel. It appears that the global oscillations always originate from the pixels where intensity is maximum, which means that the brightest star systematically corresponds to a red giant, for the systems where oscillations are detected. We even distinguish by eye the RG oscillations on the map of folded light curves across the photometric aperture through a point dispersion larger than the average noise (see Figure \ref{fig_jean}, panel (d)).


\subsection{Cleaning the time series} \label{sec_31}
Two distinct ways of processing the light curves are required for modeling eclipses or analyzing seismic properties. Eclipse modeling consists of extracting the stellar parameters from fitting the shape of the light curve during eclipses after removing instrumental biases and, ideally, other sources of stellar variability (e.g., activity, pulsations). On the contrary, studying global mode properties requires that the time series from which we calculate power spectra is free from periodic signals such as instrumental systematics and stellar eclipses. In other words: global modes are noise for eclipse modeling and vice versa.

\begin{figure}
  \epsscale{1.2}
  \plotone{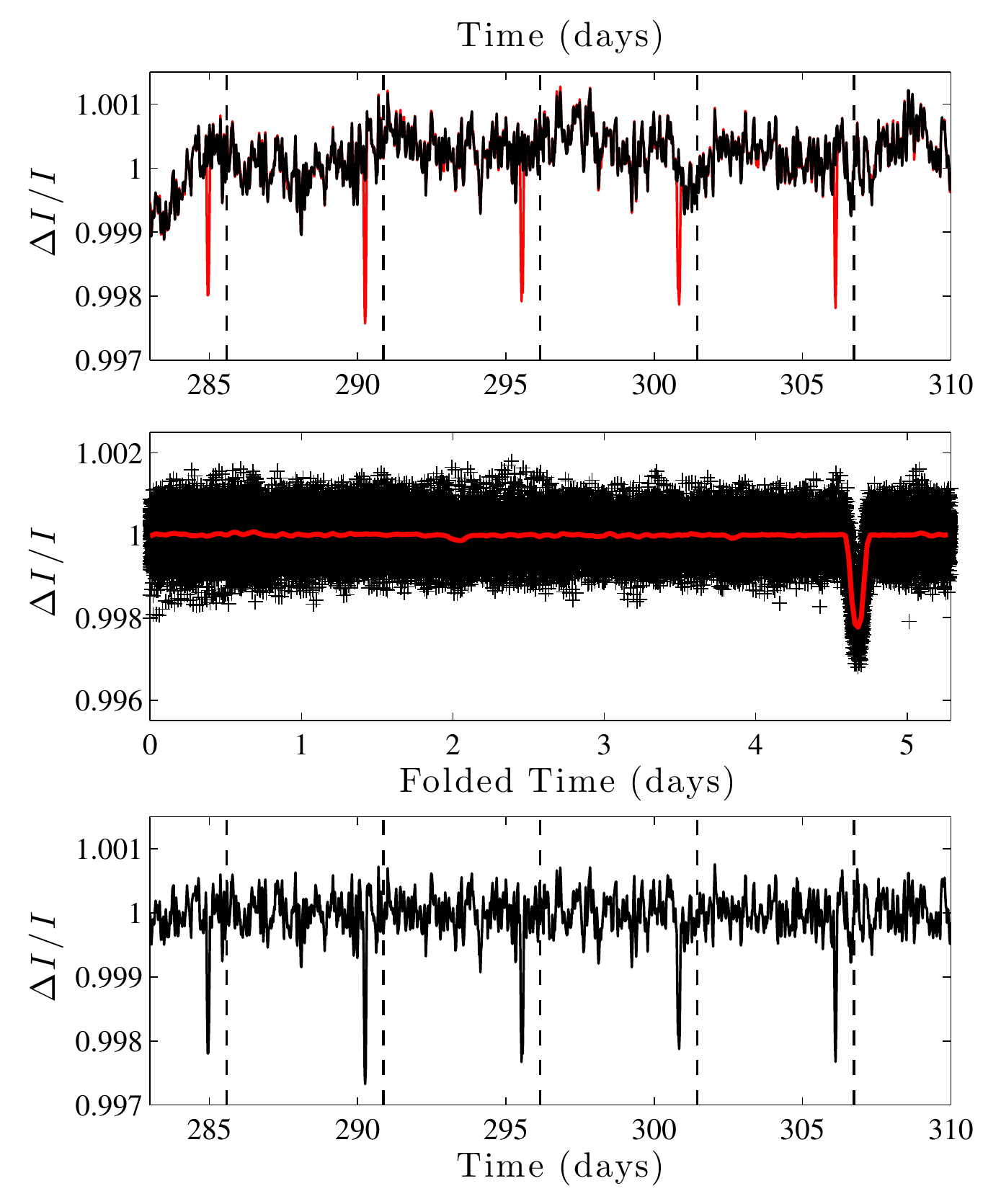}
  \caption{Cleaning the light curve of short-period binary KIC 3532985. The top panel shows a segment of the original light curve (red) and the corrected light curve (black) after eclipse removal, which is used for seismology analysis. The middle panel shows the original light curve folded on the period (black) and its rebinned version (red). The bottom panel is the result after a smoothed light curve without eclipses is subtracted, and is used for light curve modeling. The dashed vertical lines in the top and bottom panel demarcate successive orbital periods. \label{fig_clean_1}}
\end{figure}

\begin{figure}
  \centering
  \epsscale{1.2}
   \plotone{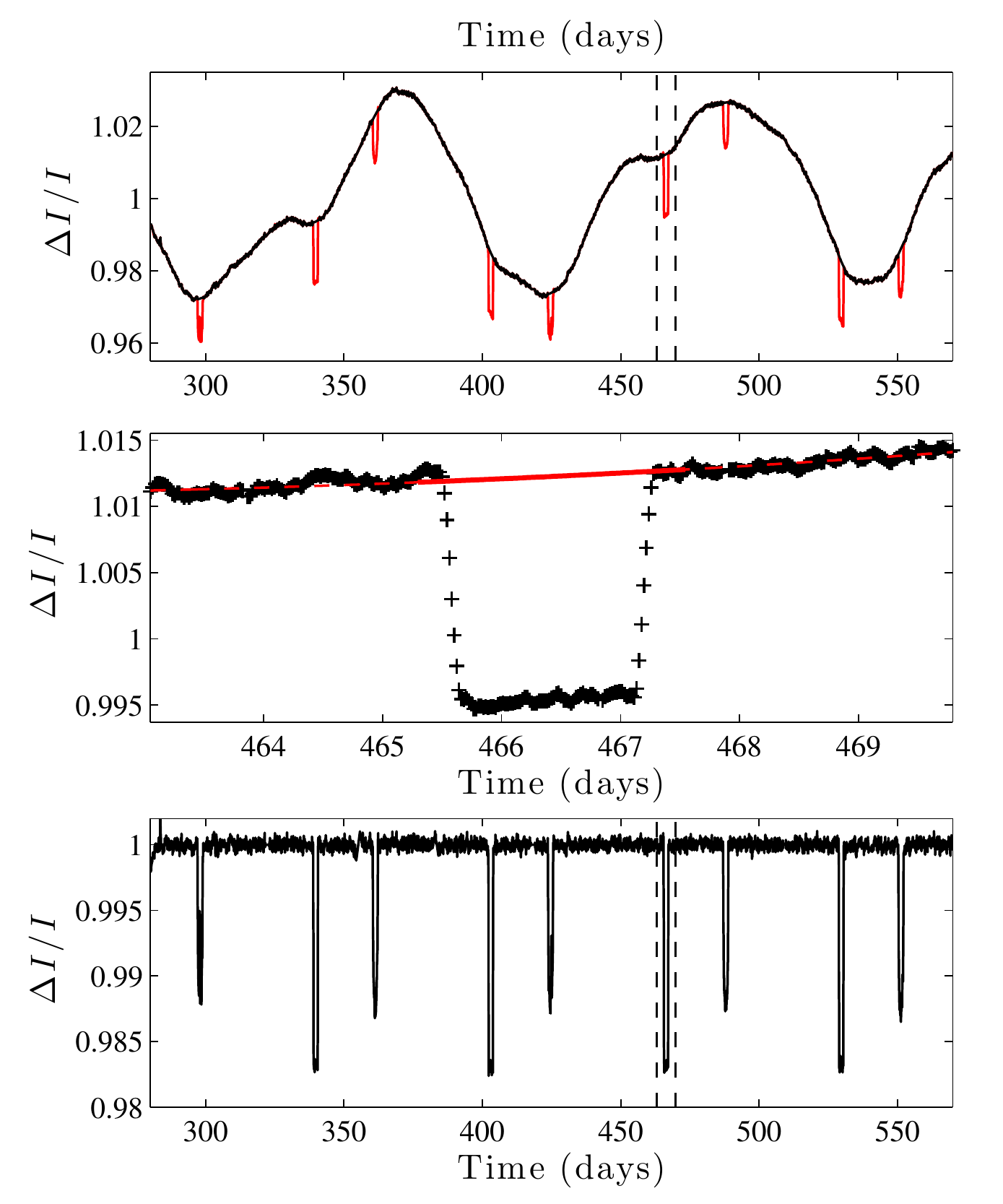}
  \caption{Cleaning the light curve of long-period binary system KIC 8430105. The top panel is the same as Figure \ref{fig_clean_1}, and shows a segment of the original light curve (red) and the corrected light curve (black) after eclipse removal, which is used for seismology analysis. The middle panel shows a zoomed view of a single eclipse from the original light curve, and the polynomial fit used to remove the eclipse signal for asteroseismology (red). The bottom panel is the result after a smoothed light curve without eclipses is subtracted, and is used for light curve modeling. The dashed vertical lines in the top and bottom panels demarcate the region highlighted in the middle panel. \label{fig_clean_2}}
\end{figure}

\begin{figure}
  \centering
  \epsscale{1.2}
   \plotone{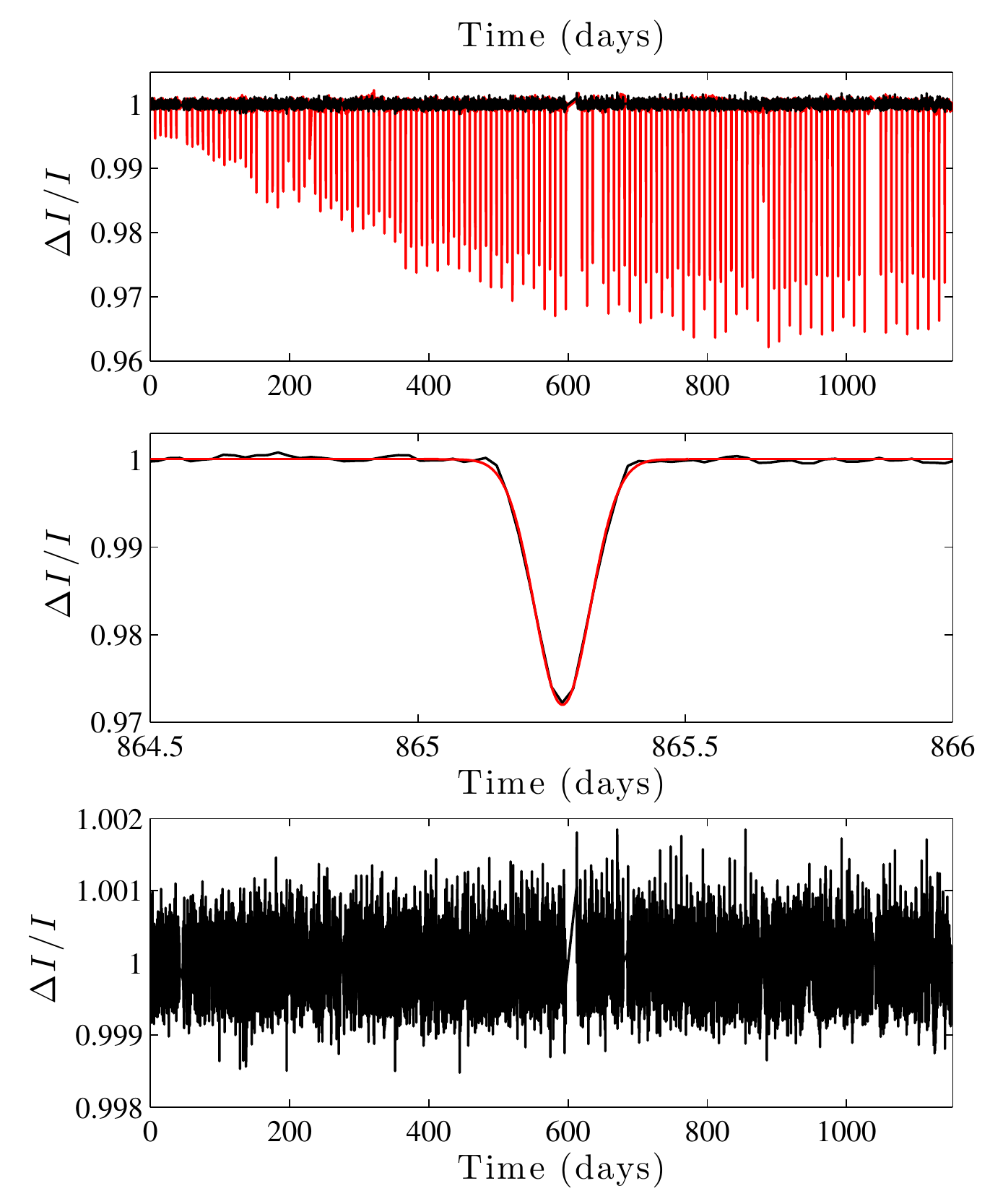}
\caption{Cleaning a light curve with variable eclipse depth and timing (KIC 7955301). The top panel shows the entire original light curve (red) and the corrected light curve (black) after eclipse removal, which is used for seismology analysis. The middle panel shows a zoomed view of one of the eclipses (black) and an example of the function used to fit each eclipse (red). The bottom panel shows the result after the eclipses are subtracted from the original light curve, and upon which asteroseismology is performed. No light curve modeling is attempted in these cases. \label{fig_clean_3}}
\end{figure}

\begin{figure}
  \centering
  \epsscale{1.2}
  \plotone{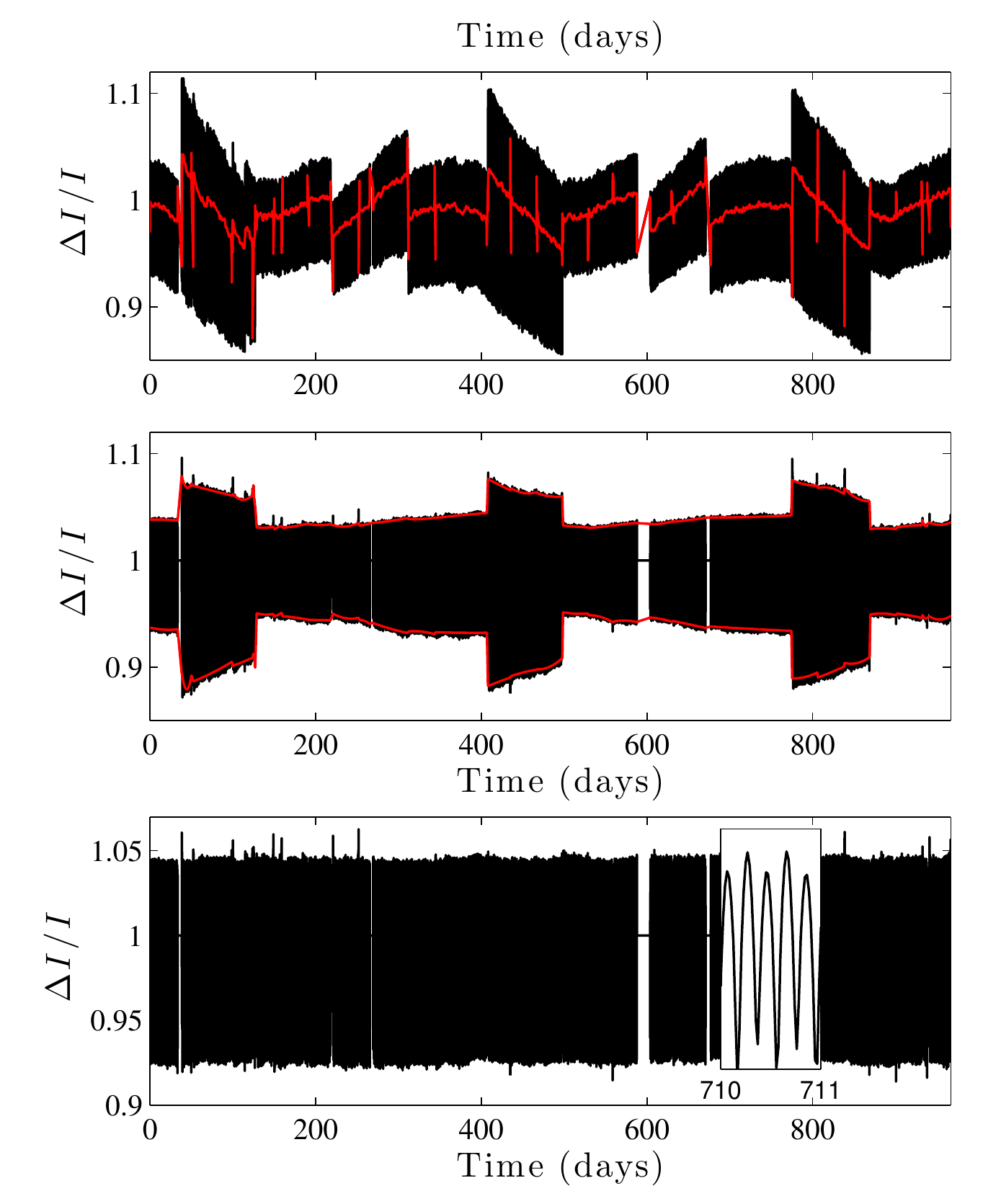}
  \caption{Cleaning an over-contact (OC) binary light curve with variable amplitude (KIC 7879404). The top panel shows the entire original light curve (black) and the smoothed version (red). Since the smoothing function is a weighted moving average, the smoothed light curve equals the signal at each interruption, giving an impression of spikes. The middle panel shows both the original light curve after subtraction by the smoothed version (black) and the local minima and maxima per half-period (red). The bottom panel is the result after the original light curve is divided by the smoothed maxima-to-minima distance. A zoomed view is is shown for time between 710 and 711 days. This final light curve is used for asteroseismology, and we attempt no light curve modeling. \label{fig_clean_4}}
\end{figure}

Nevertheless, there are initial reduction procedures common to both cases. For example, it is absolutely necessary to remove long time drifts and discontinuities, which are mostly of instrumental origin, particularly after \textit{Kepler} rotates between consecutive quarters. Note that we do not use the portion of each light curve between eclipses for our light curve models due to difficulties in disentangling instrumental flux variations, stellar activity (spots, granulation), reflection, Doppler beaming, and ellipsoidal variations. We assume that the median fluxes of each quarter's data are equal, and normalize them to 1 to work with relative fluxes. The entire light curve is then concatenated, and all outliers (selected as being out of the light curve mean dispersion by an amount evaluated individually for each target) are eliminated. The next step depends on the specifics of each light curve. Five classes of datasets and procedures are described:
\begin{itemize}

\item Short orbital period EBs (e.g., Figure \ref{fig_clean_1}). When the time series is long enough to contain more than about 30 orbits, we proceed by assuming that the signal fluctuations, be they stellar or instrumental in origin, may be averaged out by folding and rebinning the light curve. We proceed by subtracting the mean folded light curve from the original. This results in a time series with no eclipse (see the black curve in Figure \ref{fig_clean_1}, top panel) which retains the stellar variability that we use to search for global modes in the Fourier domain. Next, we smooth this time series with a moving average whose width is set on the characteristic time scale of the photometric variations to be cancelled. The smoothed time series is then subtracted from the original time series to get a light curve with a flat level between eclipses. In principle, this is the best method because it is simple and makes no assumption about the origin of any photometric fluctuations. In practice, we limited the use of this method to systems with orbital period shorter than 20 days.

\item EBs with stellar variations on the order of the orbital period, which is much longer than the eclipse duration (e.g., Figure \ref{fig_clean_2}). In this case, a stellar signal of amplitude similar to or higher than the eclipse's depth cannot be cancelled out by folding the data. We proceed by identifying the centers of the primary and secondary eclipses, bridging them with a second-order polynomial, and then subtracting the smoothed light curve from the original to yield a flattened light curve with equally deep eclipses. We use a second-order polynomial to ``fill'' each eclipse by simply fitting the short regions of the light curve on either side of the eclipse of duration equal to the eclipses. We find a second order polynomial is sufficient account for the local behavior of the light curve. Next, an asteroseismic analysis is carried out on the light curve that has the eclipses filled with polynomials to avoid periodic gaps. This technique cannot be applied if the eclipse duration is too large compared to the orbital period because we would cancel the signal by filling the eclipses.

\item Long orbital period EBs. When the eclipse duration is less than $5 \%$ of the duration of the entire light curve, we  postulate that simply removing the data where eclipses occur does not significantly change the duty cycle for the asteroseismic analysis. For eclipse modeling in such systems, we flatten the light curve after ``filling'' eclipses with second-order polynomials, as in the previous case.

\item EBs with variable eclipse depth and timing (e.g., Figure \ref{fig_clean_3}). In several cases, we encounter light curves with eclipses whose depth and timing change with time, either of astrophysical or instrumental origin (e.g., varying flux contamination at each field rotation). We do not attempt to model such eclipses in this work because it often implies accounting for a third body when there is an astrophysical cause, or it requires careful modeling of systematic errors when the cause is instrumental. For asteroseismology, however, we fit individual eclipses with simple functions and then subtract them from the time series. The functions we use to fit eclipses are Gaussian when eclipses are grazing (no flat bottom, or ``v-shaped''), or the \citet{mandel2002} exoplanetary fitting function otherwise.

\item Contact EBs with variable amplitude (e.g., Figure \ref{fig_clean_4}). As indicated in the previous cases, varying flux contamination can modify the apparent amplitude of eclipses, and the technique of fitting eclipses with Gaussian or exoplanetary functions is not possible for contact (OC) systems. Therefore, after subtracting a smooth version of the light curve, we measure positions and amplitudes of all local maxima and minima per half-orbit and smooth them. Then we normalize the time series with the smoothed amplitude of the light curve and subsequently measure asteroseismic parameters. No light-curve modeling is attempted for these systems.
\end{itemize}

\subsection{Asteroseismic analysis}\label{sec_33}

\subsubsection{Detection and properties of global pulsation modes}
We search for global oscillation modes in light curves where eclipses have been handled as described in Section \ref{sec_31}. In this paper, we only consider clear oscillatory excess power, and do not consider low signal-to-noise ratio excess power that in principle could be detected with filtered autocorrelation techniques \citep[e.g.,][]{mosser2009}. Indeed, since the light curves studied here could have remnants of eclipse features, the presence of harmonics of the orbital period would strongly perturb the computation of the autocorrelation function, leading to false mode detections. 

\begin{figure*}
  \centering
  \epsscale{1}
  \plotone{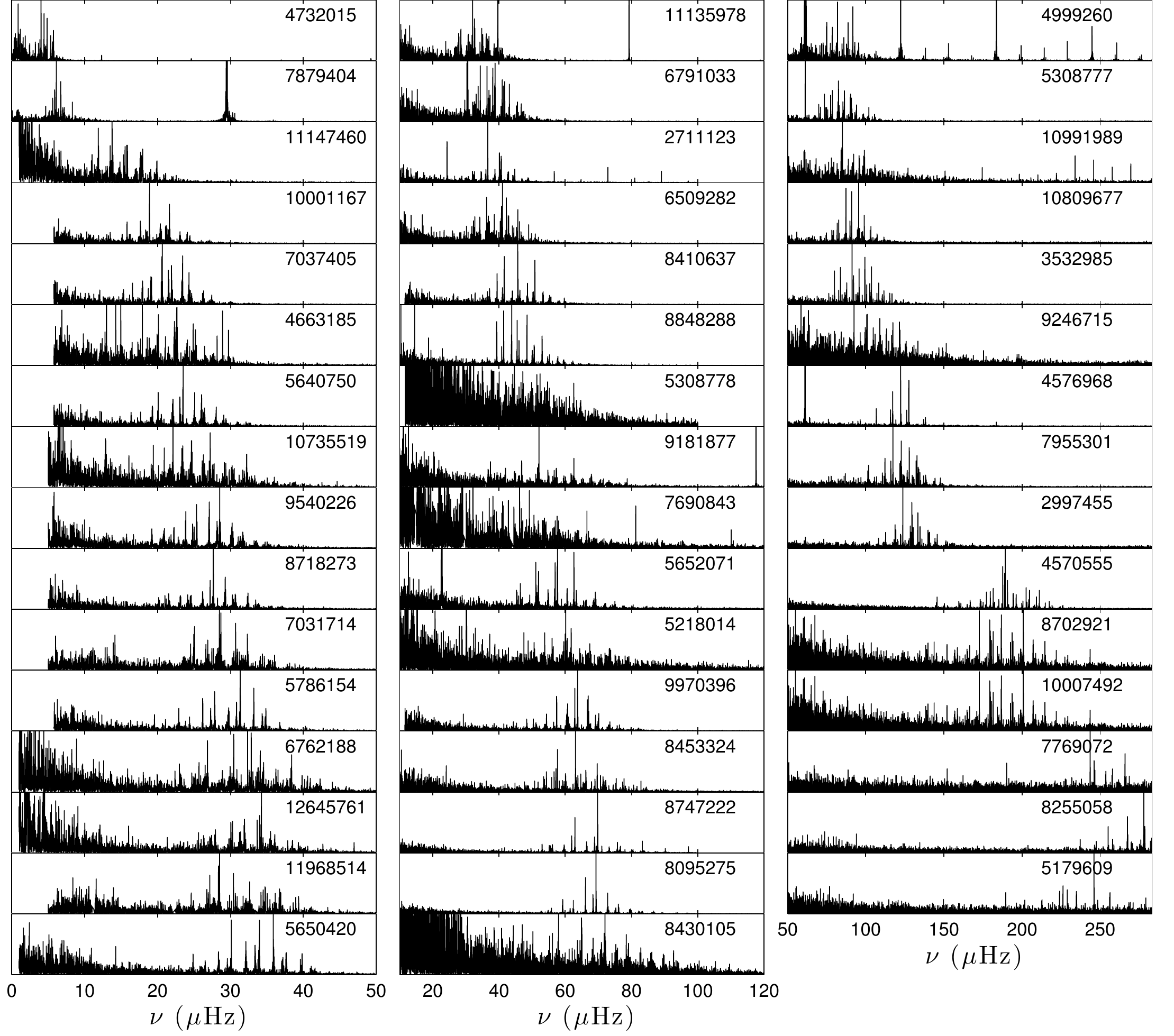}
  \caption{Power density spectra (PDS) of the 47 light curves in which global, solar-like oscillations are detected, sorted by increasing $\nu\ind{max}$ from top to bottom and left to right. Some spectra are truncated at low frequency because of the filtering applied to remove the signature of an eclipse. Note the different frequency $x$-axis scales in each column. Each system is labeled by its \textit{Kepler} ID number.\label{fig_PDS}}
\end{figure*}

Power density spectra (PDS) are computed with the discrete Fourier transform with no oversampling up to the Nyquist frequency (283~$\mu$Hz). We detect global modes in 47 of the 70 RG/EB candidates. The breakdown by binary classification is 37~D, two SD, five OC, and two ELV. All power spectra showing solar-like oscillations are presented in Figure \ref{fig_PDS}. 

One of the asteroseismic aims of this paper is the measurement of RG global mode parameters, such as  their frequency at maximum amplitude $\nu\ind{max}$ and  mean large frequency separation $\Delta\nu$.  The pair ($\nu\ind{max}, ~\Delta\nu$) allows us to estimate the RG's mass and radius through well-known asteroseismic  relations \citep{kjeldsen1995,huber2010}.

\begin{figure}
  \centering
  \epsscale{1.2}
   \plotone{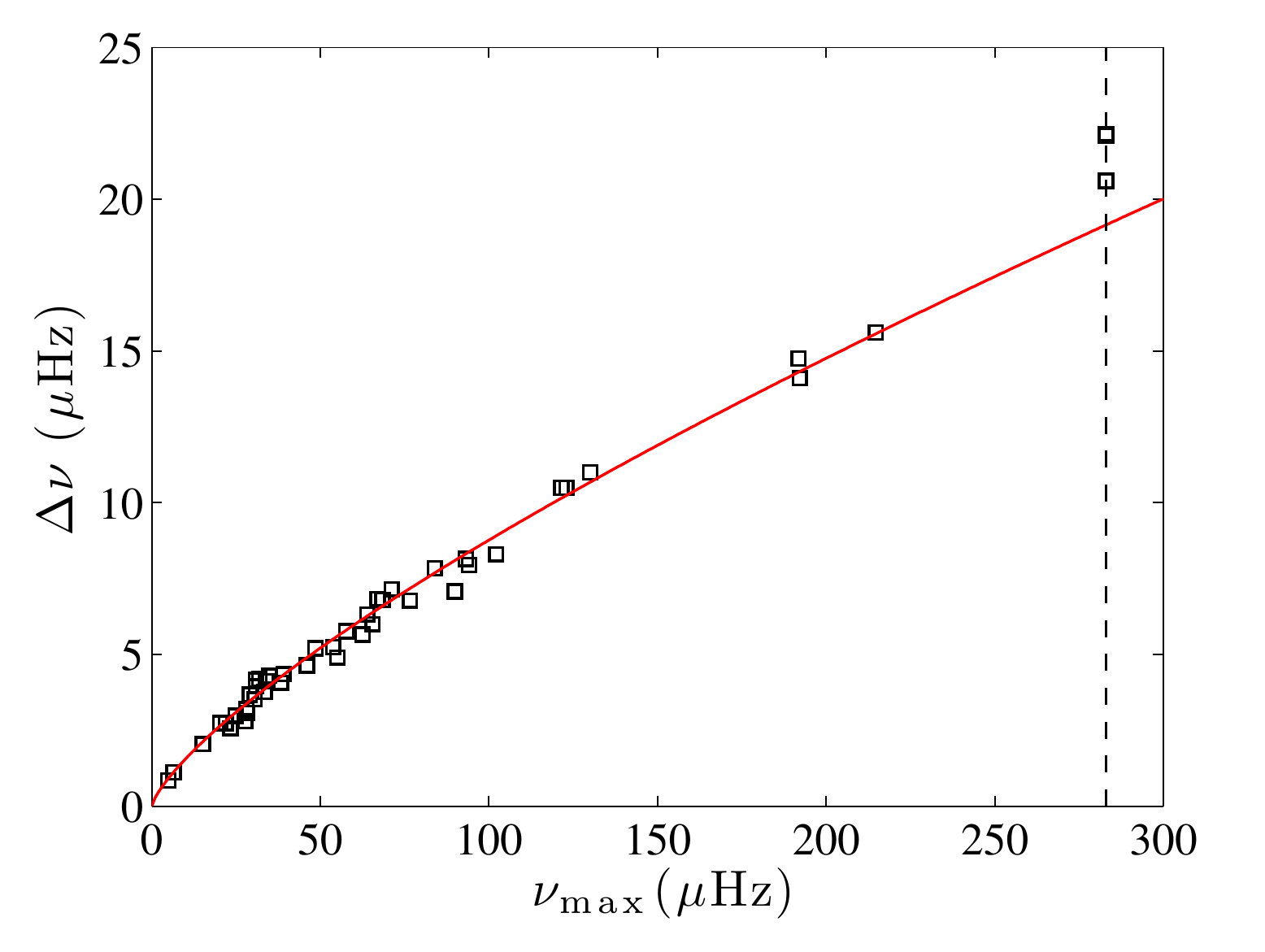}
  \caption{Large frequency separation ($\Delta\nu$) versus frequency at maximum amplitude ($\nu_{\rm max}$) for the 47 pulsating RG/EB candidates derived from asteroseismology. The red line shows the empirical relationship established from a large sample of RGs from \citet{Mosser_2012b}. The dashed vertical line denotes the Nyquist frequency for the observation cadence, above which three RGs have their $\nu_{\rm max}$. \label{fig_numax_Dnu}}
\end{figure}

\begin{figure}
  \centering
  \epsscale{1}
   \plotone{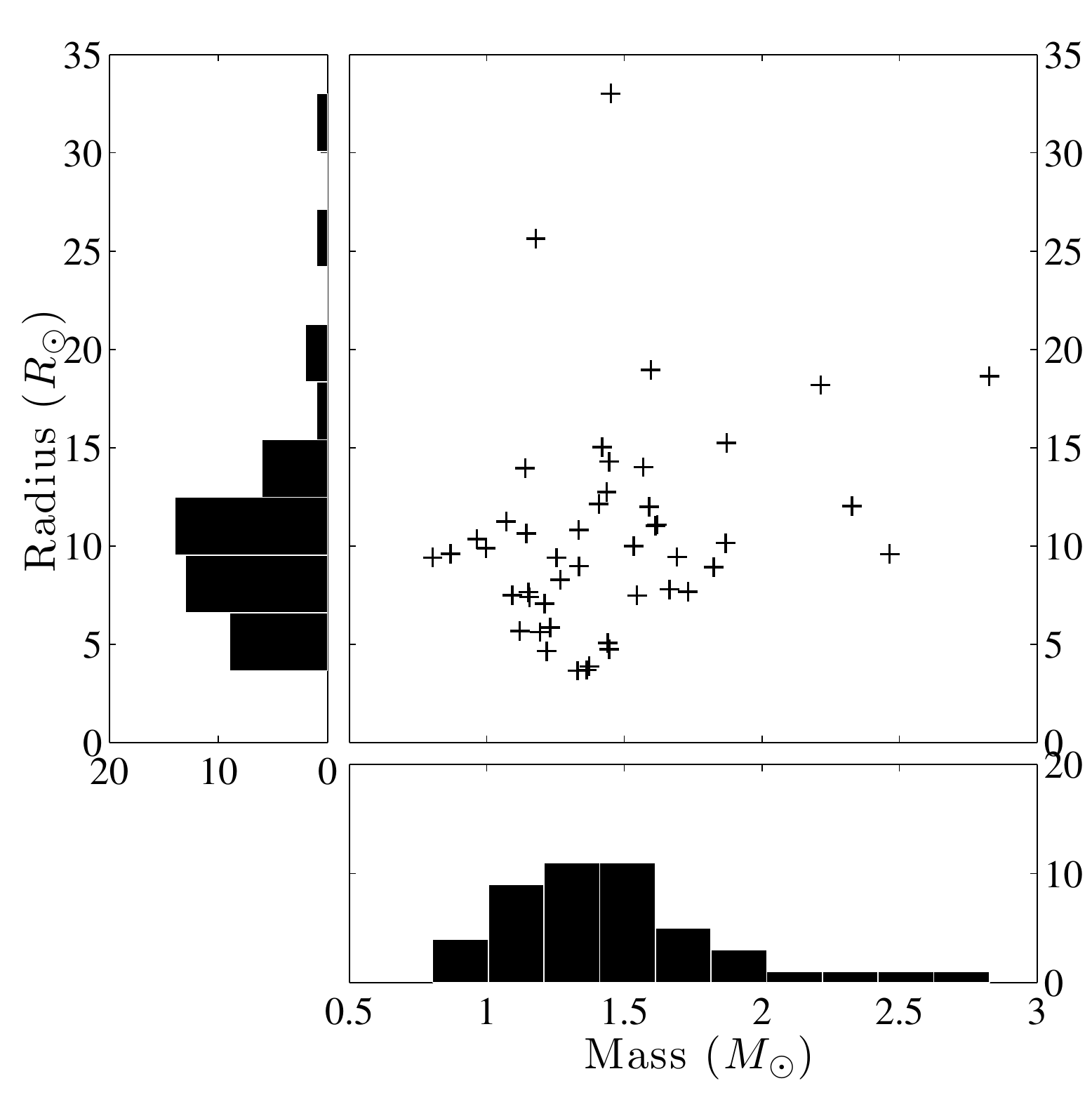}
  \caption{Histograms of masses and radii of the 47 pulsating RGs in RG/EB candidate systems. These have been derived asteroseismically and include new asymptotic scaling corrections \citep{Mosser_2012c}. The scatter plot shows the correlation between the two observables. \label{fig_mass_radius}}
\end{figure}

The frequency at maximum amplitude $\nu_{\rm max}$ is obtained by fitting the PDS with a Gaussian function for the mode envelope and a sum of semi-Lorentzian functions to model the stellar activity. This standard technique was first suggested by \citet{harvey1985} for the Sun, and is now widely used for solar-like stars \citep[e.g.,][]{chaplin2011}. 
The PDS is  divided by the semi-Lorentzian background to produce a whitened spectrum, and autocorrelation is then used to extract the mean large separation $\Delta\nu$. Figure \ref{fig_numax_Dnu} shows $\Delta\nu$ versus $\nu\ind{max}$ for the RG/EBs compared with the empirical relationship that was established on thousands of red giants, subgiants, and main sequence stars with CoRoT and {\it Kepler} data \citep{hekker2009, Stello_2009, mosser2010, huber2011, Mosser_2012b}. Only the three stars with $\nu\ind{max}$ close to the Nyquist frequency are significantly different from expectations, because the determination of $\nu\ind{max}$ is biased by the PDS truncation. 

To compute RG masses and radii, estimates of their effective temperatures $T\ind{eff}$ are required. As we concluded from analyzing the $T\ind{eff}$ distribution in Section \ref{sec_23}, we can safely assume that temperatures from the {\it Kepler} database correspond the red giants for the RG/EB sample. Using the asteroseismic parameters and these temperatures, along with updated asymptotic scaling factors for RGs \citep{Mosser_2012c}, Figure \ref{fig_mass_radius} shows the estimated masses and radii for the 47 pulsating RGs, which range from $0.8-2.8 M_\sun$ and $3.7-33.0 R_\sun$, respectively. The new scaling changes the values of mass and radius for each star by up to a few percent. These results are presented in Table \ref{table_2}.

\subsubsection{Mode identification and mixed modes}

\begin{figure*}
  \centering
  \epsscale{1.1}  
\plotone{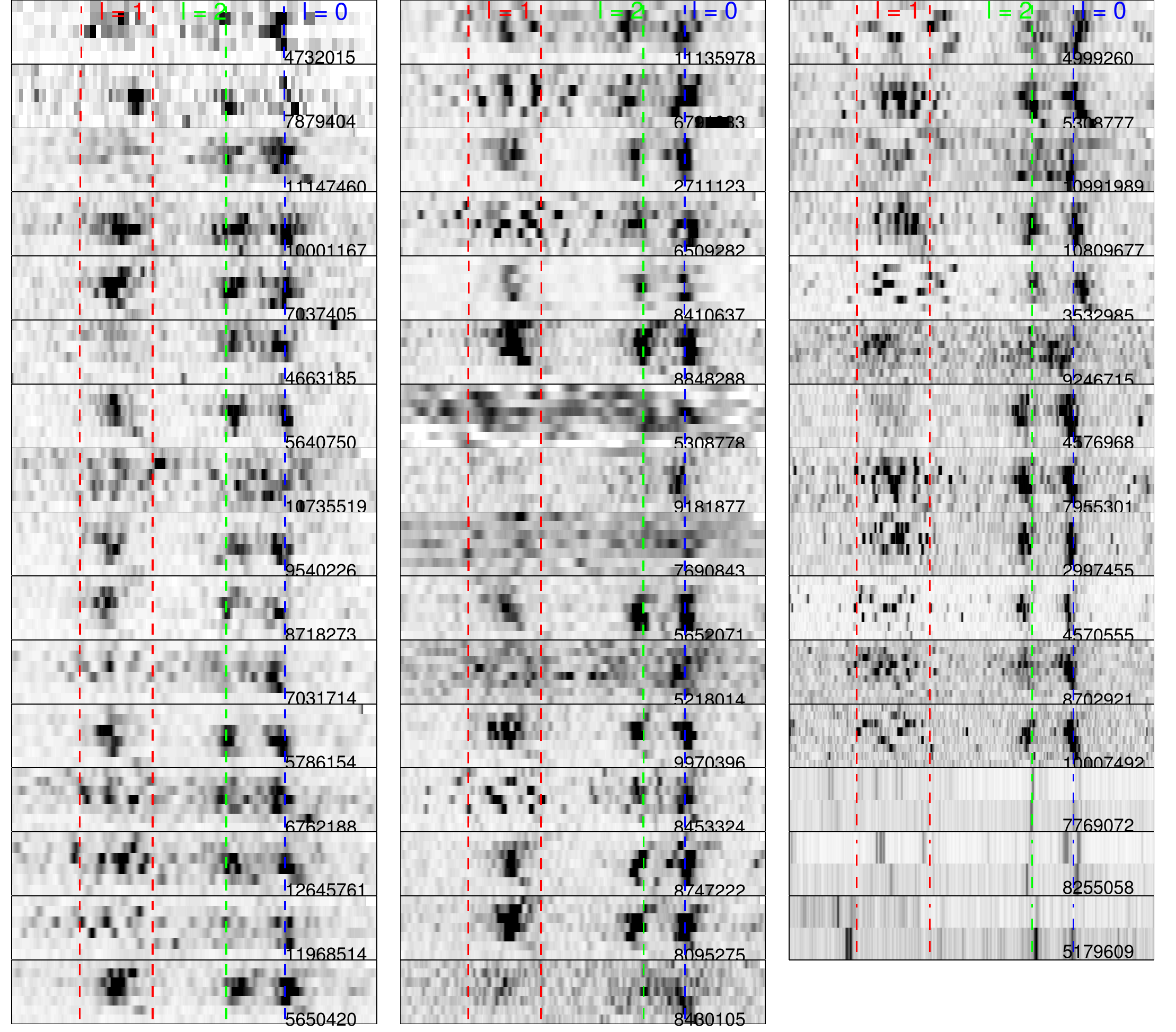}
\caption{\'Echelle diagrams of the 47 pulsating RG stars whose KIC names are labeled. The $x$-axis is the frequency modulo the large frequency spacing (i.e., from 0 to $\Delta\nu$), and the $y$-axis is frequency. The three noticeable dark mode ridges (for most stars) correspond to angular degree $\ell=1$, $\ell=2$, and $\ell=0$, from left to right, respectively indicated in dashed red, green and blue lines. Each case has been shifted so that the radial modes are roughly vertically aligned for illustration purposes. Each system is identified by its \textit{Kepler} ID number.
\label{fig_ech}}
\end{figure*}

Peaks in light curve power spectra correspond to unique stellar oscillations, and identifying these oscillations yields enormous insight into interior stellar properties. For example, modes that are a mixture of acoustic and gravity modes provide diagnostics for probing the stellar core. Such mixed modes may be used to determine if a star is on the red-giant branch (RGB) where it is still burning H in a shell surrounding the core, or if is part of the red clump (RC) after possibly experiencing a He flash and is now fusing He in its core \citep{bedding2010,bedding2011,mosser2011,mosser2012}. In addition, we may distinguish the secondary red clump (RC2) that consists of stars more massive than 1.8\,$M_\odot$. These stars have started to burn He in a non-degenerate core, which distinguishes them from low-mass stars in the main red clump \citep{Girardi_1999}. These mixed modes typically appear most clearly in the $\ell=1$ modes as a ``forest'' of peaks.

Figure~\ref{fig_ech} shows the \'echelle diagram of all 47 pulsating RGs in candidate RG/EBs. We detect the typical pair of ridges of solar-like oscillations ($\ell = 0, 2$ and $\ell = 1$ spaced by half the large separation) for 14 of the 47 oscillating RGs, while clear signatures of  mixed $\ell = 1$ modes are evident in 23 stars. The remaining targets present confusing oscillatory spectra. These few RGs that show suppressed mixed modes certainly belong to the RG group identified by \cite{Mosser_2012b}. Such low-amplitude mixed modes have been observed in a group of stars starting the ascent of the RGB, at all evolutionary stages. We note that 11 of the 14 stars with suppressed mixed modes have the longest orbital periods of all the RG/EBs.

Careful analysis of the dipole mixed-mode signatures allows us to classify red giants as either RGB or RC stars. This was first computed by measuring the bumped mixed-mode spacings \citep{bedding2010, mosser2011}. However, using the asymptotic development of mixed modes provides the exact period spacing $\DP$, which precisely characterizes the radiative core \citep{mosser2012}. It also allows access to the core rotation rate of these giants, and provides an estimate of $P_{\rm rot}$ \citep{Beck_2012,Mosser_2012c}. The values of these mixed-mode parameters for 12 of the 23 cases are given in Table \ref{table_2}. Interestingly, the cores are rotating with periods from 30 to about 385 days, seemingly uncorrelated with their masses and increasing in most cases with radius. Also provided in Table \ref{table_2} are the classifications of 27 RGs. 
We note that when neither the gravity-mode period spacing nor the bumped spacing are available, the value of the large separation $\Dnu$ can be used to classify RGs in certain cases: a RG with $\Dnu \ge 9\,\mu$Hz is on the RGB, but when $3.7 < \Dnu < 5.5 \mu$Hz, the probability of having a clump star is higher than 90\,\% \citep[see Figure 3 of][]{mosser2012}

\subsection{Light curve modeling}
\subsubsection{Modeling the light curves} \label{sec_32}

To independently compare physical parameters of a subset of the binary systems, we model their light curves using the Eclipsing Light Curve (ELC) code \citep{orosz2000} and/or the JKTEBOP code \citep[e.g.,][]{southworth2009}. ELC in particular uses a genetic algorithm and Monte Carlo Markov Chain optimizers to simultaneously solve for a suite of stellar parameters. It is well-suited to this analysis because additional data can be added as it becomes available, such as radial velocities from spectra. Using only the \textit{Kepler} light curves, we use ELC and JKTEBOP to solve for relative fractional radii $R_1/a$ and $R_2/a$, temperature ratio $T_2/T_1$, orbital inclination $i$, eccentricity $e$ and $e \cos \omega$ when applicable (where $\omega$ is the longitude of periastron), the \emph{Kepler} contamination factor, and the stellar limb darkening parameters for the quadratic limb darkening law.
The orbital period was determined previously through a simple iterative technique and held fixed during this analysis. In the limiting case where the two stars are sufficiently separated as to be spherical in shape, ELC has a ``fast analytic'' mode that uses the equations of \citet{gimenez2006}. We make this assumption here because our modeled subsample includes only well-detached binary systems.

\begin{figure}[t]
  \epsscale{1.2}
  \plotone{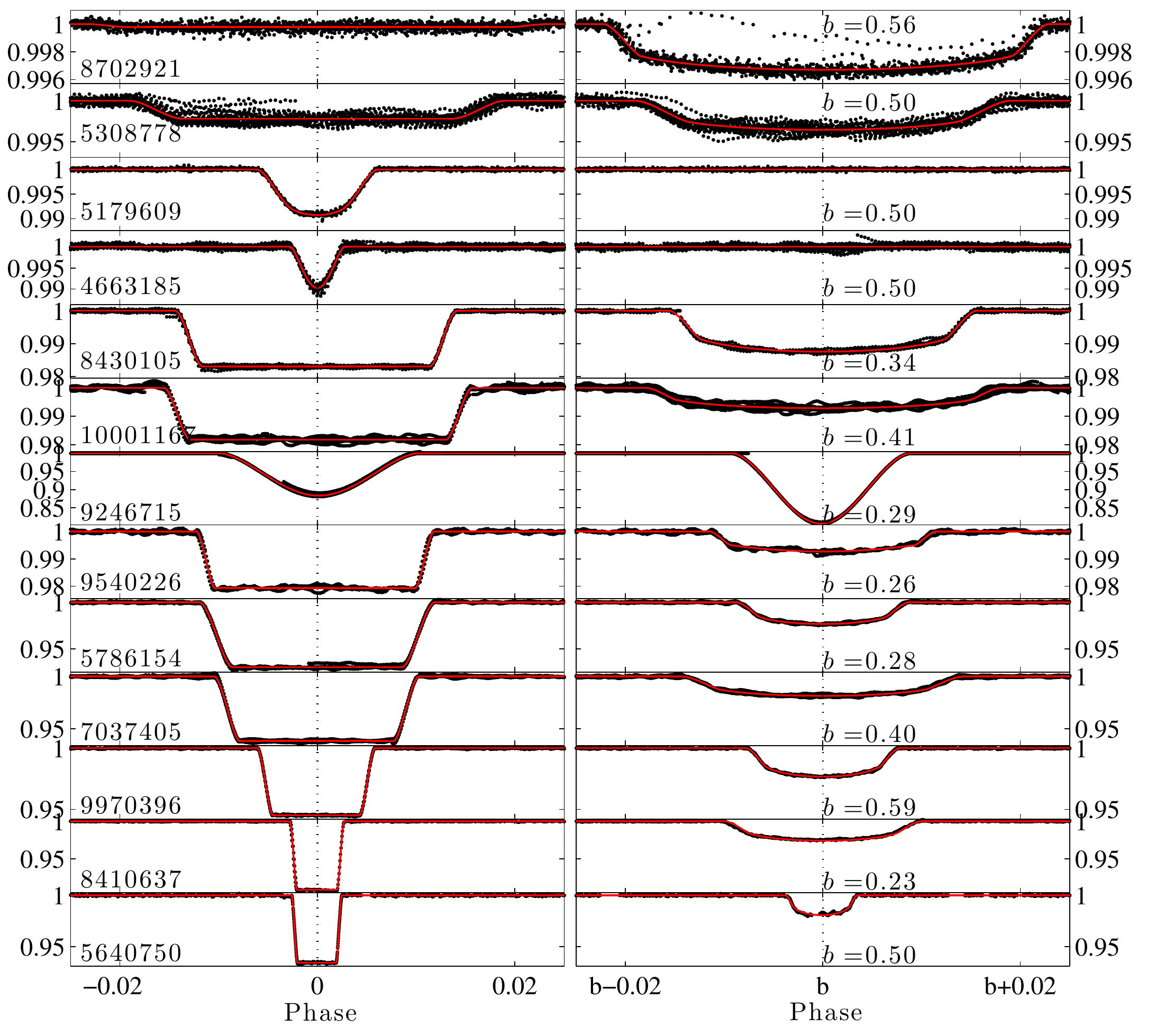}
\caption{Modeled light curves for 13 of the 14 longest-period RG/EB pulsating systems. In each case, the secondary eclipse (defined here as the RG eclipsing the secondary star) has been set to an orbital phase of zero. The primary eclipses (secondary star eclipsing the RG) have also been aligned, and the given $b$ value indicates the phase of the primary eclipse with respect to the secondary. (A EB with a circular orbit would have $b = 0.5$.) The $y$-axis for each panel is the normalized relative flux. The best-fit parameters of the models are given in Table \ref{table_1}.
\label{fig_lcmod}}
\end{figure}

\begin{figure}[t]
  \epsscale{1.2}
  \plotone{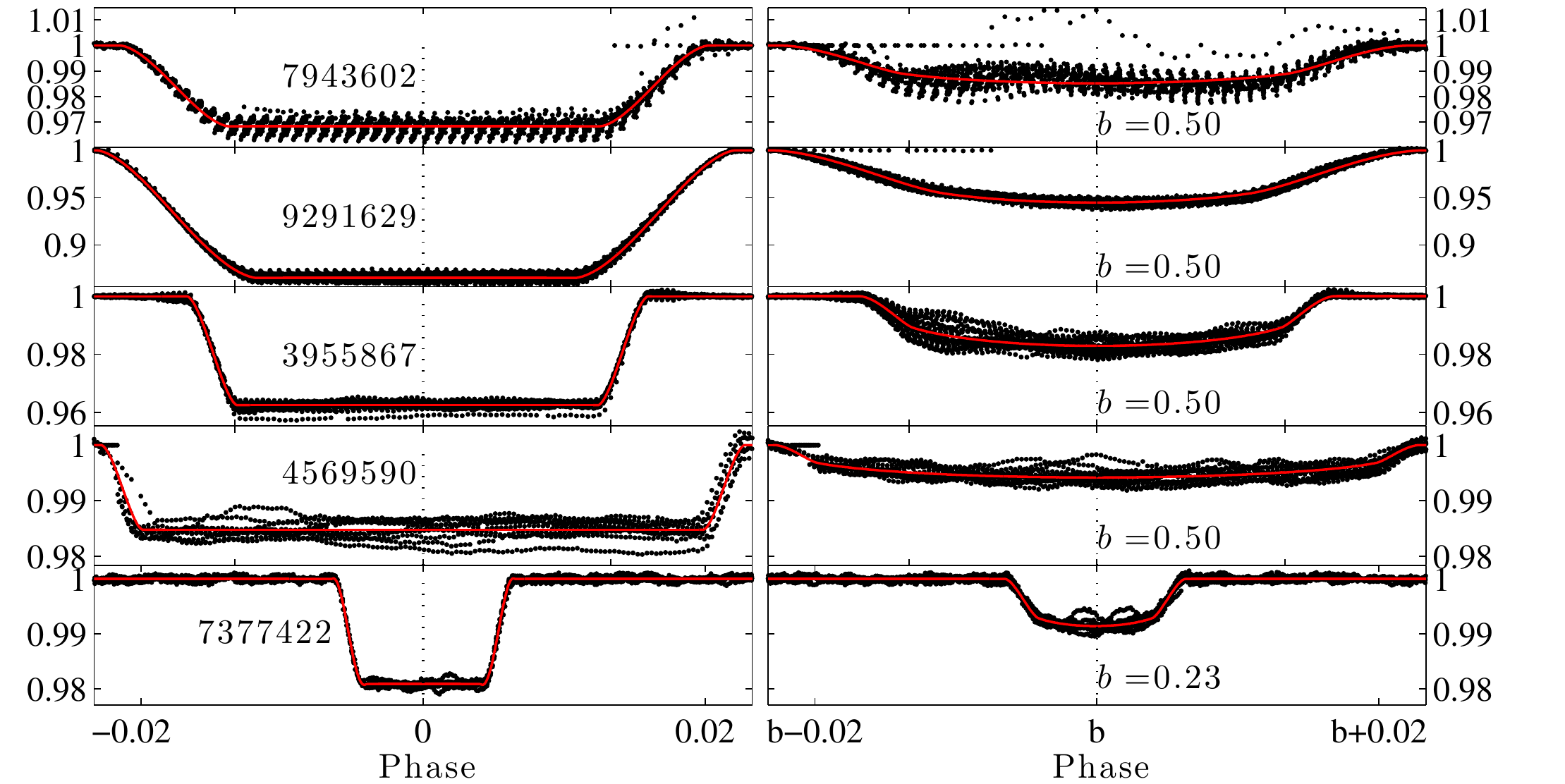}
\caption{Modeled light curves for the five longest-period RG/EBs with no RG pulsations. As in Figure \ref{fig_lcmod}, each secondary eclipse (defined here as the RG eclipsing the secondary star) has been set to an orbital phase of zero. The primary eclipses (secondary star eclipsing the RG) have also been aligned, and the given $b$ value indicates the phase of the primary eclipse with respect to the secondary. The $y$-axis for each panel is the normalized relative flux, and the best-fit parameters of the models are given in Table~\ref{table_1}.
\label{fig_lcmodbis}}
\end{figure}

The derived parameters for a subset of 18 binary systems are presented in Table \ref{table_1}.\footnote{An online sortable version of this table is available at http://nsol2.nmsu.edu/solarstorm/index.php} We have analyzed the 18 most promising detached systems where a RG likely belongs to the EB (see Section \ref{sec_41}). Thirteen of them present RG oscillations, and the other five are the longest-period systems where no detectable pulsations are observed. This subset actually corresponds to the systems with the longest periods and with no identified contamination from nearby stars (see Section \ref{sec_35}). The light curve data and models are shown in Figures \ref{fig_lcmod} and \ref{fig_lcmodbis}.

\subsubsection{Search for  eclipse timing variations} \label{sec_34}

As discussed in Section \ref{sec_23}, a significant number of the candidate RG/EBs are probably not \textit{bona-fide} RGs in binary systems, but are more likely the result of blendings from a nearly aligned RG and EB. A further possibility is the presence of RGs in triple (multiple) systems in which two ``small'' stars mutually eclipse a primary component \citep{derekas2011, carter2011}. Within the triple-system hypothesis, if the RG's orbit is elliptical and the distance of the RG to the pair of eclipsing stars is short enough to let tidal forces contribute to the evolution of the eclipsing system, we might detect eclipse timing variations (ETVs). This further requires that the RG orbital period is, at maximum, the same order of magnitude as the observation length. In addition, we may expect some variations in the eclipse depth and width if the precession of the EB system is rapid and strong enough to perturb the inclination angle of the EB orbital plane. The detection of ETVs is a way to claim a system is at least triple, but the absence of a ETV detection cannot be considered proof that a system is not multiple since variations may be undetectable or on time scales much larger than the observation length. In addition, if the RG orbit is almost circular, no ETVs are expected. The presence of ETVs in this particular sample suggests that the RG is the third body that perturbs the eclipse timing, but this can only be proven with mass determinations from detailed ETV models and/or the addition of radial velocities to the light curve analysis. We note that low-mass triple systems are common, particularly for systems with a close-in binary (e.g., \citealt{Tokovinin_2006}), so a chance alignment of a triple system with a background RG is also possible.

Modeling ETVs is a common approach to derive masses and orbital parameters for exoplanetary systems with at least two planets (e.g., \citealt{Agol_2005, Ford_2012, Fabrycky_2012, Steffen_2013}). Modeling ETVs of multiple stellar systems is beyond the scope of this paper; rather, we first determine whether ETVs are detectable and subsequently measure their amplitudes and periods. We use one of two methods to measure eclipse timings depending on the type of binary system. For D and SD systems, we fit the primary and secondary eclipses with the \citet{mandel2002} function used for fitting exoplanetary transits in the small-planet approximation, which acceptably reproduces the behavior of stellar eclipses. However, exoplanetary transit functions are not suitable for OC systems. Therefore, for the OC binaries, we measure the minimum intensity time by interpolating around the minimum of both eclipses with spline interpolation. Such an approach would not work for D systems with long orbital periods because the eclipse shape is usually strongly modulated by stellar spots, but it is suitable for SD systems. Once ETVs are detected, despite their typically asymmetric profiles, we estimate their periods by fitting a simple sine curve via the least-square method.


\begin{figure*}
  \epsscale{1.15}
  \plotone{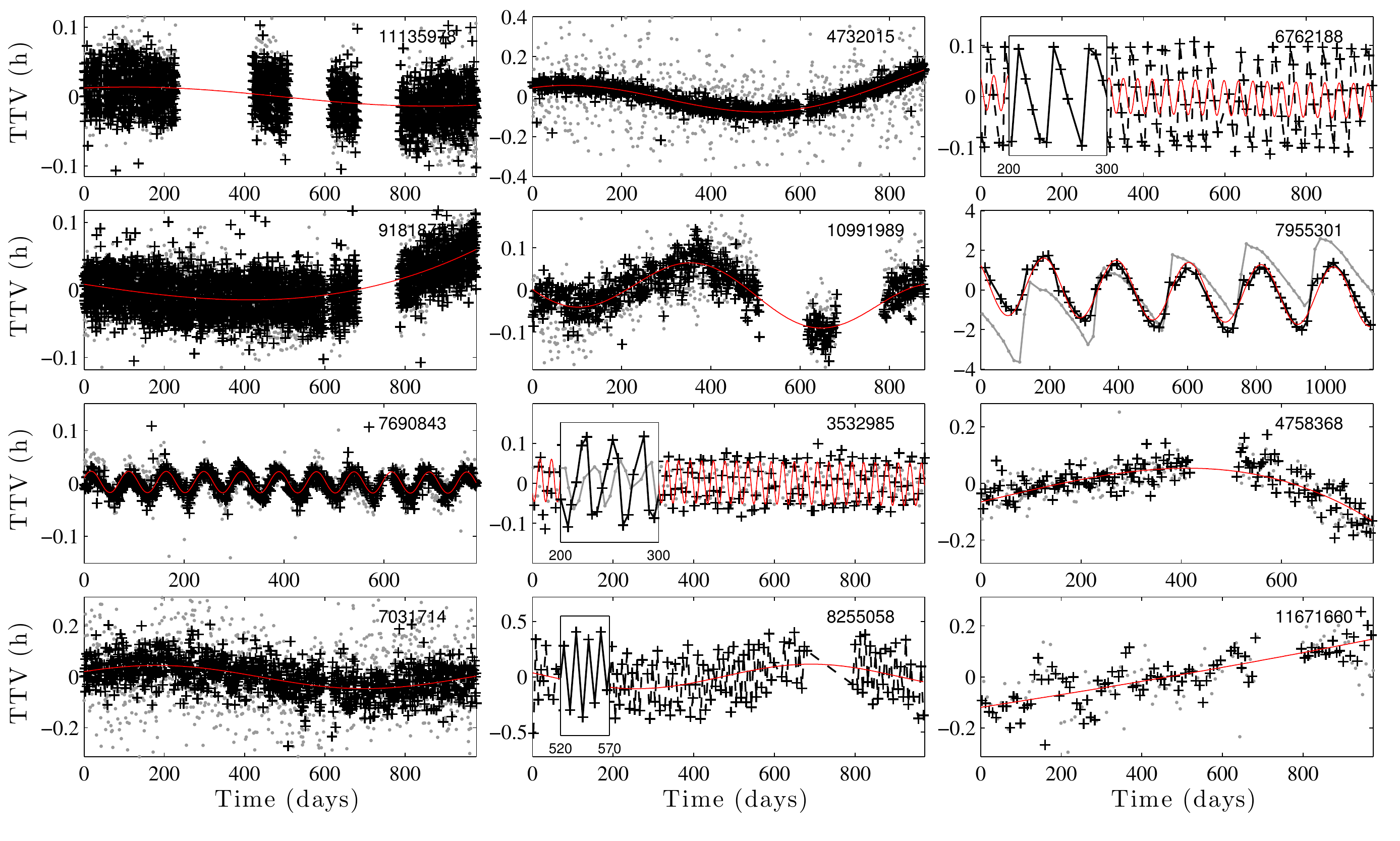}
  \caption{Eclipse timing variations (ETVs) in hours of the 12 systems that show detectable variations. The $y$-label ``O-C'' stands for (observed $-$ computed) eclipse epoch. The timings are plotted for the deepest eclipses (black plusses) and for the shallow eclipses when  detectable (grey dots). Red lines indicate sine-curve fitting on the deepest eclipses. For KICs 3532985, 8255058, and 6762188, the sub-boxes zoom in for a more detailed view of the ETVs. \label{fig_ETV}}
\end{figure*}

Eleven systems exhibit ETVs at various signal-to-noise ratio levels, as shown in Figure \ref{fig_ETV}. KIC 7990843, 7031714, 4732015, 10991989, and 7955301 show ETVs that are unambiguously coherent between primary and secondary eclipses, and for which periods are shorter than or near to the total observation length. \citet{slawson2011} note that KIC 7955301 (shown in Figure \ref{fig_clean_3}) presents eclipse depth variations, but based only on Q1--Q2 data they could not determine whether these effects were real or due to aperture jitter from quarter to quarter. With a longer light curve now available, we see this system presents the highest ETV amplitude of about four hours. Precession effects are detectable in this system through the shift between primary and secondary eclipse timings and the long time-scale evolution of the eclipse depths. We also note the unexpected ETVs  detected for contact systems 11135978 and 9181877, which show variations on time scales longer than the observation length (1400 and 5500 days, respectively) and with a mean amplitude of only about 0.83 minutes. This is much lower than the 29-minute observing cadence and is about half of the ETV standard deviation about the fitted sine curve of 1.68 minutes.

%
%
\section{Discussion}\label{sec_4}
\subsection{Most RGs do not belong to their associated EB} \label{sec_41}

\subsubsection{Testing binarity with Kepler's law }
\begin{figure}
  \centering
  \epsscale{1.2}
  \plotone{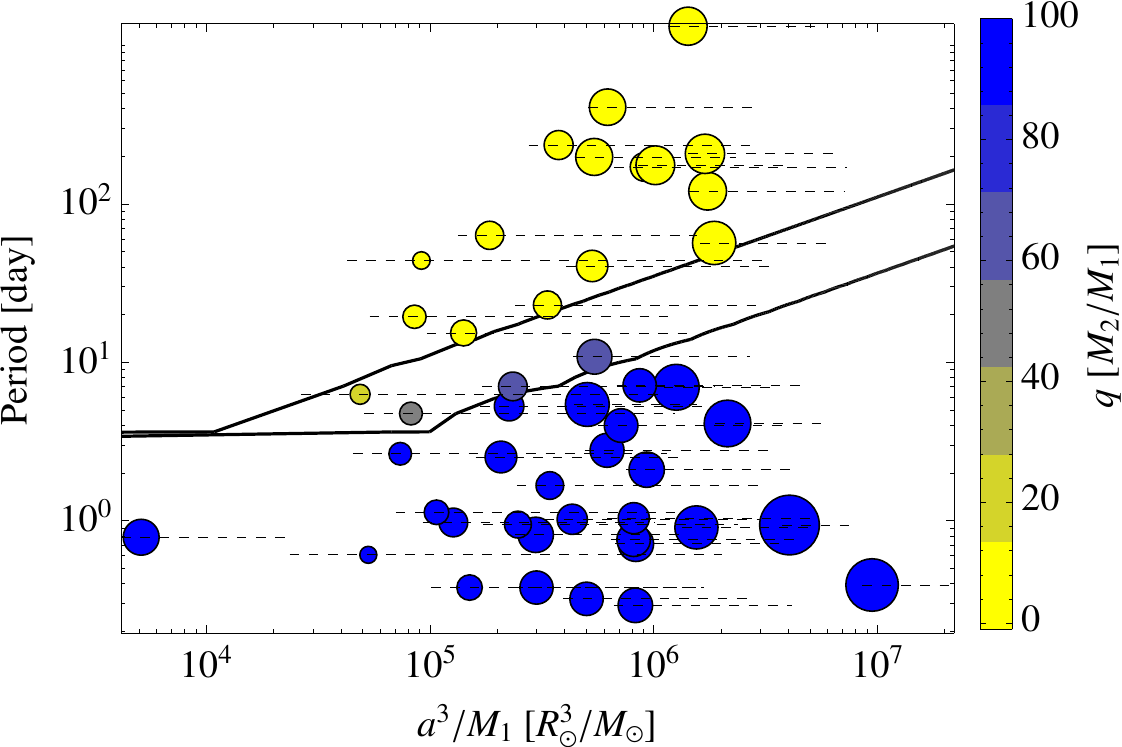}  
  \caption{Mass ratio $q$ as a function of orbital period and $a^3/M_1$ for the pulsating RG/EB candidate sample based on the orbital parameters in \citet{slawson2011} and those obtained in this work from asteroseismology. For all systems, the symbols are plotted assuming $R_2=R_\odot$, and the dashed lines show the extent of $a^3/M_1$ if the secondary star's radius spans $0.1 \ R_\odot < R_2 < 10 \ R_\odot$. The color at each system's position represents the mass ratio $q$. Contours of constant $q = 10$ and 100 are shown by the diagonal lines. The size of each symbol is proportional to the pulsating RG stellar radius, $R_1$, measured asteroseismically. The pulsating OC systems are not included here. The colors reflect the huge variation of mass ratio across the sample, and clearly distinguish which systems are likely to be a bona-fide RG in an EB (yellow, i.e., $q\sim1$) from those that are unlikely (blue, i.e., $q\geq10$).}
  \label{qfact}
\end{figure}

We have shown that contamination in the \textit{Kepler} target pixel files confirmed seven systems as false positive RG/EBs. To uncover more false positives, we can rearrange Kepler's third law to express the stellar mass ratio $q = M_2/M_1$ as a function of the orbital period $P$, the semi-major axis $a$, and the mass of the primary star $M_1$:
\begin{equation}
  q = \left(\frac{2\pi}{P}\right)^2\ \frac{a^3}{GM_1}\ -\ 1.
  \label{q}
\end{equation}
Note that our convention defines star 1 to be the RG. We use RG masses and radii from the asteroseismic analysis (see Section \ref{sec_33}) with a range of radii for the secondary star, $0.1~R_\odot < R_2 < 10~R_\odot$, and also consider the relative stellar radii $(R_1+R_2)/a$ and $P$ from eclipse fitting and \citet{slawson2011}. With all of this information taken together, we estimate the $q$ required to satisfy Kepler's third law for the pulsating sample.


Figure~\ref{qfact} shows the distribution of $q$ in the phase space where $a^3/M_1$ and $P$ are the independent variables. The symbols are plotted for the case where $R_2=1R_\odot$. The horizontal extent of possible values of $a^3/M_1$ is shown over the radius range, where larger radii shift the values to the right along the dotted lines. Upon considering modest values of $R_2$, it appears that only 12 of the pulsating RGs in candidate RG/EBs may truly belong to EBs (not including the ETV systems), as these have $q\lesssim 10$. Cases in which the RG would need to be part of a binary with a companion star $\gtrsim 10$ times more massive are considered unrealistic.  
In this analysis, several systems take on a value of $q<0$, which is of course unphysical: this is likely the result of inaccurate estimations of the relative stellar radii. The likely 13th candidate (KIC 5640750) system is not yet conclusive as its orbital period is longer than the observation time. The likely 14th candidate (KIC 8095275) does not allow us to estimate $q$ since it is a ``heartbeat system'' for which its orbital parameters are unknown and would require specific light curve modeling and radial velocities to determine (see Section \ref{sec_433} and \citealt{Thompson_2012}). However, as such ``heartbeat'' light curve features are unique, we are confident that it is a \textit{bona-fide} RG/EB system.

In summary, the analysis here and in Section \ref{sec_35} provides strong evidence that only 13 of the 47 pulsating systems are true RG/EBs in detached configurations, and one is a RG in a non-eclipsing binary system. All of these have orbital periods greater than 19~days. All of the shorter-period OC and SD systems are either contaminated or the RG is part of a multiple-star system. The rest of the D cases must be where the EB and the RG fall in the same line-of-sight but may be weakly gravitationally bound. Finally, we note that testing the likeliness of RG/EB candidates is not possible when no RG pulsations are detectable. However, true RG/EBs may belong to this sample. In particular, the five non-pulsating RGs with the longest orbital periods (shown in Figure \ref{fig_lcmodbis}) are almost certainly true RG/EBs, as they have light curves very similar in appearance to the 13 pulsating \textit{bona-fide} RG/EBs.

\subsubsection{Candidate multiple-star ($> 2$) systems}

We find eleven stars with ETVs that suggest the RG is part of a multiple system composed of a close-in EB and a more distant RG in an elliptical orbit. This configuration, a ``hierarchical triple system with two low-mass stars,'' was also detected  for two cases in \citet{carter2011} and \citet{derekas2011}. In \citet{derekas2011} the low-mass stars were co-planar with the primary component (a RG) and all eclipses were visible, while in \citet{carter2011}, each star separately eclipses the disk of the primary component (a subgiant). No solar-like oscillations were observed in either case, and so here we report the first time global $p$-mode oscillations are detected for an RG in such a system. In the eleven ETV detections, only the close-in eclipses are observed and the RG's presence is indirectly deduced as the perturber. We find this in seven D, two SD, and two OC cases. The system KIC 4758368 is the only out of the eleven where we do not detect any RG oscillations (see Table \ref{table_1}).

\subsection{EB and multiple-star candidates} \label{sec_43}

We classify the RGs displaying oscillations and presumably belonging to a double or triple system into one of three categories. First, \textit{fundamental} cases are those that deserve to be studied in more detail in the future, to be used as cornerstones for testing stellar-evolution models, and for which current \textit{Kepler} data are sufficient for accurate eclipse and asteroseismic modelings. Second, the \textit{promising} cases are those for which eclipse or asteroseismic modeling require additional \textit{Kepler} observations to be considered \textit{fundamental}. Third, \textit{intriguing} cases are those in which we cannot rule out the detected signal as a false positive, and where more data could lead to unexpected discoveries.

\subsubsection{Fundamental cases} 
The fundamental cases are EBs and one hierarchical triple system, with orbital periods over 19~days. Due to the  high signal-to-noise ratio of their oscillation patterns, these systems are suitable for precise modeling of their interior properties. The candidates with highly eccentric orbits are interesting for studying tidal interactions in multiple-star systems. Their modeled light curves are shown in Figure \ref{fig_lcmod}. We list them here, sorted by the spectral type of the companion star:

\begin{description}\itemsep 0cm
\item[M or K dwarf companion.] We identify two cases where the EB is composed of a RG and a smaller, cooler companion, rendering the primary eclipse deeper than the secondary eclipse.\\
\textbf{- KIC 8702921} is a 19-day EB, which shows clear RG oscillations typical of the RGB. Its asteroseismic radius and mass are $5.1\pm0.1\,R_\odot$ and $1.4\pm0.1\,M_\odot$. We find that the secondary star is compatible with an M dwarf, as its mass is estimated as  $0.6\pm0.3\,M_\odot$ with radius $0.44\pm0.01~R_\odot$, and an effective temperature of $2600$~K.  We see  variations in the light curve that could be due to M-star activity (i.e., spots), tidal distortions, and ellipsoidal and Doppler beaming. We observe a 97.8 day (almost precisely five times the orbital period) modulation in the light curve of rather large amplitude, which is comparable to the eclipse depth. \\
\textbf{- KIC 5308778} has a 41-day orbital period and low signal-to-noise ratio RG oscillations, corresponding to an asteroseismic radius and mass of $9.4\pm0.8~R_\odot$ and  $1.3\pm0.3~M_\odot$. From eclipse modeling, the companion's radius and mass $R_2=0.6\pm0.1~R_\odot$ and $M_2=0.3_{-0.3}^{+0.7}~M_\odot$ fit with an M dwarf, while its effective temperature $T_2 = 4140$ K fits with a K dwarf. Its light curve presents photometric variations of amplitude up to 6\,\%, which is 17 times larger than the deepest eclipse. These variations are quasi-periodic with periodicity almost equal to the orbital period. The absence of any orbital eccentricity suggests that the photometric variations are due to features on the RG, in a system that is tidally locked on a circularized orbit. The periodic fluctuations appears incompatible with spots on the companion star since its relative brightness is 0.2\,\% of the RG.

\item[G or late F dwarf companion.] Three systems are characterized by a RG with a G or late type F star on an eccentric orbit with typical $e \sim 0.2$, and light curves that present large variations.\\
\textbf{- KIC 8430105} is a 63-day period EB with clear RG oscillations that correspond to an asteroseismic radius and mass of $7.4\pm0.3~R_\odot$ and  $1.2\pm0.1~M_\odot$. No mixed modes are identified, but the large separation value and the seismic mass estimate suggest this RG cannot belong to the red clump and is more likely an RGB star. From eclipse modeling, the orbit appears to be eccentric as $e=0.26$, and we estimate the companion's radius, mass,and effective temperature to be $R_2 = 0.83\pm0.04~R_\odot$, $M_2 = 0.9\pm0.4~M_\odot$, and $T_2 = 5960$~K. Thus, we likely have an RG in orbit with a solar analog. A strong variation in the light curve of (almost precisely) two times the orbital period and five times the eclipse depth is found. At the bottom of the primary eclipse (the G star passing in front of the RG), we observe sharp peaks that can reach up to half of the eclipse depth and last about of third of the eclipse time. These could be strong flares on the solar analog, which seem unlikely, or hot spots.\\
\textbf{- KIC 10001167} is a 120-day EB with clear RG oscillations that are characteristic of an RGB star of a $14.0\pm0.8 ~R_\sun$ and $1.1\pm0.2~M_\sun$. Modeling the eclipses indicates the orbit is eccentric with $e = 0.16$, and the companion is likely a G0 or F9 star with radius $R_2=1.1\pm0.1~R_\sun$, mass $M_2=1.0\pm0.6~M_\sun$, and effective temperature $T_2=6090$~K.\\
\textbf{- KIC 9970396} presents a 235-day period and clear modes of a RG with $8.3\pm0.3~R_\odot$ and $1.3\pm0.1~M_\odot$. The orbit appears to be eccentric with $e=0.20$, and its companion is likely a late F star of radius $ R_2=1.1\pm0.1~R_\odot$, mass $M_2 = 1.2\pm0.2~M_\odot$, and effective temperature $T_2 = 6060$~K. In contrast to the two previous cases, its light curve does not exhibit high photometric variations.

\item[F type companion.] These four bona-fide RG/EBs are likely composed of an RGB star with an F type companion on an eccentric orbit ($e = 0.23-0.67$), with orbital periods ranging from 175 to 408 days. In these cases, the secondary F star's mass is estimated to be nearly these same or even slightly larger than the RG star mass.  However, the error bars on the masses are large so that the parameters are consistent with the RG mass being larger than the secondary star mass, as would be expected if the stars formed together and the more massive RG finished its main-sequence phase first.  However, it is also possible that the RG could lose several tenths of a solar mass in a wind during the RG phase, or could have transferred mass to the companion since the orbits are highly eccentric, so that at the present time the RG has a slightly lower mass than its companion. \\
\textbf{- KIC 9540226} presents a 175.5-day orbital period and clear RG oscillations that are characteristic of an AGB or RGB star with $14.0\pm0.7 ~R_\sun$ and $1.6\pm0.2~M_\sun$. Eclipse modeling indicates the orbit is eccentric with $e = 0.39$, and the companion is likely an F star of radius $R_2=1.4\pm0.1~R_\sun$, mass $M_2=1.5\pm0.7~M_\sun$, and surface temperature $T_2=6920$~K. \\
\textbf{- KIC 5786154} presents a 197.9-day period and shows clear oscillations of a a RG of $12.7\pm0.6~R_\odot$ and $1.4\pm0.2~M_\odot$. The orbit appears to be eccentric with $e = 0.38$, and the companion is likely an F star of radius $R_2=1.9\pm0.1~R_\odot$, mass $M_2 = 1.8\pm0.7~M_\odot$, and surface temperature $T_2 = 6545$~K. \\
\textbf{-  KIC 7037405} presents a 207-day period and clear modes of an RGB star with $15.0\pm0.9~R_\odot$ and $1.4\pm0.2~M_\odot$. The orbit appears to be eccentric with $e=0.23$, and the companion is likely an F star of radius $ R_2=1.9\pm0.1~R_\odot$, mass $M_2 = 1.5\pm0.7~M_\odot$, and surface temperature $T_2 = 6400$~K.\\
\textbf{- KIC 8410637:} This system was the only EB previously known to host a RG displaying global oscillations \citep{hekker2010}. Its light curve is one of the most challenging among the set of EBs to model because of a high eccentricity ($e=0.67$), coupled with relatively small stellar radii compared to the semi-major axis, $(R_1+R_2)/a = 3.67\,\%$. We confirm the system is composed of a RG and an F star, whose respective sizes and masses are $(R_1, M_1) = (11.0\pm0.5 R_\sun,1.6\pm0.2 M_\sun)$ and $(R_2, M_2) = (1.6\pm0.1 R_\sun,1.8\pm0.7 M_\sun)$. 

\item[Probable $\delta$-Scuti companion.] The system KIC 4663185 presents a 57-day orbital period and is the only from our sample that has a power spectrum with two clear sets of oscillations. On the one hand, RG oscillations are observed at $\nu\ind{max} = 23\,\mu$Hz: this corresponds to a rather massive RGB star with radius $18.2\pm1.2~R_\odot$ and mass $2.2\pm0.4~M_\odot$. On the other hand, a second oscillation spectrum is observed at $\nu\ind{max} \simeq 132~\mu$Hz (Fig. \ref{fig_de_trop}). These mode amplitudes and widths are too high and narrow, i.e., the lifetimes are too long to match with a RG oscillation spectrum. The oscillation spectrum with $\nu\ind{max} \simeq 132~\mu$Hz is consistent with a $\delta$-Scuti or $\beta$-Cep star p-mode spectrum.  However, a $\beta$-Cep star is probably too massive ($\sim10~M_\sun$) to have evolved together with an RGB star of current mass $2.2~M_\sun$. Regarding the light curve, only one type of eclipse is detected and they appear to be shallow and grazing. We associate the detected eclipses with the secondary transit, i.e., the companion star is eclipsed, since the $\delta$-Scuti is likely to present a higher surface brightness than the RG.  With this hypothesis, we put an upper on the companion's size of $R_2 < 1.5\pm0.1~R_\sun$. The companion's size estimate is consistent with a main-sequence star of mass $\sim1.6 M_\sun$, which fits with the interpretation that the secondary could be a  $\delta$-Scuti star.  In the future, we will obtain ground-based spectroscopic observations, and attempt to match the oscillation frequency spectrum with stellar pulsation models to confirm this hypothesis and better constrain the parameters for the secondary.

\item[RG companion.] The system KIC 9246715 is a 171.3-day EB showing RG oscillations with a low signal-to-noise ratio compared to the other oscillating RGs, despite it being the brightest system (K$_p = 9.266$). From eclipse modeling it appears that this system has an eccentric orbit ($e =0.35$) and is composed of a pair of RGs of similar radii and temperature ($R_2/R_1 = 0.77$, $T_2/T_1 = 1.03$). However, we detect only one oscillation pattern ($\nu\ind{max} = 102~\mu$Hz), whose asteroseismic parameters correspond to a RG of $7.7\pm0.4~R_\sun$ and $1.7\pm0.3~M_\sun$. The large separation indicates it is a RGB star. From this, we deduce the companion's radius and mass to be $R_2=5.3\pm0.3~R_\sun$ ad $M_2 = 0.8\pm0.7~R_\sun$. No pulsations from the companion star are detected: it likely pulsates with a lower signal-to-noise ratio in the same frequency range, since the companion's frequency at maximum amplitude is expected to be $\nu\ind{max} = 96 _{-84}^{+133}~\mu$Hz. In addition, the autocorrelation of the light curve reveals the presence of a sine modulation with about a 3-day period with a mean peak-to-peak amplitude of about 100 ppm.

\item[Hierachical triple system.] The system KIC 7955301 displays high signal-to-noise ratio ETVs (4-h amplitude, Figure \ref{fig_ETV}) 
and clear RG pulsations, typical of an RGB star, with $1.2\pm0.1~M_\sun$, $5.9\pm0.2~R_\sun$, and a core rotation period of 30 days. From the eclipse timings, we are able to infer that the RG takes 210 days to orbit the more compact 15-day period EB. The components of this system are likely to interact strongly as indicated by the high and complex amplitude of their ETVs. The asymmetrical shape of the ETV curve with respect to a sine curve suggests the RG orbit is highly eccentric. In addition, the variable eclipse depth as function of time, on a period of about 1800 days, suggests that the orbital plane of the pair of EB is precessing due to tidal interactions with the RG.

\end{description}

\subsubsection{Promising cases}  \label{sec_433}
\begin{figure}
  \epsscale{2}
  \plottwo{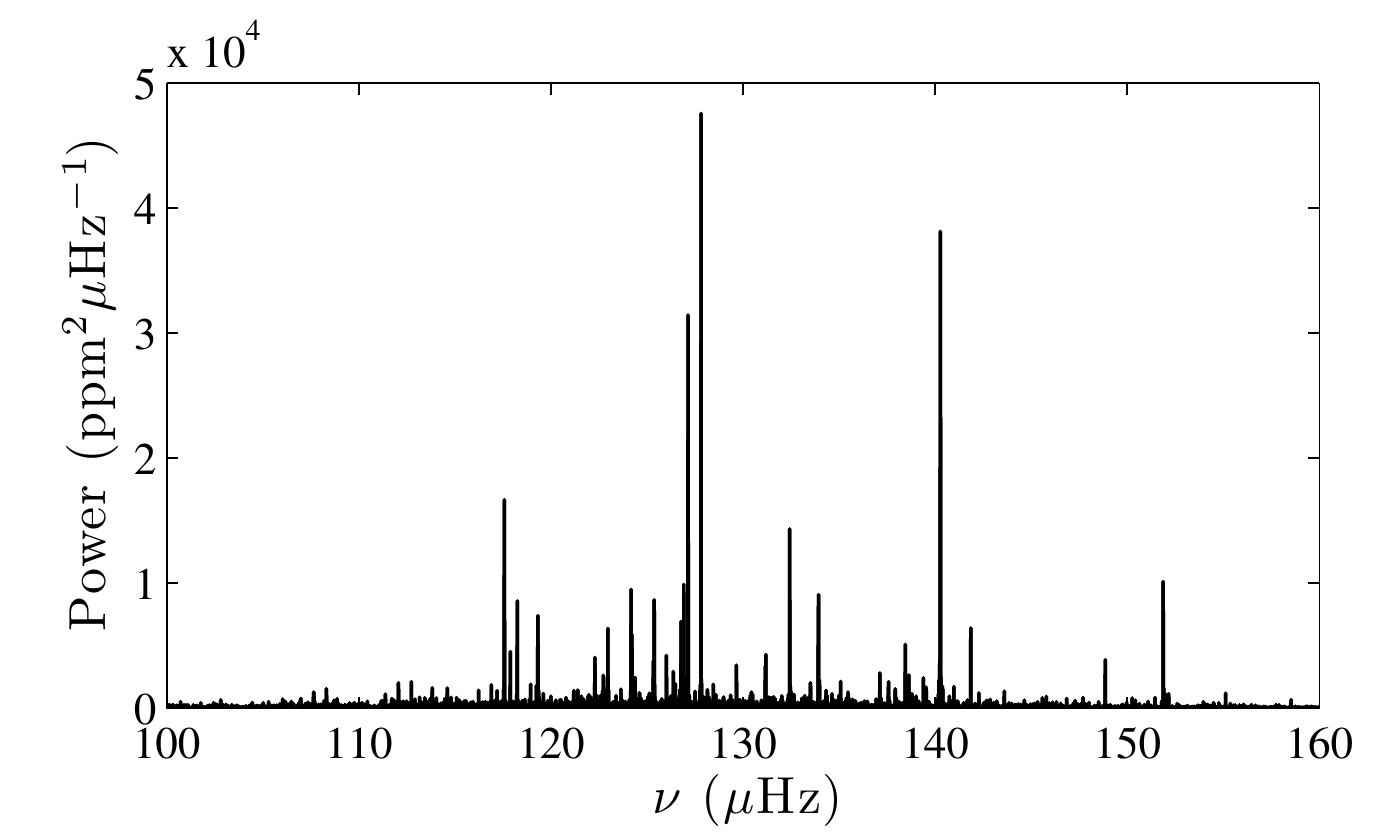}{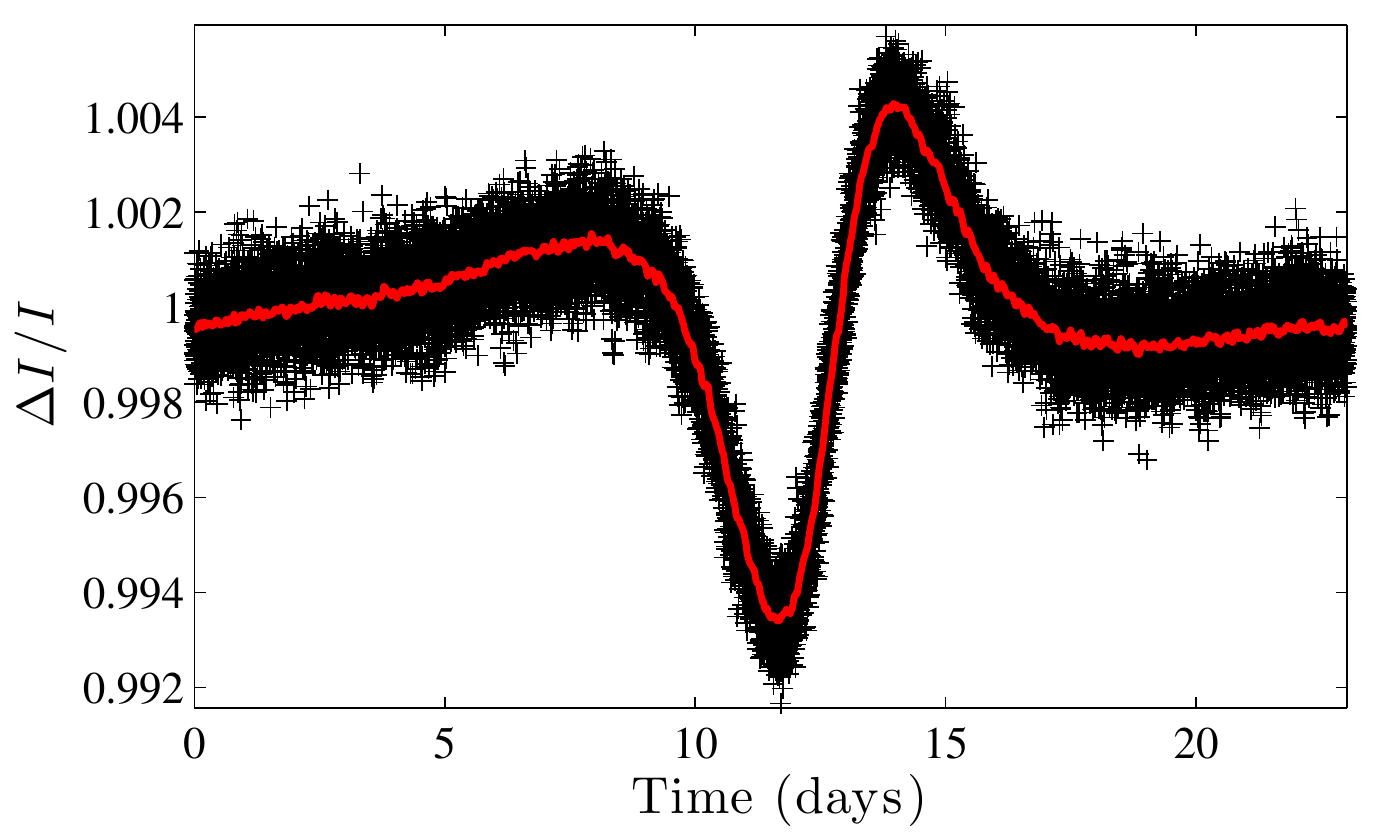}  
  \caption{Top: oscillation spectrum of the RG's companion of KIC 4663185, which is likely a $\delta$-Scuti star. Bottom: folded light curve of the ``heartbeat'' star KIC 8095275, which is the only from our sample to be a RG in a non-eclipsing binary. Black crosses indicate the folded signal, while the red line indicates the same rebinned in 30-min bins.}
  \label{fig_de_trop}
\end{figure}

We sort these potentially fundamental cases by increasing orbital period. 
\begin{description}\itemsep 0cm

\item[KIC 5179609:] This system has a value of $q$ in a plausible range (0.6) with a 43.9-day period. Unfortunately, the global modes of the RG are largely at frequencies higher than the Nyquist frequency, so we are only able to properly measure the mean large separation. Asteroseismic scaling laws infer the RG's mass and radius to be $1.3\pm0.1~M_\sun$ and $3.7\pm0.1~R_\sun$. In addition, we do not find a detectable signature of the secondary eclipse in this EB. Given that the standard deviation of the light curve is 164 ppm and the primary eclipse depth is rather significant (0.92\,\%), the ratio  $R_2^2 T_2^4/(R_1^2 T_1^4 + R_2^2 T_2^4)$ is lower than 1.78\,\%. This is a result of a small or cool companion star. For simplicity, we can model the system as an exoplanetary system, i.e., as a dark planet eclipsing a star. With such an assumption, the radius of the companion would be $0.4 R_\sun$, which could then be an M, L, or brown dwarf. One quarter or \textit{Kepler} data at short cadence would allow us to better constrain this system and make it a cornerstone of for testing asteroseismology. It deserves further study as it hosts the smallest companion star of our sample.

\item[KIC 8095275:] Highly eccentric but non-eclipsing binary systems have been observed by \textit{Kepler}, and are commonly referred to as ``heartbeat stars'' due to the resemblance of their light curves to an electrocardiogram (Fig. \ref{fig_de_trop}, \citealt{Welsh_2011,Thompson_2012}. In these cases, the photometric variability is due to tidally induced distortions generated by the companion star as it passes close to the pericenter. The system KIC 8095275 presents a 23-day orbital period, a clear heartbeat signal of about 1\,\% amplitude, and RG oscillations that correspond to a radius of $7.6\pm0.2 ~R_\sun$ and mass of $1.2\pm0.1~M_\sun$. A more detailed model of the light curve coupled with radial velocity measurements should aid us in characterizing this system. We note that it was also detected in parallel to this work and publicly reported on the Internet, even though it is not yet mentioned in any peer-reviewed paper.\footnote{http://keplerlightcurves.blogspot.com/2012/09/three-giant-heartbeats.html}

\item[KIC 4732015 and 10991989:] These D systems both present clear ETVs with periods on the order of the observation duration (1050 and 544 days, respectively) with a rather small amplitude (6.9 and 3.9 min). Both systems are probably hierarchical triple systems that have a more distant RG orbiting a pair of close main-sequence stars (0.93 and 0.97 day periods, respectively). However, these systems are significantly different physically. KIC 4732015 is an RGB star and is the biggest RG in our sample $(R_1, M_1) = (33.0\pm5.1~R_\sun,1.5\pm0.5~M_\sun)$, while KIC~10991989 is a quiet, massive RC2 star with $(R_1, M_1) = (9.6\pm0.5~R_\sun,2.5\pm0.4~M_\sun)$.

\item[KIC 5640750:] This system is the only one whose orbital period is longer than the total observation time (Q0--Q13; $P\ge 1149.5$~days). It appears that the eclipse shape is similar to those of RGs orbiting F type companions. Modeling the light curve yields a degeneracy between the orbital period and the eccentricity because we see only one primary and one secondary eclipse so far. For our purpose, however, the temperature and radius ratios may still be estimated even if the mass and semi-major axis are not yet. To model the light curve, we assumed the orbital period to be twice the time between the primary and secondary eclipses and set the argument of the periastron to $\omega = \pi/2$, because such an assumption leads to the lower boundary of the eccentricity. From asteroseismic measurements, the RG has radius and mass of $(R_1, M_1) = (14.3\pm0.8~R_\sun,1.4\pm0.2~M_\sun)$. Coupling this result with the preliminary eclipse modeling indicates that the companion seems to be an F star with radius $R_2 = 1.8\pm0.1~R_\sun$  and temperature $T_2=6497$~K, on an elliptical orbit with $e>0.18$. The high value of the RGs asteroseismic mass ($3.2\pm1.0~M_\sun$) suggests its period and semi-major axis are largely underestimated, which also suggests a higher eccentricity.

\end{description}

\subsubsection{Intriguing cases} 
\begin{description}\itemsep 0cm
\item[KIC 3532985 and 6762188:] These D systems both present ETVs with short periods (29 and 35 days, respectively) of small amplitude (3.2 and 6.5 min, respectively), with asymmetrical shapes. At first glance, the ETVs look like artifacts, but several methods to determine the eclipse timing were used and all yielded ETVs in these systems but not in the other targets. The RGs associated with these systems are red clump stars: KIC 3532985 is a RC2 with $(R_1, M_1) = (7.8\pm0.2~R_\sun,1.7\pm0.1~M_\sun)$, while KIC 6762188 is a solar-mass RC1 with $(R_1, M_1) = (10.3\pm0.5~R_\sun,1.0\pm0.1~M_\sun)$.

\item[KIC 8255058:] This D system presents the most puzzling ETV features from our sample. It appears similar to KICs 3532985 and 6762188, but its ETVs actually split into two components. We observe a rapid oscillation of period equal to the EB orbital period, with 5-minute amplitude, as well as a long period (868 days) of similar amplitude (6.7 min). The RG is an RGB star and is one of the smallest RGs in our sample with $(R_1, M_1) = (3.9\pm0.1~R_\sun,1.4\pm0.1~M_\sun)$. 

\item[KIC 7690843 and 7031714:] These SD systems both present clear ETVs with 75 and $> 1000$ day periods, respectively. The amplitude of the ETVs are weak and on the order of a few minutes, i.e., shorter than Kepler's long-cadence sampling. However, KIC 7031714 presents evidence for stellar crowding and blending.

\item[KIC 11135978 and 9181877:] These OC systems display very slowly-varying  ETV trends ($>1400$ day) with low signal-to-noise ratios in amplitude. One would not expect that a third body could significantly perturb the orbital equilibrium of the two contact stars.

\item[KIC 7377422, 4569590, 3955867, 9291629, \& 7943602:] These detached systems present orbital periods, eclipse depths, and eclipse shapes similar to the eight strongest RG/EB candidates where we detect RG pulsations (see Figures \ref{fig_lcmod} and \ref{fig_lcmodbis}). Oddly, the light curves of these four systems do not show pulsations. Their peculiarity, with respect to the pulsating RG/EBs candidates, consists of high quasi-periodic photometric variability (between 10 and 30\,\%). This suggests that either their modes are buried in the noise of photometric variations, or that their mode frequencies are above the Nyquist frequency. Alternatively, the fact that no solar-like oscillations are detected in these objects that display significant stellar variability agrees with \citet{Chaplin_2011a}, which suggests that stellar activity inhibits the amplitude of solar-like oscillations. This scenario does assume that the stellar variability originates from the RG and not the companion, which is unknown.

\end{description}

\section{Conclusion and prospects}\label{sec_5}

We have identified a set of potentially useful targets to test theories of stellar evolution. We are confident that 13 (of 70) systems cross-listed from the RG and EB \textit{Kepler} catalogs are bona-fide pulsating red giants in eclipsing binaries. One is a red giant in a non-eclipsing binary system, and an additional five are likely red giants in eclipsing binaries even though we do not detect their pulsations. It is likely that at least 11 other systems are candidates for belonging to three-body configurations composed of a pair of eclipsing main-sequence stars and a red giant.

Oddly, we detect no oscillations in 23 of the red giant stars in candidate RG/EB systems across all classes of binaries. We identify several reasons that may explain this. First, the star could be misidentified and is not a giant. Second, the treatment used for removing the eclipse signature may suppress any oscillations at very low frequency. Also, the giants could actually be subgiants or low-luminosity red giants pulsating at frequencies beyond the Nyquist frequency. Finally, some phenomena in the interactions of a multiple system may yield a large damping effect of the global oscillations in the RG \citep{Fuller_2013}. Therefore, in total, we believe many of the RG/EB candidates deserve to be followed-up with short-cadence \textit{Kepler} observations. This would allow global modes to be detected in a red giant for which $\nu\ind{max}$ is larger than the long-cadence Nyquist frequency. Ideally, this would allow for pulsations of the main-sequence companions to be measured as well; however, the contribution to the total light from the companion star is only several percent in the best cases, making such detections very challenging in practice.

Spectroscopic measurements from the ground will certainly help in understanding these systems too. First, the identification of overlapping spectra could indicate whether the system is indeed composed of a RG and a main-sequence star. Second, the measurement of radial velocities is a way of extracting accurate masses for the system that can be cross-checked with the asteroseismic inferences. Third, it is a way to refine the estimate of the stellar parameters which are currently based on color photometry. We have already started observations of a subset of these 70 stars with the ARCES \'echelle spectrometer (resolution $R = 30\,000$) at the Apache Point Observatory (APO), New Mexico. Other targets from our sample are currently being observed by the APOGEE spectrometer on the SLOAN Digital Sky Survey (SDSS) telescope at APO, in the context of the APOKASC program to support \textit{Kepler} observations in asteroseismology. This effort is coordinated by teams from the APOGEE project and the Kepler Asteroseismic Science Consortium (KASC). In addition, the KASC Working Group 8 is studying in detail the systems KIC 8410637, 5640750, and 9540226. If more precise estimates of the stellar parameters can be obtained and coupled with eclipse information and asteroseismology, these RG/EB systems have the potential to become some of the most accurately studied stars.

\acknowledgements

We gratefully acknowledge support from the Los Alamos National Laboratory Institute of Geophysics and Planetary Physics subcontract 150623-1. Part of this work was also funded by the \textit{Kepler Guest Observer Program} Cycle 2 project GO 20011, and by a NASA EPSCoR award to NMSU under contract \#~NNX09AP76A. We thank J.~Orosz for guidance using ELC, G. Vigeesh for guidance using JKTEBOP, and S. A. T. Appourchaux for science reviewing.

\bibliographystyle{apj}

\begin{thebibliography}{}

\bibitem[{{Agol} {et~al.}(2005){Agol}, {Steffen}, {Sari}, \&
  {Clarkson}}]{Agol_2005}
{Agol}, E., {Steffen}, J., {Sari}, R., \& {Clarkson}, W. 2005, \mnras, 359, 567

\bibitem[{{Baglin} {et~al.}(2009){Baglin}, {Auvergne}, {Barge}, {Deleuil},
  {Michel}, \& {The CoRoT Exoplanet Science Team}}]{baglin2009}
{Baglin}, A., {Auvergne}, M., {Barge}, P., {Deleuil}, M., {Michel}, E., \& {The
  CoRoT Exoplanet Science Team}. 2009, in IAU Symposium, Vol. 253, IAU
  Symposium, 71--81

\bibitem[{{Beck} {et~al.}(2012){Beck}, {Montalban}, {Kallinger}, {De Ridder},
  {Aerts}, {Garc{\'{\i}}a}, {Hekker}, {Dupret}, {Mosser}, {Eggenberger},
  {Stello}, {Elsworth}, {Frandsen}, {Carrier}, {Hillen}, {Gruberbauer},
  {Christensen-Dalsgaard}, {Miglio}, {Valentini}, {Bedding}, {Kjeldsen},
  {Girouard}, {Hall}, \& {Ibrahim}}]{Beck_2012}
{Beck}, P.~G. {et~al.} 2012, \nat, 481, 55

\bibitem[{{Bedding} {et~al.}(2010){Bedding}, {Huber}, {Stello}, {Elsworth},
  {Hekker}, {Kallinger}, {Mathur}, {Mosser}, {Preston}, {Ballot}, {Barban},
  {Broomhall}, {Buzasi}, {Chaplin}, {Garc{\'{\i}}a}, {Gruberbauer}, {Hale}, {De
  Ridder}, {Frandsen}, {Borucki}, {Brown}, {Christensen-Dalsgaard},
  {Gilliland}, {Jenkins}, {Kjeldsen}, {Koch}, {Belkacem}, {Bildsten}, {Bruntt},
  {Campante}, {Deheuvels}, {Derekas}, {Dupret}, {Goupil}, {Hatzes}, {Houdek},
  {Ireland}, {Jiang}, {Karoff}, {Kiss}, {Lebreton}, {Miglio}, {Montalb{\'a}n},
  {Noels}, {Roxburgh}, {Sangaralingam}, {Stevens}, {Suran}, {Tarrant}, \&
  {Weiss}}]{bedding2010}
{Bedding}, T.~R. {et~al.} 2010, \apjl, 713, L176

\bibitem[{{Bedding} {et~al.}(2011){Bedding}, {Mosser}, {Huber},
  {Montalb{\'a}n}, {Beck}, {Christensen-Dalsgaard}, {Elsworth},
  {Garc{\'{\i}}a}, {Miglio}, {Stello}, {White}, {De Ridder}, {Hekker}, {Aerts},
  {Barban}, {Belkacem}, {Broomhall}, {Brown}, {Buzasi}, {Carrier}, {Chaplin},
  {di Mauro}, {Dupret}, {Frandsen}, {Gilliland}, {Goupil}, {Jenkins},
  {Kallinger}, {Kawaler}, {Kjeldsen}, {Mathur}, {Noels}, {Aguirre}, \&
  {Ventura}}]{bedding2011}
---. 2011, \nat, 471, 608

\bibitem[{{Belkacem} {et~al.}(2011){Belkacem}, {Goupil}, {Dupret}, {Samadi},
  {Baudin}, {Noels}, \& {Mosser}}]{Belkacem_2011}
{Belkacem}, K., {Goupil}, M.~J., {Dupret}, M.~A., {Samadi}, R., {Baudin}, F.,
  {Noels}, A., \& {Mosser}, B. 2011, \aap, 530, A142

\bibitem[{{Borucki} {et~al.}(2010){Borucki}, {Koch}, {Basri}, {Batalha},
  {Brown}, {Caldwell}, {Caldwell}, {Christensen-Dalsgaard}, {Cochran},
  {DeVore}, {Dunham}, {Dupree}, {Gautier}, {Geary}, {Gilliland}, {Gould},
  {Howell}, {Jenkins}, {Kondo}, {Latham}, {Marcy}, {Meibom}, {Kjeldsen},
  {Lissauer}, {Monet}, {Morrison}, {Sasselov}, {Tarter}, {Boss}, {Brownlee},
  {Owen}, {Buzasi}, {Charbonneau}, {Doyle}, {Fortney}, {Ford}, {Holman},
  {Seager}, {Steffen}, {Welsh}, {Rowe}, {Anderson}, {Buchhave}, {Ciardi},
  {Walkowicz}, {Sherry}, {Horch}, {Isaacson}, {Everett}, {Fischer}, {Torres},
  {Johnson}, {Endl}, {MacQueen}, {Bryson}, {Dotson}, {Haas}, {Kolodziejczak},
  {Van Cleve}, {Chandrasekaran}, {Twicken}, {Quintana}, {Clarke}, {Allen},
  {Li}, {Wu}, {Tenenbaum}, {Verner}, {Bruhweiler}, {Barnes}, \&
  {Prsa}}]{borucki2010}
{Borucki}, W.~J. {et~al.} 2010, Science, 327, 977

\bibitem[{{Brown} {et~al.}(2011){Brown}, {Latham}, {Everett}, \&
  {Esquerdo}}]{Brown_2011}
{Brown}, T.~M., {Latham}, D.~W., {Everett}, M.~E., \& {Esquerdo}, G.~A. 2011,
  \aj, 142, 112

\bibitem[{{Carter} {et~al.}(2011){Carter}, {Fabrycky}, {Ragozzine}, {Holman},
  {Quinn}, {Latham}, {Buchhave}, {Van Cleve}, {Cochran}, {Cote}, {Endl},
  {Ford}, {Haas}, {Jenkins}, {Koch}, {Li}, {Lissauer}, {MacQueen}, {Middour},
  {Orosz}, {Rowe}, {Steffen}, \& {Welsh}}]{carter2011}
{Carter}, J.~A. {et~al.} 2011, Science, 331, 562

\bibitem[{{Chaplin} {et~al.}(2011){Chaplin}, {Kjeldsen},
  {Christensen-Dalsgaard}, {Basu}, {Miglio}, {Appourchaux}, {Bedding},
  {Elsworth}, {Garc{\'{\i}}a}, {Gilliland}, {Girardi}, {Houdek}, {Karoff},
  {Kawaler}, {Metcalfe}, {Molenda-{\.Z}akowicz}, {Monteiro}, {Thompson},
  {Verner}, {Ballot}, {Bonanno}, {Brand{\~a}o}, {Broomhall}, {Bruntt},
  {Campante}, {Corsaro}, {Creevey}, {Do{\u g}an}, {Esch}, {Gai}, {Gaulme},
  {Hale}, {Handberg}, {Hekker}, {Huber}, {Jim{\'e}nez}, {Mathur}, {Mazumdar},
  {Mosser}, {New}, {Pinsonneault}, {Pricopi}, {Quirion}, {R{\'e}gulo},
  {Salabert}, {Serenelli}, {Silva Aguirre}, {Sousa}, {Stello}, {Stevens},
  {Suran}, {Uytterhoeven}, {White}, {Borucki}, {Brown}, {Jenkins}, {Kinemuchi},
  {Van Cleve}, \& {Klaus}}]{chaplin2011}
{Chaplin}, W.~J. {et~al.} 2011, Science, 332, 213

\bibitem[{{Chaplin} et~al.(2011){Chaplin}, {Bedding}, {Bonanno}
  et~al.}]{Chaplin_2011a}
{Chaplin} W.J., {Bedding} T.R., {Bonanno} A., et~al., May 2011, \apjl, 732, L5+


\bibitem[{{Coughlin} {et~al.}(2011){Coughlin}, {L{\'o}pez-Morales}, {Harrison},
  {Ule}, \& {Hoffman}}]{coughlin2011}
{Coughlin}, J.~L., {L{\'o}pez-Morales}, M., {Harrison}, T.~E., {Ule}, N., \&
  {Hoffman}, D.~I. 2011, \aj, 141, 78

\bibitem[{{Derekas} {et~al.}(2011){Derekas}, {Kiss}, {Borkovits}, {Huber},
  {Lehmann}, {Southworth}, {Bedding}, {Balam}, {Hartmann}, {Hrudkova},
  {Ireland}, {Kov{\'a}cs}, {Mez{\H o}}, {Mo{\'o}r}, {Niemczura}, {Sarty},
  {Szab{\'o}}, {Szab{\'o}}, {Telting}, {Tkachenko}, {Uytterhoeven}, {Benk{\H
  o}}, {Bryson}, {Maestro}, {Simon}, {Stello}, {Schaefer}, {Aerts}, {ten
  Brummelaar}, {De Cat}, {McAlister}, {Maceroni}, {M{\'e}rand}, {Still},
  {Sturmann}, {Sturmann}, {Turner}, {Tuthill}, {Christensen-Dalsgaard},
  {Gilliland}, {Kjeldsen}, {Quintana}, {Tenenbaum}, \& {Twicken}}]{derekas2011}
{Derekas}, A. {et~al.} 2011, Science, 332, 216

\bibitem[{{Duquennoy} \& {Mayor}(1991)}]{duquennoy1991}
{Duquennoy}, A., \& {Mayor}, M. 1991, \aap, 248, 485

\bibitem[{{Fabrycky} et~al.(2012){Fabrycky}, {Ford}, {Steffen}
  et~al.}]{Fabrycky_2012}
{Fabrycky} D.C., {Ford} E.B., {Steffen} J.H., et~al., May 2012, \apj, 750, 114

\bibitem[{{Ford} et~al.(2012){Ford}, {Fabrycky}, {Steffen} et~al.}]{Ford_2012}
{Ford} E.B., {Fabrycky} D.C., {Steffen} J.H., et~al., May 2012, \apj, 750, 113

\bibitem[{{Fuller} et~al.(2013){Fuller}, {Derekas}, {Borkovits}
  et~al.}]{Fuller_2013}
{Fuller} J., {Derekas} A., {Borkovits} T., et~al., Mar. 2013, \mnras, 429, 2425

\bibitem[{{Gim{\'e}nez}(2006)}]{gimenez2006}
{Gim{\'e}nez}, A. 2006, \aap, 450, 1231

\bibitem[{{Girardi}(1999)}]{Girardi_1999}
{Girardi} L., Sep. 1999, \mnras, 308, 818

\bibitem[{{Harvey}(1985)}]{harvey1985}
{Harvey}, J. 1985, in ESA Special Publication, Vol. 235, Future Missions in
  Solar, Heliospheric \& Space Plasma Physics, ed. E.~{Rolfe} \& B.~{Battrick},
  199--208

\bibitem[{{Hekker} {et~al.}(2010){Hekker}, {Debosscher}, {Huber}, {Hidas}, {De
  Ridder}, {Aerts}, {Stello}, {Bedding}, {Gilliland}, {Christensen-Dalsgaard},
  {Brown}, {Kjeldsen}, {Borucki}, {Koch}, {Jenkins}, {Van Winckel}, {Beck},
  {Blomme}, {Southworth}, {Pigulski}, {Chaplin}, {Elsworth}, {Stevens},
  {Dreizler}, {Kurtz}, {Maceroni}, {Cardini}, {Derekas}, \&
  {Suran}}]{hekker2010}
{Hekker}, S. {et~al.} 2010, \apjl, 713, L187

\bibitem[{{Hekker} {et~al.}(2009){Hekker}, {Kallinger}, {Baudin}, {De Ridder},
  {Barban}, {Carrier}, {Hatzes}, {Weiss}, \& {Baglin}}]{hekker2009}
---. 2009, \aap, 506, 465

\bibitem[{Huber {et~al.}(2011)Huber, Bedding, Stello, Hekker, Mathur, Mosser,
  Verner, Bonanno, Buzasi, Campante, Elsworth, Hale, Kallinger, Aguirre,
  Chaplin, Ridder, García, Appourchaux, Frandsen, Houdek, Molenda-Żakowicz,
  Monteiro, Christensen-Dalsgaard, Gilliland, Kawaler, Kjeldsen, Broomhall,
  Corsaro, Salabert, Sanderfer, Seader, \& Smith}]{huber2011}
Huber, D. {et~al.} 2011, The Astrophysical Journal, 743, 143

\bibitem[{{Huber} {et~al.}(2010){Huber}, {Bedding}, {Stello}, {Mosser},
  {Mathur}, {Kallinger}, {Hekker}, {Elsworth}, {Buzasi}, {De Ridder},
  {Gilliland}, {Kjeldsen}, {Chaplin}, {Garc{\'{\i}}a}, {Hale}, {Preston},
  {White}, {Borucki}, {Christensen-Dalsgaard}, {Clarke}, {Jenkins}, \&
  {Koch}}]{huber2010}
{Huber}, D. {et~al.} 2010, \apj, 723, 1607

\bibitem[{{Jenkins} {et~al.}(2010{\natexlab{a}}){Jenkins}, {Caldwell},
  {Chandrasekaran}, {Twicken}, {Bryson}, {Quintana}, {Clarke}, {Li}, {Allen},
  {Tenenbaum}, {Wu}, {Klaus}, {Middour}, {Cote}, {McCauliff}, {Girouard},
  {Gunter}, {Wohler}, {Sommers}, {Hall}, {Uddin}, {Wu}, {Bhavsar}, {Van Cleve},
  {Pletcher}, {Dotson}, {Haas}, {Gilliland}, {Koch}, \&
  {Borucki}}]{jenkins2010a}
{Jenkins}, J.~M. {et~al.} 2010{\natexlab{a}}, \apjl, 713, L87

\bibitem[{{Jenkins} {et~al.}(2010{\natexlab{b}}){Jenkins}, {Caldwell},
  {Chandrasekaran}, {Twicken}, {Bryson}, {Quintana}, {Clarke}, {Li}, {Allen},
  {Tenenbaum}, {Wu}, {Klaus}, {Van Cleve}, {Dotson}, {Haas}, {Gilliland},
  {Koch}, \& {Borucki}}]{jenkins2010b}
---. 2010{\natexlab{b}}, \apjl, 713, L120

\bibitem[{{Kinemuchi} {et~al.}(2012){Kinemuchi}, {Barclay}, {Fanelli},
  {Pepper}, {Still}, \& {Howell}}]{kinemuchi2012}
{Kinemuchi}, K., {Barclay}, T., {Fanelli}, M., {Pepper}, J., {Still}, M., \&
  {Howell}, S.~B. 2012, \pasp, 124, 963

\bibitem[{{Kjeldsen} \& {Bedding}(1995)}]{kjeldsen1995}
{Kjeldsen}, H., \& {Bedding}, T.~R. 1995, \aap, 293, 87

\bibitem[{{Lada}(2006)}]{lada2006}
{Lada}, C.~J. 2006, \apjl, 640, L63

\bibitem[{{Mandel} \& {Agol}(2002)}]{mandel2002}
{Mandel}, K., \& {Agol}, E. 2002, \apjl, 580, L171

\bibitem[{{Matijevi{\v c}} {et~al.}(2012){Matijevi{\v c}}, {Pr{\v s}a},
  {Orosz}, {Welsh}, {Bloemen}, \& {Barclay}}]{matijevic2012}
{Matijevi{\v c}}, G., {Pr{\v s}a}, A., {Orosz}, J.~A., {Welsh}, W.~F.,
  {Bloemen}, S., \& {Barclay}, T. 2012, \aj, 143, 123

\bibitem[{{Mosser} \& {Appourchaux}(2009)}]{mosser2009}
{Mosser}, B., \& {Appourchaux}, T. 2009, \aap, 508, 877

\bibitem[{{Mosser} {et~al.}(2011){Mosser}, {Barban}, {Montalb{\'a}n}, {Beck},
  {Miglio}, {Belkacem}, {Goupil}, {Hekker}, {De Ridder}, {Dupret}, {Elsworth},
  {Noels}, {Baudin}, {Michel}, {Samadi}, {Auvergne}, {Baglin}, \&
  {Catala}}]{mosser2011}
{Mosser}, B. {et~al.} 2011, \aap, 532, A86

\bibitem[{{Mosser} {et~al.}(2010){Mosser}, {Belkacem}, {Goupil}, {Miglio},
  {Morel}, {Barban}, {Baudin}, {Hekker}, {Samadi}, {De Ridder}, {Weiss},
  {Auvergne}, \& {Baglin}}]{mosser2010}
---. 2010, \aap, 517, A22

\bibitem[{{Mosser} {et~al.}(2012{\natexlab{b}}){Mosser}, {Goupil}, {Belkacem},
  {Marques}, {Beck}, {Bloemen}, {De Ridder}, {Barban}, {Deheuvels}, {Elsworth},
  {Hekker}, {Kallinger}, {Ouazzani}, {Pinsonneault}, {Samadi}, {Stello},
  {Garcia}, {Klaus}, {Li}, {Mathur}, \& {Morris}}]{Mosser_2012c}
---. 2012{\natexlab{b}}, ArXiv e-prints

\bibitem[{{Mosser} {et~al.}(2012{\natexlab{c}}){Mosser}, {Goupil}, {Belkacem},
  {Michel}, {Stello}, {Marques}, {Elsworth}, {Barban}, {Beck}, {Bedding}, {De
  Ridder}, {Garc{\'{\i}}a}, {Hekker}, {Kallinger}, {Samadi}, {Stumpe},
  {Barclay}, \& {Burke}}]{mosser2012}
---. 2012{\natexlab{c}}, \aap, 540, A143

\bibitem[{{Mosser} et~al.(2012{\natexlab{a}}){Mosser}, {Goupil}, {Belkacem}
  et~al.}]{Mosser_2012b}
{Mosser} B., {Goupil} M.J., {Belkacem} K., et~al., Dec. 2012{\natexlab{a}},
  \aap, 548, A10

\bibitem[{{Mosser} et~al.(2012{\natexlab{b}}){Mosser}, {Michel}, {Belkacem}
  et~al.}]{Mosser_2012c}
{Mosser} B., {Michel} E., {Belkacem} K., et~al., Dec. 2012{\natexlab{b}}, ArXiv
  e-prints

\bibitem[{{Orosz} \& {Hauschildt}(2000)}]{orosz2000}
{Orosz}, J.~A., \& {Hauschildt}, P.~H. 2000, \aap, 364, 265

\bibitem[{{Pr{\v s}a} {et~al.}(2011){Pr{\v s}a}, {Batalha}, {Slawson}, {Doyle},
  {Welsh}, {Orosz}, {Seager}, {Rucker}, {Mjaseth}, {Engle}, {Conroy},
  {Jenkins}, {Caldwell}, {Koch}, \& {Borucki}}]{prsa2011}
{Pr{\v s}a}, A. {et~al.} 2011, \aj, 141, 83

\bibitem[{{Pr{\v s}a} {et~al.}(2008){Pr{\v s}a}, {Guinan}, {Devinney},
  {DeGeorge}, {Bradstreet}, {Giammarco}, {Alcock}, \& {Engle}}]{prsa2008}
{Pr{\v s}a}, A., {Guinan}, E.~F., {Devinney}, E.~J., {DeGeorge}, M.,
  {Bradstreet}, D.~H., {Giammarco}, J.~M., {Alcock}, C.~R., \& {Engle}, S.~G.
  2008, \apj, 687, 542

\bibitem[{{Smith} et~al.(2012){Smith}, {Stumpe}, {Van Cleve}
  et~al.}]{Smith_2012}
{Smith} J.C., {Stumpe} M.C., {Van Cleve} J.E., et~al., Sep. 2012, \pasp, 124,
  1000
  
\bibitem[{{Slawson} {et~al.}(2011){Slawson}, {Pr{\v s}a}, {Welsh}, {Orosz},
  {Rucker}, {Batalha}, {Doyle}, {Engle}, {Conroy}, {Coughlin}, {Gregg},
  {Fetherolf}, {Short}, {Windmiller}, {Fabrycky}, {Howell}, {Jenkins}, {Uddin},
  {Mullally}, {Seader}, {Thompson}, {Sanderfer}, {Borucki}, \&
  {Koch}}]{slawson2011}
{Slawson}, R.~W. {et~al.} 2011, \aj, 142, 160

\bibitem[{{Southworth} {et~al.}(2009){Southworth}, {Hinse}, {Dominik},
  {Glitrup}, {J{\o}rgensen}, {Liebig}, {Mathiasen}, {Anderson}, {Bozza},
  {Browne}, {Burgdorf}, {Calchi Novati}, {Dreizler}, {Finet}, {Harps{\o}e},
  {Hessman}, {Hundertmark}, {Maier}, {Mancini}, {Maxted}, {Rahvar}, {Ricci},
  {Scarpetta}, {Skottfelt}, {Snodgrass}, {Surdej}, \&
  {Zimmer}}]{southworth2009}
{Southworth}, J. {et~al.} 2009, \apj, 707, 167

\bibitem[{{Steffen} et~al.(2013){Steffen}, {Fabrycky}, {Agol}
  et~al.}]{Steffen_2013}
{Steffen} J.H., {Fabrycky} D.C., {Agol} E., et~al., Jan. 2013, \mnras, 428,
  1077
  
\bibitem[{{Stumpe} et~al.(2012){Stumpe}, {Smith}, {Van Cleve}
  et~al.}]{Stumpe_2012}
{Stumpe} M.C., {Smith} J.C., {Van Cleve} J.E., et~al., Sep. 2012, \pasp, 124,
  985


\bibitem[{{Stello} {et~al.}(2009){Stello}, {Chaplin}, {Basu}, {Elsworth}, \&
  {Bedding}}]{Stello_2009}
{Stello}, D., {Chaplin}, W.~J., {Basu}, S., {Elsworth}, Y., \& {Bedding}, T.~R.
  2009, \mnras, 400, L80

\bibitem[{{Thompson} et~al.(2012){Thompson}, {Everett}, {Mullally}
  et~al.}]{Thompson_2012}
{Thompson} S.E., {Everett} M., {Mullally} F., et~al., Jul. 2012, \apj, 753, 86

\bibitem[{{Tokovinin} et~al.(2006){Tokovinin}, {Thomas}, {Sterzik}, \&
  {Udry}}]{Tokovinin_2006}
{Tokovinin} A., {Thomas} S., {Sterzik} M., {Udry} S., May 2006, \aap, 450, 681

\bibitem[{{Welsh} et~al.(2011){Welsh}, {Orosz}, {Aerts} et~al.}]{Welsh_2011}
{Welsh} W.F., {Orosz} J.A., {Aerts} C., et~al., Nov. 2011, \apjs, 197, 4

\end{thebibliography}


\clearpage

\LongTables

\begin{landscape}
\begin{deluxetable}{lccccc|ccccccccccc}
\tabletypesize{\tiny}
\tablecaption{Main properties of the 70 systems from light-curve modeling and analysis.\label{table_1}}
\tablewidth{0pt}
\tablehead{
\multicolumn{6}{|c|}{KIC parameters}&
\multicolumn{10}{|c}{Light-curve modeling}\\
\hline\hline
\colhead{KIC} & 
\colhead{Type} &
\colhead{$K\ind{p}$} &
\colhead{$T_{\rm eff}$} & 
\colhead{$\log g$} & 
\colhead{[Fe/[H]} & 
\colhead{$P\ind{orb}$} & 
\colhead{$\frac{\Delta I}{I}$} & 
\colhead{DC\tablenotemark{a}} & 
\colhead{$\frac{T_2}{T_1}$\tablenotemark{b}} &
\colhead{$\frac{R_1}{a}$} & 
\colhead{$\frac{R_2}{a}$} &  
\colhead{$e$} & 
\colhead{$\sin i$} & 
\colhead{ETV (amp.)\tablenotemark{c}} & 
\colhead{ETV (per.)\tablenotemark{d}} & 
\colhead{Result\tablenotemark{e}}\\
&&&      K   &  dex & dex &  day &\%  &\%  & &&&&&min  &day &\\
}
\startdata
  \hline
  \multicolumn{17}{c}{Systems where pulsations are detected}\\
  \hline
 5640750 & D &     11.565 &       4557 &       2.56 &      -0.49 & 1324.260000 &       6.54 &       92.3 &       1.43 &     0.0169 &     0.0021  &$\geq0.18$&      1.000 &     \nodata &     \nodata &RG/EB \\ 
   8410637 & D &     10.771 &       4682 &       2.76 &      -0.34 & 408.350000 &       9.20 &       91.1 &       1.44 &     0.0318 &     0.0046  &       0.67 &      1.000 &     \nodata &     \nodata&RG/EB\\ 
   9970396 & D &     11.447 &       4716 &       2.98 &      -0.25 & 235.299398 &       5.60 &       77.3 &       1.29 &     0.0385 &     0.0053  &       0.20 &      1.000 &     \nodata &     \nodata&RG/EB\\ 
   7037405 & D &     11.875 &       4605 &       2.72 &      -0.40 & 207.108258 &       6.30 &       93.0 &       1.39 &     0.0715 &     0.0091  &       0.23 &      1.000 &     \nodata &     \nodata&RG/EB\\ 
   5786154 & D &     13.534 &       4743 &       3.07 &      -0.26 & 197.920781 &       1.50 &       92.9 &       1.38 &     0.0601 &     0.0090  &       0.38 &      1.000 &     \nodata &     \nodata&RG/EB\\ 
   9540226 & D &     11.672 &       4584 &       2.40 &      -0.42 & 175.458820 &       2.20 &       72.1 &       1.51 &     0.0729 &     0.0073  &       0.39 &      1.000 &     \nodata &     \nodata&RG/EB\\ 
   9246715 & D &      9.266 &       4699 &       2.42 &      -0.39 & 171.277690 &      19.40 &       83.2 &       1.03 &     0.0431 &     0.0334  &       0.35 &      0.999 &     \nodata &     \nodata&RG/EB\\ 
  10001167 & D &     10.050 &       4683 &       2.53 &      -0.50 & 120.395967 &       1.95 &       92.9 &       1.30 &     0.1054 &     0.0081  &       0.16 &      0.999 &     \nodata &     \nodata&RG/EB\\ 
   8430105 & D &     10.420 &       4965 &       2.78 &      -0.60 &  63.327106 &       1.70 &       83.2 &       1.20 &     0.0870 &     0.0098  &       0.26 &      1.000 &     \nodata &     \nodata&RG/EB\\ 
   4663185 & D &     11.356 &       4638 &       2.48 &       0.07 &  56.699076 &       1.00 &       77.3 &       0.00 &     0.1330 &     0.0110  &       0.00 &      0.990 &     \nodata &     \nodata&RG/EB\\ 
   5179609 & D &     12.776 &       4777 &       3.11 &       0.14 &  43.931100 &       0.92 &       93.0 &       0.00 &     0.0536 &     0.0056  &       0.00 &      0.999 &     \nodata &     \nodata&RG/EB\\ 
   5308778 & D &     11.777 &       4812 &       2.57 &      -0.16 &  40.567337 &       0.35 &       84.6 &       0.86 &     0.1629 &     0.0099  &       0.00 &      0.992 &     \nodata &     \nodata&RG/EB\\ 
   8095275 &HB &     13.606 &       4683 &       2.93 &       0.03 &  23.014500 &       1.05 &       92.9 &     \nodata&     \nodata&     \nodata &     \nodata&     \nodata&     \nodata &     \nodata&RG/EB\\ 
   8702921 & D &     11.980 &       4824 &       2.84 &       0.16 &  19.384900 &       0.35 &       83.2 &      0.54 &     0.1320 &     0.0116  &       0.10 &      1.000 &     \nodata &     \nodata&RG/EB\\ 
   7955301 & D &     12.672 &       4821 &       3.12 &      -0.07 &  15.326400 &       3.50 &       92.9 &      (0.71) &     \nodata&     \nodata &     \nodata&     (0.999)&       89.5  &        210 &Triple\\ 
   5218014 & D &     12.923 &       4752 &       2.97 &      -0.09 &  10.845310 &       0.80 &       93.1 &      (0.79)&     \nodata&     \nodata &      (0.13)&     (0.999)&     \nodata &     \nodata & FP\\ 
   6762188 & D &     13.672 &       4801 &       2.97 &       0.01 &   7.155500 &       0.26 &       93.1 &       0.77 &     \nodata&     \nodata &       0.14 &      0.997 &        6.5  &         35 &Triple\\ 
  10809677 & D &     13.942 &       4995 &       2.98 &      -0.07 &   7.042220 &       0.57 &       71.9 &       0.77 &     \nodata&     \nodata &       0.13 &      0.997 &     \nodata &     \nodata & FP\\ 
   8718273 & D &     10.565 &       4577 &       2.13 &      -0.15 &   6.959050 &       0.34 &       82.0 &       0.77 &     \nodata&     \nodata &       0.14 &      0.997 &     \nodata &     \nodata & FP\\ 
   8255058 & D &     13.285 &       4878 &       3.13 &       0.01 &   6.279969 &       0.42 &       83.2 &       0.77 &     \nodata&     \nodata &       0.14 &      0.997 &        6.7  &        868 &Triple\\ 
  12645761 & D &     13.368 &       4844 &       3.18 &      -0.17 &   5.419190 &       1.30 &       83.2 &       0.79 &     \nodata&     \nodata &       0.12 &      0.990 &     \nodata &     \nodata & FP\\ 
   3532985 & D &     11.317 &       4810 &       2.55 &      -0.16 &   5.288530 &       0.22 &       93.0 &       0.77 &     \nodata&     \nodata &       0.14 &      0.997 &        3.2  &         29 &Triple \\ 
   4570555 & D &     11.540 &       4883 &       3.05 &      -0.19 &   4.750300 &       0.07 &       74.6 &       0.76 &     \nodata&     \nodata &       0.15 &      0.997 &     \nodata &     \nodata & FP\\ 
  11147460 & D &     13.912 &       4855 &       3.23 &      -0.43 &   4.107720 &       0.53 &       83.2 &       0.79 &     \nodata&     \nodata &       0.16 &      0.996 &     \nodata &     \nodata & FP\\ 
   6509282 & D &     13.560 &       4812 &       3.11 &       0.01 &   3.989049 &       0.04 &       83.6 &       0.76 &     \nodata&     \nodata &       0.15 &      0.997 &     \nodata &     \nodata & FP\\ 
   8848288 & D &      9.837 &       4624 &       2.54 &      -0.03 &   2.783130 &       0.03 &       92.9 &       0.96 &     \nodata&     \nodata &     \nodata&      0.322 &     \nodata &     \nodata & FP\\ 
  10007492 & D &     12.375 &       5071 &       3.28 &      -0.26 &   2.645600 &       0.18 &       83.2 &       0.77 &     \nodata&     \nodata &       0.14 &      0.997 &     \nodata &     \nodata & FP\\ 
   8453324 & D &     11.516 &       4733 &       2.40 &      -0.32 &   2.524545 &       0.81 &       83.2 &       0.80 &     \nodata&     \nodata &       0.12 &      0.993 &     \nodata &     \nodata & FP\\ 
   5650420 & D &     12.387 &       4611 &       2.59 &      -0.54 &   2.098827 &       0.04 &       85.5 &       0.76 &     \nodata&     \nodata &       0.15 &      0.997 &     \nodata &     \nodata & FP\\ 
   8747222 & D &     12.882 &       4777 &       2.79 &       0.04 &   1.667374 &       0.04 &       92.9 &       0.76 &     \nodata&     \nodata &       0.15 &      0.997 &     \nodata &     \nodata & FP\\ 
   2997455 & D &     13.800 &       4795 &       3.17 &      -0.04 &   1.129852 &       0.49 &       93.1 &       0.78 &     \nodata&     \nodata &       0.14 &      0.996 &     \nodata &     \nodata & FP\\ 
  11968514 & D &     11.449 &       4940 &       3.06 &      -0.51 &   1.036602 &       0.33 &       83.2 &       0.77 &     \nodata&     \nodata &       0.15 &      0.997 &     \nodata &     \nodata & FP\\ 
   5652071 &OC &     13.299 &       4679 &       2.89 &      -0.10 &   1.020465 &       0.26 &       49.1 &       0.96 &     \nodata&     \nodata &     \nodata&      0.341 &     \nodata &     \nodata & FP\\ 
  10991989 & D &     10.282 &       5021 &       2.67 &       0.18 &   0.974480 &       0.92 &       72.1 &       0.78 &     \nodata&     \nodata &       0.14 &      0.993 &        3.9  &        544 & Triple \\ 
   5308777 & D &     13.199 &       4705 &       2.85 &      -0.07 &   0.944740 &       0.02 &       85.6 &       0.95 &     \nodata&     \nodata &     \nodata&      0.320 &     \nodata &     \nodata & FP\\ 
   4732015 & D &     10.147 &       4185 &       1.53 &      -0.13 &   0.938840 &       1.46 &       92.9 &       0.77 &     \nodata&     \nodata &       0.03 &      0.991 &        6.6  &       1050 & Triple  \\ 
  10735519 & D &     11.780 &       4881 &       2.58 &       0.14 &   0.907060 &       0.20 &       83.2 &       0.75 &     \nodata&     \nodata &       0.13 &      0.996 &     \nodata &     \nodata & FP\\ 
   7031714 &SD &     12.126 &       4793 &       2.94 &      -0.24 &   0.814132 &       0.94 &       93.0 &       0.76 &     \nodata&     \nodata &       0.06 &      0.990 &        2.6  &       1015 & Triple  \\ 
   7690843 &SD &     11.083 &       4827 &       3.18 &      -0.15 &   0.786260 &       5.77 &       92.4 &       0.87 &     \nodata&     \nodata &       0.06 &      0.889 &        1.2  &         75 & Triple  \\ 
   6791033 &ELV&     12.385 &       4833 &       2.77 &      -0.30 &   0.758194 &       0.40 &       92.9 &       0.96 &     \nodata&     \nodata &     \nodata&      0.350 &     \nodata &     \nodata & FP\\ 
   2711123 & D &     12.529 &       4723 &       2.91 &      -0.04 &   0.714758 &       0.02 &       93.2 &     \nodata&     \nodata&     \nodata &     \nodata&     \nodata&     \nodata &     \nodata & FP\\ 
   7769072 & D &     13.886 &       4858 &       3.31 &      -0.17 &   0.608864 &       0.21 &       93.1 &       0.74 &     \nodata&     \nodata &       0.14 &      0.997 &     \nodata &     \nodata & FP\\ 
   7879404 &OC &     11.835 &       4291 &       2.10 &      -0.39 &   0.392697 &      12.42 &       93.1 &       1.07 &     \nodata&     \nodata &     \nodata&      0.795 &     \nodata &     \nodata & FP\\ 
   4576968 &OC &     12.537 &       4646 &       3.34 &      -0.87 &   0.378417 &       0.06 &       74.5 &     \nodata&     \nodata&     \nodata &     \nodata&     \nodata&     \nodata &     \nodata & FP\\ 
   4999260 &OC &      9.333 &       5048 &       2.81 &      -0.12 &   0.378369 &       2.27 &       93.0 &       1.02 &     \nodata&     \nodata &     \nodata&      0.476 &     \nodata &     \nodata & FP\\ 
   9181877 &OC &     11.698 &       4599 &       1.93 &      -0.01 &   0.321007 &       1.12 &       83.2 &       1.00 &     \nodata&     \nodata &     \nodata&      0.386 &       30.3  &       5503 & Triple  \\ 
  11135978 &OC &     12.331 &       5004 &       2.55 &      -0.06 &   0.292060 &       0.76 &       56.1 &       0.98 &     \nodata&     \nodata &     \nodata&      0.358 &        0.7  &       1400 & Triple  \\ 
  \hline
  \multicolumn{17}{c}{Systems where no pulsations are detected}\\
  \hline
   7377422 & D &     13.562 &       4488 &       2.79 &      -0.33 & 107.622847 &       7.00 &       92.9 &     \nodata&     \nodata&     \nodata &       1.00 &     \nodata&     \nodata &     \nodata & FP\\ 
   4569590 & D &     12.799 &       4588 &       2.79 &      -0.18 &  41.366979 &       1.60 &       77.2 &       1.33 &     \nodata&     \nodata &       0.00 &      1.000 &     \nodata &     \nodata & FP\\ 
   3955867 & D &     13.547 &       4706 &       3.27 &      -0.11 &  33.662031 &       3.50 &       92.9 &       0.71 &     \nodata&     \nodata &       0.44 &      0.927 &     \nodata &     \nodata & FP\\ 
   9291629 & D &     13.957 &       4629 &       3.10 &       0.09 &  20.686513 &      12.80 &       81.9 &       0.93 &     \nodata&     \nodata &       0.50 &      0.993 &     \nodata &     \nodata & FP\\ 
   4649440 & M &     12.956 &       5109 &       3.27 &      -0.12 &  19.370610 &     \nodata&       93.0 &     \nodata&     \nodata&     \nodata &     \nodata&     \nodata&     \nodata &     \nodata & FP\\ 
   7943602 & D &     13.988 &       4889 &       3.40 &      -0.56 &  14.692016 &       3.20 &       93.1 &       0.45 &     \nodata&     \nodata &       0.50 &      0.859 &     \nodata &     \nodata & FP\\ 
  11671660 & D &     13.350 &       4867 &       3.30 &      -0.29 &   8.710200 &       2.00 &       83.2 &      (0.86)&     \nodata&     \nodata &      (0.51)&     (0.918)&       49.5  &      14200 & Triple  \\ 
   9489411 &SD &     13.960 &       4499 &       2.22 &      -0.18 &   6.689000 &      63.00 &       83.2 &      (0.99)&     \nodata&     \nodata &      (0.06)&     (0.956)&     \nodata &     \nodata & FP\\ 
   8893936 &SD &     13.973 &       4697 &       3.07 &      -0.29 &   4.244440 &      18.70 &       55.7 &      (0.67)&     \nodata&     \nodata &      (0.03)&     (0.960)&     \nodata &     \nodata & FP\\ 
   4758368 & D &     10.805 &       4594 &       2.62 &      -0.47 &   3.749990 &       3.66 &       81.2 &      (0.90)&     \nodata&     \nodata &      (0.02)&     (0.968)&      113.2  &       4266 & Triple  \\ 
   8719897 & D &     12.392 &       4905 &       3.05 &       0.05 &   3.151430 &      20.00 &       83.2 &      (0.79)&     \nodata&     \nodata &      (0.06)&     (0.987)&     \nodata &     \nodata & FP\\ 
   5823121 &SD &     13.890 &       5283 &       3.37 &      -0.28 &   2.297910 &       2.10 &       93.1 &      (0.53)&     \nodata&     \nodata &      (0.01)&     (0.973)&     \nodata &     \nodata & FP\\ 
   8846978 & D &     13.371 &       5191 &       3.24 &      -1.09 &   1.379000 &       1.80 &       83.2 &      (1.04)&     \nodata&     \nodata &      (0.11)&     (0.980)&     \nodata &     \nodata & FP\\ 
   8452840 &OC &     12.594 &       6473 &       3.08 &      -0.07 &   1.201347 &       0.30 &       83.2 &      (0.97)&     \nodata&     \nodata &     \nodata&     (0.315)&     \nodata &     \nodata & FP\\ 
   7286410 &ELV&     13.053 &       4667 &       3.32 &      -0.58 &   0.997170 &      16.00 &       93.0 &     \nodata&     \nodata&     \nodata &     \nodata&     \nodata&     \nodata &     \nodata & FP\\ 
   5820209 &OC &     13.495 &       4750 &       3.17 &      -0.95 &   0.656084 &      10.10 &       93.1 &      (0.94)&     \nodata&     \nodata &     \nodata&     (0.605)&     \nodata &     \nodata & FP\\ 
   3241248 &OC &     13.211 &       4887 &       3.37 &      -0.11 &   0.489025 &       0.14 &       92.9 &      (0.96)&     \nodata&     \nodata &     \nodata&     (0.330)&     \nodata &     \nodata & FP\\ 
   9091810 &SD &     12.755 &       4807 &       3.35 &       0.06 &   0.479722 &      52.10 &       83.2 &      (0.90)&     \nodata&     \nodata &      (0.01)&     (0.992)&     \nodata &     \nodata & FP\\ 
   4909422 &OC &     12.135 &       5404 &       2.91 &       0.32 &   0.395103 &      10.29 &       93.2 &      (1.08)&     \nodata&     \nodata &     \nodata&     (0.784)&     \nodata &     \nodata & FP\\ 
   9833806 &SD &     13.497 &       4942 &       3.38 &       0.16 &   0.393176 &       1.60 &       73.7 &       (0.89)&     \nodata&     \nodata &      (0.06)&     (0.975)&     \nodata &     \nodata & FP\\ 
  11716688 &ELV&     13.647 &       5130 &       3.28 &       0.31 &   0.301214 &      14.00 &       83.2 &      (0.95)&     \nodata&     \nodata &     \nodata&     (0.322)&     \nodata &     \nodata & FP\\ 
   9300595 &OC &     13.294 &       5035 &       3.08 &      -0.23 &   0.252354 &       0.37 &       93.0 &      (0.97)&     \nodata&     \nodata &     \nodata&     (0.312)&     \nodata &     \nodata & FP\\ 
   3731292 &OC &     12.494 &       6662 &       3.47 &      -0.10 &   0.175117 &       1.05 &       93.0 &      (0.96)&     \nodata&     \nodata &     \nodata&     (0.352)&     \nodata &     \nodata & FP\\ 
\enddata
\tablenotetext{a}{Duty cycle}
\tablenotetext{b}{Our convention is that $T_1$ is always the RG. For the values cited from \citet{slawson2011}, we suspect their convention is the inverse, as they likely take the hotter (non-RG) star to be $T_1$. }
\tablenotetext{c}{Amplitude of measured ETV}
\tablenotetext{d}{Period of measured ETV}
\tablenotetext{e}{Results from the complete analysis: ``RG/EB'' indicate systems that are likely EBs where at least one component is a RG, ``Triple'' indicate candidate triple systems, and ``F P' indicate false positives. }
\tablecomments{Stars below the horizontal line have not been confirmed to be pulsating. All parameters in parenthesis (\ldots) are reproduced from \citet{slawson2011}. The star with Type ``M'' is likely a flaring M dwarf. Two stars have temperature over 6000~K and are likely misidentified RGs. See also http://nsol2.nmsu.edu/solarstorm/index.php.}
\end{deluxetable}
\clearpage
\end{landscape}

\clearpage
\begin{deluxetable}{c|lc|cccccccccc}
\tabletypesize{\scriptsize}
\tablecaption{Derived properties of the 47 pulsating systems from combined asteroseismic and light-curve analysis.\label{table_2}}
\tablewidth{0pt}
\tablehead{
Result&
\multicolumn{2}{|c|}{KIC parameters}
&
\multicolumn{10}{|c}{Asteroseismic analysis and combined output}\\
\hline\hline
\colhead{} & 
\colhead{KIC} & 
\colhead{Type} &
\colhead{$\nu\ind{max}$}& 
\colhead{$\Delta\nu$}& 
\colhead{$\Delta\Pi_1$}&
\colhead{$P_{\rm rot}$\tablenotemark{a}} &
\colhead{${M_1}$} & 
\colhead{${R_1}$} & 
\colhead{${M_2}$} & 
\colhead{${R_2}$} & 
\colhead{$a$} &
\colhead{Class.\tablenotemark{b}} \\
&&       
\colhead{$\mu$Hz} &
\colhead{$\mu$Hz} &
\colhead{s} &
\colhead{day} &
\colhead{$M_\odot$} &
\colhead{$R_\odot$}&
\colhead{$M_\odot$}&
\colhead{$R_\odot$}&
\colhead{$R_\odot$} &
}
\startdata
  \multirow{14}{*}{RG/EB} &    5640750 & D  &   24.9   &    3.0 &  \nodata& \nodata& $    1.4 \pm    0.2$ & $   14.3 \pm    0.8$ & \nodata & $    1.8 \pm    0.1 $ &  845.2 \\ 
&      8410637 & D  &   46.0   &    4.6 &  \nodata& \nodata& $    1.6 \pm    0.2$ & $   11.0 \pm    0.5$ & $    1.8 \pm    0.7 $ & $    1.6 \pm    0.1 $ &  347.0& RGB \\ 
 &     9970396 &D  &   63.9   &    6.3 &  \nodata& \nodata& $    1.3 \pm    0.1$ & $    8.3 \pm    0.3$ & $    1.2 \pm    0.3 $ & $    1.1 \pm    0.1 $ &  215.2 \\ 
 &     7037405 & D  &   22.0   &    2.7 &  \nodata& \nodata& $    1.4 \pm    0.2$ & $   15.0 \pm    0.9$ & $    1.5 \pm    0.7 $ & $    1.9 \pm    0.1 $ &  210.1 \\ 
 &     5786154 & D  &   30.6   &    3.5 &  \nodata& \nodata& $    1.4 \pm    0.2$ & $   12.7 \pm    0.6$ & $    1.8 \pm    0.7 $ & $    1.9 \pm    0.1 $ &  211.9 \\ 
 &     9540226 & D  &   28.1   &    3.2 &  \nodata& \nodata& $    1.6 \pm    0.2$ & $   14.0 \pm    0.7$ & $    1.5 \pm    0.7 $ & $    1.4 \pm    0.1 $ &  192.1 & RGB/AGB \\ 
 &      9246715 & D  &  102.2   &    8.3 &  \nodata& \nodata& $    1.7 \pm    0.3$ & $    7.7 \pm    0.4$ & $    0.8 \pm    0.7 $ & $    5.9 \pm    0.3 $ &  177.9 & RGB \\ 
 &    10001167 & D  &   20.3   &    2.7 &  \nodata& \nodata& $    1.1 \pm    0.2$ & $   14.0 \pm    0.8$ & $    1.0 \pm    0.6 $ & $    1.1 \pm    0.1 $ &  132.4&RGB/AGB \\ 
 &     8430105 & D  &   71.2   &    7.2 &  \nodata& \nodata& $    1.2 \pm    0.1$ & $    7.4 \pm    0.3$ & $    0.9 \pm    0.4 $ & $    0.8 \pm    0.1 $ &   85.0 & RGB\\ 
 &    4663185 & D  &   23.4   &    2.6 &  \nodata& \nodata& $    2.2 \pm    0.4$ & $   18.2 \pm    1.2$ & $    8.5 \pm    2.7 $ & $    1.5 \pm    0.1 $ &  136.7 \\ 
 &     5179609 & D  &  342.5   &   22.1 &  \nodata& \nodata& $    1.3 \pm    0.1$ & $    3.7 \pm    0.1$ &   \nodata & $    0.4 \pm    0.1 $ &  \nodata & RGB\\ 
 &     5308778 & D  &   48.6   &    5.2 &  69.4& 61& $    1.3 \pm    0.3$ & $    9.4 \pm    0.8$ & $    0.3_{-0.3}^{+0.7} $ & $    0.6 \pm    0.1 $ &   57.7 & RGB\\ 
 &     8095275 &HB  &   68.5   &    6.8 &  \nodata& \nodata& $    1.2 \pm    0.1$ & $    7.6 \pm    0.2$   & \nodata     & \nodata  &    \nodata \\ 
 &     8702921 & D  &  192.2   &   14.1 &  \nodata& \nodata& $    1.4 \pm    0.1$ & $    5.1 \pm    0.1$ & $    0.6 \pm    0.3 $ & $    0.4 \pm    0.1 $ &   38.4 & RGB\\ 
 \hline
 \multirow{8}{*}{Triple} &      7955301 & D  &  123.0   &   10.5 &   76.5 &   30 & $    1.2 \pm    0.1$ & $    5.9 \pm    0.2$ &     \nodata          &     \nodata          &\nodata & RGB\\ 
 &      6762188 & D  &   30.9   &    4.0 &  300.0  &  144& $    1.0 \pm    0.1$ & $   10.3 \pm    0.5$ &     \nodata          &     \nodata          &\nodata & RC1\\ 
&      8255058 & D  &  309.8   &   20.5 &  \nodata& \nodata& $    1.4 \pm    0.1$ & $    3.9 \pm    0.1$ &     \nodata          &     \nodata          &\nodata & RGB \\ 
&      3532985 & D  &   94.1   &    8.0 &  235.0  &  165 & $    1.7 \pm    0.1$ & $    7.8 \pm    0.2$ &     \nodata          &     \nodata          &\nodata  & RC2\\ 
&     10991989 & D  &   89.9   &    7.1 &  \nodata& \nodata& $    2.5 \pm    0.4$ & $    9.6 \pm    0.5$ &     \nodata          &     \nodata          &\nodata & RC2\\ 
&      4732015 & D  &    4.9   &    0.8 &  \nodata& \nodata& $    1.5 \pm    0.5$ & $   33.0 \pm    5.1$ &     \nodata          &     \nodata          &\nodata & RGB\\ 
&      7031714 &SD  &   29.1   &    3.7 &  318.8  &  257 & $    1.1 \pm    0.1$ & $   11.2 \pm    0.5$ &     \nodata          &     \nodata          &\nodata & RC1\\ 
&      7690843 &SD  &   55.1   &    4.9 &  \nodata& \nodata& $    2.3 \pm    0.3$ & $   12.0 \pm    0.6$ &     \nodata          &     \nodata          &\nodata \\ 
\hline
Triple &      9181877 &OC  &   53.8   &    5.2 &  \nodata& \nodata& $    1.5 \pm    0.2$ & $   10.0 \pm    0.4$ &     \nodata          &     \nodata          &\nodata \\ 
\& FP&     11135978 &OC  &   34.0   &    4.1 &  296.0  &  165 & $    1.1 \pm    0.1$ & $   10.6 \pm    0.4$ &     \nodata          &     \nodata          &\nodata & RC1\\ 
\hline
 \multirow{23}{*}{FP}&      5218014 & D  &   62.5   &    5.7 &  \nodata& \nodata& $    1.9 \pm    0.3$ & $   10.1 \pm    0.6$  &     \nodata          &     \nodata          &\nodata \\ 
&     10809677 & D  &   93.1   &    8.2 &  \nodata& \nodata& $    1.5 \pm    0.1$ & $    7.5 \pm    0.2$ &     \nodata          &     \nodata          &\nodata \\ 
&      8718273 & D  &   28.3   &    3.1 &  \nodata& \nodata& $    1.9 \pm    0.3$ & $   15.2 \pm    0.8$ &     \nodata          &     \nodata          &\nodata \\ 
&     12645761 & D  &   31.0   &    4.2 &  \nodata& \nodata& $    0.8 \pm    0.1$ & $    9.4 \pm    0.5$ &     \nodata          &     \nodata          &\nodata \\ 
&      4570555 & D  &  191.9   &   14.8 &   84.0 &   31 & $    1.2 \pm    0.1$ & $    4.6 \pm    0.1$ &     \nodata          &     \nodata          &\nodata  & RGB\\ 
&     11147460 & D  &   15.2   &    2.0 &  \nodata& \nodata& $    1.6 \pm    0.3$ & $   19.0 \pm    1.6$ &     \nodata          &     \nodata          &\nodata \\ 
&      6509282 & D  &   39.1   &    4.3 &  279.2& 385& $    1.3 \pm    0.1$ & $   10.8 \pm    0.5$ &      \nodata           &      \nodata          &\nodata & RC1\\ 
&      8848288 & D  &   46.1   &    4.6 &  \nodata& \nodata& $    1.6 \pm    0.2$ & $   11.1 \pm    0.4$ &     \nodata           &      \nodata          & \nodata \\ 
&     10007492 & D  &  214.7   &   15.6 &  \nodata& \nodata& $    1.4 \pm    0.1$ & $    4.7 \pm    0.1$ &     \nodata          &     \nodata          &\nodata & RGB\\ 
&      8453324 & D  &   65.5   &    6.0 &  214.0  &  116 & $    1.7 \pm    0.2$ & $    9.4 \pm    0.3$ &     \nodata          &     \nodata          &\nodata \\ 
&       5650420 & D  &   33.5   &    3.8 &  \nodata& \nodata& $    1.4 \pm    0.2$ & $   12.1 \pm    0.6$ &     \nodata          &      \nodata           &\nodata \\ 
&      8747222 & D  &   67.0   &    6.8 &  \nodata& \nodata& $    1.1 \pm    0.1$ & $    7.5 \pm    0.2$ &      \nodata           &      \nodata          & \nodata \\ 
&      2997455 & D  &  130.1   &   11.0 &   77.98 &   46 & $    1.2 \pm    0.1$ & $    5.6 \pm    0.1$ &     \nodata          &     \nodata          &\nodata & RGB\\ 
&     11968514 & D  &   31.9   &    4.2 &  \nodata& \nodata& $    0.9 \pm    0.1$ & $    9.6 \pm    0.4$ &     \nodata          &     \nodata          &\nodata & RC1\\ 
&       5652071 &OC  &   57.7   &    5.8 &  \nodata& \nodata& $    1.3 \pm    0.1$ & $    9.0 \pm    0.3$ &      \nodata           &     \nodata          &\nodata \\ 
 &       5308777 & D  &   84.0   &    7.8 &  \nodata& \nodata& $    1.2 \pm    0.1$ & $    7.1 \pm    0.2$ &      \nodata          &     \nodata          &\nodata \\ 
&     10735519 & D  &   27.7   &    2.8 &  \nodata& \nodata& $    2.8 \pm    0.4$ & $   18.6 \pm    1.1$ &     \nodata          &     \nodata          &\nodata & RGB\\ 
&      6791033 &ELV &   34.9   &    4.3 &  284& 289& $    1.0 \pm    0.1$ & $    9.9 \pm    0.4$ &     \nodata          &     \nodata          &  \nodata & RC1 \\ 
&      2711123 & D  &   38.3   &    4.1 &  \nodata& \nodata& $    1.6 \pm    0.2$ & $   12.0 \pm    0.5$ &     \nodata          &     \nodata           &  \nodata \\ 
&      7769072 & D  &  340.4   &   22.0 &  \nodata& \nodata& $    1.4 \pm    0.1$ & $    3.7 \pm    0.1$ &     \nodata          &     \nodata          &\nodata & RGB\\ 
&      7879404 &OC  &    6.5   &    1.1 &  \nodata& \nodata& $    1.2 \pm    0.3$ & $   25.6 \pm    3.1$ &     \nodata          &     \nodata          &\nodata & RGB\\ 
&      4576968 &OC  &  121.4   &   10.5 &  \nodata& \nodata& $    1.1 \pm    0.1$ & $    5.7 \pm    0.1$ &     \nodata          &     \nodata          &\nodata & RGB\\ 
&      4999260 &OC  &   76.6   &    6.8 &  296.5  &  148.4 & $    1.8 \pm    0.2$ & $    8.9 \pm    0.3$ &     \nodata          &     \nodata          &\nodata & RC2\\ 
\enddata
\tablenotetext{a}{Core rotation period determined from mixed-mode analysis}
\tablenotetext{b}{Asteroseismic RG classification: Red-giant branch (RGB); Asymptotic red-giant branch (AGB); Red clump 1 (RC1); Red clump 2 (RC2)}
\tablecomments{Only the first 13 EB light curves were modeled, as these are the most likely true RG-EB systems. See also http://nsol2.nmsu.edu/solarstorm/index.php.}
\end{deluxetable}


\clearpage

\end{document}